\documentclass[twocolumn]{revtex4-2}
\usepackage{comment}
\newcommand{\id}{\mathbbm{1}}

\usepackage{graphicx}
\usepackage{braket}
\usepackage{amsmath,amssymb,mathtools}
\usepackage{bm}
\usepackage{gensymb}
\usepackage{adjustbox}
\usepackage{float}
\usepackage[T1]{fontenc}
\usepackage[utf8]{inputenc}
\usepackage{parskip}
\usepackage[normalem]{ulem}
\usepackage{bbm}
\usepackage[colorlinks=true,allcolors=blue]{hyperref}

\renewcommand{\vec}[1]{\mathbf{#1}}
\usepackage{placeins}

\begin{document}
\title{Escaping the Composite Fermi Sea: An Incompressible State at Half Filling in Graphene’s Lowest Landau Level}

\author{Domagoj Perković}
\email{domagoj.perkovic@physics.ox.ac.uk}
\author{Ziwei Wang}
\altaffiliation{Present address: Condensed Matter Theory Center,
Department of Physics, University of Maryland, College Park, Maryland 20742, USA}
\author{Steven H. Simon}

\affiliation{Rudolf Peierls Centre for Theoretical Physics, Parks Road, Oxford, OX1 3PU, UK}
\date{\today}
\begin{abstract}
Recent experiments in monolayer and trilayer graphene have observed incompressible even-denominator fractional quantum Hall states within narrow regions of parameter space associated with crossings of zeroth Landau levels. This is surprising because at half filling the Coulomb interaction in the zeroth Landau level is expected to favour a compressible composite Fermi liquid. Here we provide the first quantitative energetic study of the candidate incompressible phases and of the microscopic mechanisms that can stabilise them near such Landau-level crossings. Using exact diagonalization and infinite density-matrix renormalisation group, we first study ultra-short-range corrections to the Coulomb interaction arising from finite layer separation, lattice-scale effects, and electron–phonon coupling. We find that a sufficiently strong attractive short-range interaction drives the composite Fermi liquid first into a weakly paired \(d\)-wave pseudospin-singlet state and subsequently into a strongly paired state. However, microscopic estimates of these short-range corrections are far too small to stabilise either incompressible phase. We then incorporate Landau-level mixing through screening within the random-phase approximation. Screening substantially reshapes the effective interaction, suppresses the weak-pairing regime, and strongly enhances the tendency toward strong pairing. For a realistic range of microscopic parameters, the strongly paired phase is stabilised over a substantial fraction of that range. Our results identify the interplay of Landau-level mixing and ultra-short-range interactions as a plausible microscopic mechanism for the experimentally observed incompressible states in monolayer and trilayer graphene and point toward a strongly paired composite-fermion phase as their candidate ground state.
\end{abstract}

\maketitle

\section{Introduction}

The fractional quantum Hall effect (FQHE) provides a particularly rich setting for studying strongly correlated phases of matter, with graphene offering a unique platform due to the interplay of its Landau-level structure and internal spin and valley degrees of freedom
\cite{Bolotin2009,TokeJain2007}. The latter gives rise to an approximate SU(4) symmetry within each Landau level \cite{TokeJain2007,SodemannMacDonald2014}, enabling a variety of correlated states with different spin and valley polarization structures \cite{Zibrov2018,Huang2022,trilayer_expt}. In multilayer systems, the additional layer degree of freedom further enriches this internal structure and will be of particular importance in the cases considered below.  These internal degrees of freedom can be selectively controlled by external fields and other experimentally tunable parameters, providing a versatile setting in which to explore multicomponent quantum Hall physics. Moreover, because both the Landau-level spacing in graphene and the characteristic Coulomb interaction energy scale as $\sqrt{B}$, Landau-level mixing remains relevant even at high magnetic fields \cite{PetersonNayak2014,Sokolik2017} and can significantly modify the effective interactions between electrons. Together, these features have led to the observation of a broad range of fractional quantum Hall states in graphene-based systems \cite{Bolotin2009}. Of particular interest are even-denominator fractional quantum Hall states \cite{Zibrov2018,Kim2019,Huang2022,trilayer_expt}, which lie outside the usual composite-fermion hierarchy and are therefore natural candidates for more exotic correlated phases.

In this work, we focus on a set of recent experiments in graphene at half-integer filling, in particular the trilayer graphene experiment of Refs.~\cite{chanda,trilayer_expt} and the monolayer graphene experiment of Ref.~\cite{Zibrov2018}. 
A striking feature of these measurements is the appearance of incompressible, apparently gapped states only within relatively narrow regions of the parameter space. In the trilayer system, for example, these regions can be accessed by tuning the displacement-field parameter $\Delta_1$. Notably, no analogous incompressible state was observed in bilayer graphene~\cite{Bilayer}.

Understanding why the gapped phase is confined to such narrow windows provides an important clue to its microscopic origin.
In the bare single-particle Landau-level spectrum, the observed gapped states occur at half filling of a zeroth Landau level (0LL). This is unusual, as the natural state at half filling for an isolated 0LL with approximately Coulomb interactions is generally expected to be a gapless composite Fermi liquid (CFL)  \cite{HalperinLeeRead1993,Morf1998,Geraedts2016,CFL_Haldane_rezayi}.

The experimental observation of a robust gap therefore suggests that the physics cannot be understood simply in terms of an isolated Landau level with an unscreened Coulomb interaction.
Previous theoretical work has attributed the emergence of the gapped state to a crossing between distinct zeroth Landau levels \cite{chanda,trilayer_expt}. Near such a crossing, additional internal degrees of freedom become relevant and can qualitatively alter the competition between many-body phases. Nevertheless, for a model containing only the Coulomb interaction, the expected ground state near the crossing is an unpolarised/partially polarised CFL, which is gapless \cite{CFL_N_8,CFL_Onoda2001,CFL_RezayiRead1994,sarma_CFL_Monte_carlo,BalramJain2017}.

The experimentally observed incompressible phase must therefore rely on physics beyond a model of two crossing 0LLs interacting solely through the Coulomb interaction.
Several candidate states \cite{chanda,trilayer_expt,parton,Zibrov2018} have been proposed for the observed gapped phase, including the Moore-Read Pfaffian \cite{MooreRead1991} and anti-Pfaffian states \cite{PHPfaffian1,PHpfaffian2}, weakly and strongly paired $d$-wave pseudospin-singlet states \cite{JSS}, and the Halperin 331 state \cite{Halperin_states}. Henceforth, we refer to the Moore–Read Pfaffian state simply as the Pfaffian state.

 The $d$-wave states considered here are paired states of composite fermions in which the two components form a pseudospin singlet, with pairing in the $d$-wave channel \cite{milovanovic,Read_2000,kun_yang_weak_to_strong}.
 Two topologically distinct Abelian phases can arise: a weak-pairing phase, topologically equivalent to the Jain singlet state \cite{JSS}, and a strong-pairing phase characterised by tightly bound pairs.

A number of microscopic mechanisms have likewise been suggested as possible sources for perturbations to the Hamiltonian that could favour these pairing phases. Despite this variety of proposals, the magnitudes of the corresponding microscopic effects have not, to our knowledge, been systematically estimated and compared on equal footing.
A central goal of this work is to determine which of these effects are quantitatively important, how they modify the effective interaction near the Landau-level crossing, and which of the candidate trial states is ultimately realised as the ground state. 

We organise these corrections into two classes. The first consists of short-range modifications to the interaction arising from electron--phonon coupling, the finite separation between graphene layers, the finite spatial extent of the electron wave function associated with a lattice site, and the discrete nature of the graphene lattice. At the length scales relevant to the quantum Hall problem, these contributions can be represented approximately as point-like perturbations to the dominant long-range Coulomb interaction \cite{parton,Alicea_2006, kharitonov,zaletel_phonons}. Although small in bare magnitude, such perturbations can play an important role in selecting between competing states that are nearly degenerate under the Coulomb interaction alone.
A second class of corrections arises from Landau-level mixing. We treat this effect by including screening from the filled and empty Landau levels using the random-phase approximation (RPA) \cite{kharitonov, snizhko2015,abanin_papic,Yuan_RPA,Ando2006Screening_RPA,NomuraMacDonald2006_RPA,PyatkovskiyGusynin2011_RPA,Luo_2022}.

We find that RPA screening produces the largest modification to the effective interaction. Its role is twofold. First, it reshapes the momentum dependence of the dominant interaction itself and therefore changes the pseudopotentials governing the competition between correlated states. Second, it renormalises the various short-range perturbations differently from the long-range Coulomb interaction. Consequently, the relative strength of the pairing-favouring terms can be substantially enhanced or suppressed after screening is taken into account.

To determine the many-body phases stabilised by the resulting effective interactions, we employ complementary numerical approaches: infinite density matrix renormalization group (iDMRG) calculations on an infinite cylinder and exact diagonalization (ED) on a torus. We generally find that the Halperin 331 and Pfaffian/anti-Pfaffian states are unlikely to be realised. Instead, the ground state is consistent with either a weakly- or strongly-paired $d$-wave pseudospin-singlet state, depending on the microscopic parameters.

For most of the paper we focus on the trilayer case, and discuss the
analogous monolayer case, along with its extra difficulties, near the
end in Section~\ref{sec:extension_monolayer}.

The remainder of this paper is organised as follows. In Sec.~\ref{sec:model}, we introduce the effective Hamiltonian and describe the numerical methods used: exact diagonalisation and the infinite density-matrix renormalisation group (iDMRG).
 In Sec.~\ref{sec:topo_phases_identification}, we introduce the numerical diagnostics used to identify different topological phases and review the topological properties of the candidate states.

 In Sec.~\ref{sec:toy_model}, we study a minimal two-band model supplemented by a short-range point-like interaction $\Delta V_0$.
We show that an attractive $\Delta V_0$ can stabilise both weak- and strong-paired $d$-wave singlet phases, with a transition between the two pairing regimes analogous to one studied previously in the context of cold atoms \cite{kun_yang_weak_to_strong}. However, the magnitude of $\Delta V_0$ required to access these phases is substantially larger than what can be generated by the bare microscopic perturbations considered here.  In Sec.~\ref{sec:realistic_interaction}, we determine the magnitudes of the ultra-short-range corrections to the Hamiltonian and incorporate screening within the random-phase approximation (RPA). We find that RPA screening can either enhance or suppress these short-range perturbations relative to the dominant zeroth pseudopotential, depending sensitively on the structure of the polarization matrix.
In Sec.~\ref{sec:realistic_phase_diagram}, we combine RPA screening with the microscopic perturbations and study the resulting effective Hamiltonian. We show that, for realistic parameter values, a gapped $d$-wave strongly paired state can plausibly be stabilised. In Section~\ref{sec:extension_monolayer} we discuss
the extension of these results to the monolayer case. 
In Conclusion~\ref{sec:outlook}, we discuss directions for future work, in particular how Landau-level mixing can be treated more accurately beyond the RPA. Furthermore, we discuss possible experimental probes of our theory.

\medskip
\noindent

\section{Model and Numerical Methods}\label{sec:model}

From now on, unless stated otherwise, we consider the trilayer graphene setting realised in the experiment of Ref.~\cite{chanda}. There, the filling factor $\nu$ is fixed while the displacement field $D$ is varied, and conductivity is measured to locate the gapped regions. The experiment finds a gapped state at $\nu =5/2, 7/2, 9/2$, each confined to a narrow window in $D$. In this section, we first model the single-particle dispersion using the Slonczewski--Weiss--McClure tight-binding model \cite{Slonczewski1958,McClure1957}. We then project the many-body problem onto the two relevant bands and write down the form of  effective Hamiltonian.

\subsection{Single particle dispersion}

\begin{figure}
\includegraphics[width=\linewidth]{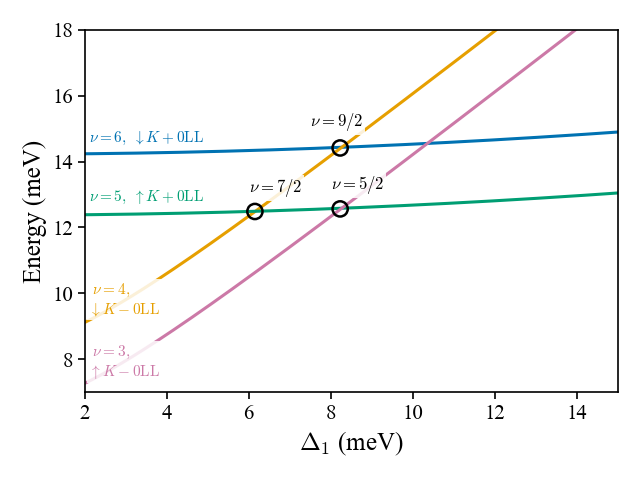}
   \caption{\textit{Single-particle dispersion and 0LL crossings.} The dispersion of the four relevant zeroth Landau levels (0LLs) is shown as a function of the interlayer potential $\Delta_1$ at $B=16$~T. The spin and valley character of each level is indicated together with the corresponding integer filling factor. Open circles mark the 0LL crossings at half-integer fillings $\nu=5/2$, $7/2$, and $9/2$, where incompressible states are observed experimentally. We note that even though there are two crossings for $\nu=7/2$, only one of them observes an incompressible state.}

        \label{fig:dispersion}
\end{figure}

We compute the Landau level spectrum using the Slonczewski--Weiss--McClure tight-binding parametrization at $B = 16$~T; the resulting bare spectrum is shown in Fig.~\ref{fig:dispersion}. In this model, the low-energy electronic states arise from the $p_z$ orbitals of the carbon atoms in Bernal-stacked trilayer graphene. The system has twenty-four degrees of freedom in the continuum limit, corresponding to two valleys ($K_+$ and $K_-$), two spin orientations ($\uparrow$ and $\downarrow$), three graphene layers, and two sublattices per layer.

In the mirror-symmetric limit of zero displacement field, $D=0$, the Hamiltonian decomposes into a single-layer graphene (mirror-odd) sector $H_{\rm slg}$ and a bilayer graphene (mirror-even) sector $H_{\rm blg}$. In this basis, the Hamiltonian takes the form
\begin{equation}
    \hat{H}=\begin{pmatrix}
        H_{\rm slg}& H_{\Delta_1}\\
        H^{\dagger}_{\Delta_1}&H_{\rm blg}\\
    \end{pmatrix},
\end{equation}
where the two sectors are coupled by the block $H_{\Delta_1}$, the matrix representation of the interlayer potential difference $\Delta_1=AD$ induced by the applied displacement field $D$, with $A$ a constant set by the permittivity of graphene. We note that the tight-binding model includes nearest-neighbour intralayer hopping, but neglects next-nearest-neighbour intralayer hopping, which we assume to be negligible.
Details of the full Hamiltonian and the parameter values are presented in Appendix~\ref{app:tight_binding}.

We consider the Landau levels corresponding to fillings $\nu=3,\ldots,6$. Hybridization between different Landau levels is weak, so that all four levels considered here are 0LL to high accuracy (approximately $98\%$) over a wide range of parameters. As $\Delta_1$ increases, the Landau levels change their ordering, with the relevant levels immediately before and after each crossing summarised in Tab.~\ref{tab:LL_crossings}.

\begin{table}[t]
\centering
\begin{tabular}{c|cc}
\hline
& Before crossing & After crossing \\
\hline

\multicolumn{3}{c}{\boldmath$\nu=5/2$} \\
\hline
$\nu=3$ & $0\mathrm{LL},\uparrow K-$ & $0\mathrm{LL},\uparrow K+$ \\
$\nu=4$ & $0\mathrm{LL},\uparrow K+$ & $0\mathrm{LL},\uparrow K-$ \\
\hline
\noalign{\vskip 2pt}

\multicolumn{3}{c}{\boldmath$\nu=7/2$} \\
\hline
$\nu=4$ & $0\mathrm{LL},\downarrow K-$   & $0\mathrm{LL},\uparrow K+$ \\
$\nu=5$ & $0\mathrm{LL},\uparrow K+$ & $0\mathrm{LL},\downarrow K-$ \\
\hline
\noalign{\vskip 2pt}

\multicolumn{3}{c}{\boldmath$\nu=9/2$} \\
\hline
$\nu=5$ & $0\mathrm{LL},\downarrow K-$ & $0\mathrm{LL},\downarrow K+$ \\
$\nu=6$ & $0\mathrm{LL},\downarrow K+$ & $0\mathrm{LL},\downarrow K-$ \\
\hline
\end{tabular}

\caption{Assignment of the spin- and valley-resolved 0LLs to integer filling factors immediately before and after the relevant Landau-level crossings at $\nu=5/2$, $7/2$, and $9/2$.}
\label{tab:LL_crossings}
\end{table}

The gapped states seem to occur precisely at the relevant single-particle band crossings indicated in Fig.~\ref{fig:dispersion}. The main evidence comes from Ref.~\cite{chanda}, which shows that the charge gap of the half-integer states coincides in $\Delta_1$ with the IQHE gap at the nearest integer filling factor. Since this is a strongly interacting problem, however, the effective dispersion at half-integer filling need not coincide with that at closest integer filling, as the Hartree-Fock contributions can differ. We find that the effective band crossings at half-integer filling and the closest integer filling are nevertheless fairly close, as shown in the Appendix~\ref{app:dispersion_renormalisation}. As a further comparison with experiment, we consider in Appendix~\ref{app:incompressible_widths} the widths $\Delta D$ of the incompressible regions.

\subsection{Effective Hamiltonian}

As noted earlier, we are interested in the regime where two Landau levels cross, in the presence of two types of perturbations: (i) ultra-short-range perturbations generated by lattice-scale effects (including short range Hubbard terms) and electron--phonon coupling, and (ii) screening due to Landau-level mixing, treated within the random-phase approximation (RPA). In order to separate the effects of the ultra-short-range perturbations from those of RPA screening, we consider two models: one including only the ultra-short-range perturbations, and another including both the ultra-short-range perturbations and RPA screening. For the first model, we start from the many-body Hamiltonian including all bands and the ultra-short-range perturbations, and project it onto the two crossing bands. For the second model, we additionally include RPA screening before carrying out the projection, so that particle--hole excitations involving bands outside the projected two-band subspace contribute to the effective interaction. The details of the RPA calculation and the determination of the magnitude of the ultra-short-range perturbations are deferred to Sec.~\ref{sec:realistic_interaction}.

Furthermore, since Landau-level mixing is incorporated only within the RPA, we neglect potentially important three-body interaction terms generated by Landau-level mixing. Such terms can play a role in lifting the degeneracy between the Pfaffian and anti-Pfaffian states. Including such terms would entail substantial additional numerical complexity, and we do not expect them to qualitatively affect the competition between the CFL and incompressible phases that is the focus of this work.

One can show that in both cases, the Hamiltonian takes the same functional form:
\begin{equation}
\label{eq:hamiltonian_main_ref}
\begin{split}
\hat{H} = \frac{1}{2}\sum_{s,s'}
\int \frac{d^2\vec q}{(2\pi)^2}V_I(\vec q)\, \rho^{0}_s(\vec q)\, \rho^{0}_{s'}(-\vec q)
\\+ \Delta V_0 \int \frac{d^2\vec q}{(2\pi)^2} \rho^0_1(\vec{q})\, \rho^0_2(-\vec{q}) 
+ \frac{\mu}{2} \big(\rho_{1}(0)-\rho_{2}(0)\big),
\end{split}
\end{equation}
where $s=1,2$ label the two relevant bands, $\rho^0_s(\vec q)$ is the density operator projected onto band $s$, $\mu$ is the splitting between the two bands, $V_I(q)$ is the relevant electron-electron Coulomb interaction, and $\Delta V_0$ is the point-like perturbation. We note that $\rho_s(0)$ is simply the total number of particles in band $s$. The distinction between the two models is encoded in $V_C(q)$ and $\Delta V_0$: RPA screening modifies the functional form of $V_C(q)$ and renormalises the value of $V_0$.

Throughout this work, we regard the two-band degree of freedom as a pseudospin-$1/2$, with the two components identified as $s=1,2$. Furthermore, the Hamiltonian in Eq.~\eqref{eq:hamiltonian_main_ref} is SU(2) symmetric under rotations in pseudospin space. In particular, it commutes with the total pseudospin operator $S^2$, so its eigenstates can be labelled by a definite total pseudospin $S$. We find that the SU(2)-breaking terms are approximately two orders of magnitude smaller than $V_0$, and therefore neglect them throughout this work.

For the model without RPA screening, the Coulomb interaction takes the form
\begin{equation}
\label{eq:gated_coulomb}
V_C(q)
=
\frac{e^2}{4\pi\epsilon_0\epsilon_{\rm hBN}}
\frac{2\pi\tanh(qd)}{q},
\end{equation}
where $d$ is the effective gate distance, typically
$d\approx 25$--$30~\mathrm{nm}$, or $4$--$6\ell_B$ for a typical
magnetic field strength of $B=15~\mathrm{T}$, and
$\epsilon_{\rm hBN}\approx 4.6$ is the effective dielectric constant of the hBN
environment. We define the characteristic Coulomb energy scale as $E_C=e^2/(4\pi\epsilon_0\epsilon_{\rm hBN}\ell_B)$.
In the model with RPA screening, the Coulomb interaction instead takes the form
\begin{equation}
V_C^{\rm RPA}(q)
=
\frac{V_C(q)}
{1+\Pi_0(q) V_C(q)},
\label{eq:RPA_Coulomb}
\end{equation}
where $\Pi_0(q)=A_{\Pi}/{(E_C\ell^2_B)}\tanh{(bq^2\ell_B^2)}$ is the static screening function, parametrised by dimensionless $A_{\Pi}$ and $b$ and obtained from the full Landau-level spectrum. The procedure for determining $V_0$, $A_{\Pi}$, and $b$, together with their values is discussed in Sec.~\ref{sec:realistic_interaction}.

An additional sufficiently strong inter-pseudospin tunnelling term could, in principle, qualitatively alter the physics of the two-band model.
In the present system, however, such pseudospin tunnelling is negligibly small, as we argue below. The corresponding single-particle term can be written as
\[
H_t=-t\sum_k c^\dagger_{1,k}c_{2,k}+\mathrm{H.c.},
\]
where $t$ denotes the inter-pseudospin tunnelling amplitude. A sufficiently large $t$ hybridises the two bands, producing an avoided crossing and thereby allowing an effectively one-component state to be stabilised. Indeed, such a mechanism has been shown to stabilise Pfaffian/Anti-Pfaffian phases in double quantum wells \cite{Zhu2016BilayerTunneling}.
In that setting, however, stabilisation requires not only a finite $t$ but also a sufficiently large interlayer separation $d$, which itself leads to appreciable SU(2)-symmetry breaking. By contrast, in our system both the inter-pseudospin tunnelling and the SU(2)-symmetry breaking are negligible.
Since the two pseudospin sectors have opposite physical spin, one mechanism by which $t$ could arise is spin-orbit coupling. However, spin-orbit coupling in graphene is very weak, giving $t\sim 3\text{--}30\,\mu\mathrm{eV}$, or only $10^{-5}\text{--}10^{-4}$ of the interaction energy scale \cite{Konschuh2012BLG,Gmitra2009SOC}.
Thus, inter-pseudospin tunnelling is far too weak to qualitatively alter the band crossing or stabilise an effectively one-component Pfaffian/Anti-Pfaffian state.

Another possible effect we should consider is band inversion. Since the energy spacing between the bands is comparable to the interaction energy, interactions may reverse their effective ordering. Instead of fully filling the 1LL and then half-filling the 0LL, the system could fully fill the 0LL—which lies higher in single-particle energy but has a larger Fock contribution—and then half-fill the 1LL. The resulting half-filled 1LL would then support a Pfaffian/Anti-Pfaffian state. However, this requires significant fine-tuning. 
First, obtaining the Pfaffian only within a narrow range of $D$ would require the single-particle 1LL and 0LL to be close in energy within this range, while being well separated elsewhere to suppress band inversion. This is because, if the two bands remained close in energy over a larger range of $D$, we would expect band inversion to occur over a correspondingly larger region of parameter space, contrary to the narrow region in which it was observed experimentally. Although this is in principle possible, we would additionally require the charge gap in the IQHE case \cite{chanda} to attain its minimum at the same values of $D$.
The IQHE gap is very small---smaller even than the Fock term---which indicates that the bands are very close in energy and are most likely crossing. Moreover, since there is no discontinuous jump, we would expect this to correspond to either (1) an avoided crossing, which we have already ruled out, or (2) a crossing between Landau levels. In the second case, this would imply that a band inversion somehow occurs only at $\nu=1/2$, but not at integer filling. Such a scenario would require fine-tuning involving three bands and therefore appears unlikely. We study the possibility of such a band inversion in the Appendix~\ref{app:dispersion_renormalisation}.

Finally, generic Landau-level mixing beyond the RPA could, in principle, stabilise a gapped state, such as the Pfaffian. To investigate this possibility, we performed four-band infinite density-matrix renormalization group (iDMRG) calculations, including the two crossing Landau levels together with the two nearest empty Landau levels. Unlike the RPA treatment, this calculation incorporates all Landau-level-mixing processes within the four-band subspace, including the three-body terms generated by mixing between these bands. We also considered several other choices of nearby Landau levels. However, we found no indication of gapped behaviour or of any of the candidate phases, although convergence of the four-band calculations was generally poor. We therefore regard these results as inconclusive; the corresponding data and further details are presented in the Appendix~\ref{app:explicit_ll_mixing}.

\subsection{Numerical methods}\label{subsec:numerical_methods}

We study the system using a combination of infinite density-matrix
renormalization group (iDMRG) and exact diagonalization (ED). In iDMRG, we work
in the Landau gauge, $\mathbf{A}=\ell_B^{-2}(y,0)$, where the single-particle
orbitals with momentum
$k_x=\frac{2\pi m}{L},  m\in\mathbb{Z},$ are localised around $y=k_x\ell_B^2$. The two-dimensional problem thus
maps onto a one-dimensional chain of orbitals with long-range
interactions.

For ED, we impose periodic boundary conditions in both directions, such
that the system has the topology of a torus. Magnetic translation
symmetry is used to block diagonalise the Hamiltonian into momentum
sectors, allowing us to obtain the low-energy spectrum and its
ground-state manifold. We consider systems of $N=8$ and $10$ electrons, with the torus aspect ratio fixed to unity.

In iDMRG, directly simulating the long-range Coulomb interaction is impractical due to the computational cost of representing interactions extending over arbitrarily large distances, so the interaction must be truncated or screened at large distances. In our system, the gate distance $d$ provides a natural cutoff. However, $d$ is relatively large, and simulations using this cutoff exhibit poor convergence. We therefore replace the gate-screened Coulomb interaction with the more rapidly decaying screened interaction
\begin{equation}
\label{eq:screening_DMRG}
V(r) = \frac{e^{-r^2/(2\zeta^2)}}{r},
\end{equation}
where we choose $\zeta = 4\ell_B$. This shorter-ranged interaction substantially improves the convergence of the iDMRG calculations.

The majority of our calculations are performed using iDMRG
\cite{Zaletel_2013,Zaletel_2015,tenpy,papic_dmrg}, where the Hamiltonian is placed on
an infinitely long cylinder of circumference $L_y$. The ground state is
variationally represented as a matrix-product state (MPS) with bond dimension $\chi$. The entanglement that can be represented by the MPS is bounded by $S \leq \log\chi$.
In DMRG, the computational complexity scales with the bond dimension, which must itself grow exponentially with the cylinder circumference to maintain a fixed accuracy.
However, the circumferences accessible in iDMRG correspond to substantially larger systems than those accessible in ED.

In this paper, we use circumferences $L_y\leq 20 \ell_B$, and $\chi=5000-8000$.

\section{Identification of Topological Phases}
\label{sec:topo_phases_identification}
\subsection{Diagnostic tools}\label{subsec:diagnostics}

The combination of ED and iDMRG gives access to several complementary
diagnostics of topological order. Since iDMRG can reach substantially
larger system sizes and directly probe the bulk of the system, we begin
with the diagnostics available on the infinite cylinder.

\paragraph{Correlation length.}
The MPS representation gives direct access to the correlation
length $\xi$ through the transfer matrix. A finite correlation length
that converges with increasing bond dimension is consistent with a
gapped bulk phase. Conversely, a correlation length that grows strongly
with $\chi$ signals a gapless phase or a continuous
phase transition. The correlation length therefore provides a useful
first diagnostic to distinguish incompressible phases from
compressible ones and for locating phase boundaries.

\paragraph{Ground-state sectors.}
Topologically ordered phases possess several degenerate ground states
on a torus and, correspondingly, several minimally entangled states on
an infinite cylinder. These states can be associated with the different
anyon sectors of the topological theory. By initializing iDMRG in
different states, one can attempt to obtain distinct members of this ground-state manifold. The number of resulting sectors and their quantum numbers therefore provide information about the underlying topological order.
In practice, however, iDMRG is not guaranteed to find every member of
the ground-state manifold, so the number of states obtained numerically
should generally be regarded as a lower bound on the full topological
degeneracy.

\paragraph{The Topological shift.}
One of the most useful quantities accessible in iDMRG is the
topological shift $\mathcal{S}$ \cite{topo_shift_wen_1,topo_shift_wen_2,topo_shift3}. Although the shift is conventionally
defined on the sphere through
\begin{equation}
    N_{\phi}=\nu^{-1}N_e-\mathcal{S},
\end{equation}
it can also be extracted directly on the infinite cylinder from the
momentum polarization. For a minimally entangled state in topological
sector $a$, the momentum polarization takes the asymptotic form \cite{Zaletel_2013,Zaletel_2015}
\begin{equation}
    M_a =
    -\frac{\nu\mathcal{S}}{(4\pi \ell_B)^2}L_y^2
    + h_a-\frac{c}{24}
    +\mathcal{O}(e^{-L_y/\tilde{\xi}}) \quad\mathrm{mod }\, 1. 
\end{equation}
Here $h_a$ is the topological
spin and $c$ is the chiral central charge. The coefficient of the
leading $L_y^2$ contribution therefore provides a direct measurement of
$\mathcal{S}$.

\paragraph{The Topological entropy.}
For a minimally entangled state $|a\rangle$ associated with anyon type
$a$, the entanglement entropy \cite{KitaevPreskill2006_topo_entropy, LevinWen2006_topo_entropy} across the cylinder scales as
\begin{equation}
    S_a(L_y)
    =
    \alpha L_y
    -\log\left(\frac{\mathcal{D}}{d_a}\right)
    +\mathcal{O}(e^{-L_y/\tilde{\xi}}),
\end{equation}
where $d_a$ is the quantum dimension of the anyon and
\begin{equation}
    \mathcal{D}=\sqrt{\sum_a d_a^2}
\end{equation}
is the total quantum dimension. The constant contribution $\gamma_a\equiv\log\left(\frac{\mathcal{D}}{d_a}\right)$ is the topological entanglement entropy.

For an Abelian topological phase all anyons have $d_a=1$. Consequently,
all minimally entangled ground states have the same entanglement entropy, independent of the topological sector $a$. A difference in the
entanglement entropies of distinct sectors therefore provides evidence
for non-Abelian anyons, for which $d_a>1$ can occur.

\paragraph{Entanglement spectrum.}
Another powerful diagnostic is provided by the low-lying
entanglement spectrum. Dividing the infinite cylinder into two
semi-infinite halves, the ground state can be Schmidt decomposed as
\begin{equation}
    |\Psi\rangle
    =
    \sum_{\alpha}
    \lambda_{\alpha}
    |\alpha\rangle_L\otimes|\alpha\rangle_R ,
\end{equation}
and the corresponding entanglement energies are defined as
\begin{equation}
    \epsilon_\alpha=-\log\lambda_\alpha^2 .
\end{equation}
The levels can be resolved according to their conserved quantum numbers, in particular momentum along the cylinder. We focus on the low-lying levels, whose counting and chirality reflect the structure of the edge theory and therefore provide a characteristic fingerprint of the underlying topological order \cite{Li_2008}.

\paragraph{Local pair-correlation function.} A further diagnostic available in iDMRG is the zero-separation density correlation function between the two pseudospin species,
\begin{equation}
g: = \langle \rho_1(0)\,\rho_2(0)\rangle,
\end{equation}
which has previously been used to characterise phase transitions in related quantum Hall problems \cite{papic_dmrg}. In particular, its behaviour as a function of a tuning parameter distinguishes continuous from first-order transitions: a discontinuous jump in $g(0)$ signals a first-order transition, whereas a smooth, continuous variation is consistent with a continuous one. Because $g(0)$ is built from local density operators, it must take the same value in every minimally entangled state belonging to a given topological phase; a difference in $g(0)$ between nearly-degenerate sectors therefore indicates that those sectors are not topologically degenerate, but instead belong to distinct, locally distinguishable ground states.

Now we turn to the diagnostics available from exact diagonalization on
the torus.

\paragraph{Topological ground-state degeneracy.}
On the torus, topological order \cite{WenNiu1990,Kitaev2003,Simon2023_book} determines both the ground-state degeneracy and the momentum sectors \(K\) in which these states appear. We therefore use the number and momentum sectors of the low-lying states as a first diagnostic of the candidate phase.

\paragraph{Overlap with trial states.}
A final diagnostic is the overlap between the ED ground
states and the corresponding trial states in each momentum sector,
\begin{equation}
    \mathcal{O}
    =
    \left|\langle\Psi_{\rm trial}|\Psi_{\rm ED}\rangle\right|^2.
\end{equation}
Although such overlaps are expected to vanish in the thermodynamic
limit, $N\rightarrow\infty$, their magnitude for the small finite
systems accessible to ED is generally a useful indicator of the
underlying phase.

\subsection{Candidate states}\label{subsec:trial_states}

We consider five candidate states at $\nu=1/2$: the Halperin $(331)$ \cite{Halperin_states}
state, the weakly paired $d$-wave state \cite{JSS}, which is topologically
equivalent to the Jain pseudospin-singlet (JSS) state \cite{Read_2000}, the strongly paired
pseudospin-singlet state, the Pfaffian \cite{MooreRead1991}, and the anti-Pfaffian \cite{PHPfaffian1,PHpfaffian2}. The first three are two-component states with equal occupation of the two pseudospins, $N_1=N_2=\frac{N}{2},$
whereas the Pfaffian and anti-Pfaffian are single-component, fully
polarised states.

Among the two-component candidates, the weak- and strong-pairing states are pseudospin singlets and eigenstates of the pseudospin $S^2$ operator, whereas the Halperin $(331)$ state is not an eigenstate of $S^2$.

The Halperin $(331)$ state is an Abelian two-component state with disk
wavefunction
\begin{equation}
    \Psi_{331}
    =
    \prod_{i<j}(z_i-z_j)^3
    \prod_{i<j}(w_i-w_j)^3
    \prod_{i,j}(z_i-w_j),
    \label{eq:psi331}
\end{equation}
where $z_i$ and $w_i$ denote particles in the two components.

The second candidate is the weakly-paired $d$-wave state. This phase
has the same topological order as the Jain pseudospin-singlet state, for which
a convenient disk trial wavefunction is
\begin{equation}
    \Psi_{\mathrm{JSS}}
    =
    \mathcal P_{\mathrm{LLL}}
    \left[
        \chi_2(\{z_i,w_j\})\,\chi_{110}(\{z_i,w_j\})\,\chi_1(\{z_i,w_j\})
    \right],
    \label{eq:psiJSS}
\end{equation}
with
\begin{equation}
    \chi_{110}
    =
    \prod_{i<j}(z_i-z_j)
    \prod_{i<j}(w_i-w_j),
\end{equation}
where $z_i$ and $w_i$ denote the coordinates of the two pseudospin species, respectively. Here, $\chi_n$ denotes the wavefunction of $n$ filled Landau levels involving all particles, irrespective of their pseudospin, while $\chi_{110}$ consists of separate $\nu=1$ Jastrow factors for each pseudospin species. Finally, $\mathcal P_{\mathrm{LLL}}$ denotes projection onto the lowest Landau level. The relation between the JSS state and weak $d$-wave pairing can
be made explicit by rewriting the state in a paired form of composite fermions \cite{milovanovic,Read_2000}.

The third two-component candidate is the strongly paired pseudospin-singlet state. This phase can be understood within the composite-fermion picture, in which two flux quanta, $2\Phi_0$, are attached to each electron to form composite fermions. In the strong-pairing phase, the composite fermions form tightly bound pairs, corresponding to charge-$2e$ objects, and the resulting Abelian topological order is described by $K_{\mathrm{SP}}=(8)$ \cite{daughter_states}. An explicit wavefunction can be constructed within the composite-fermion description \cite{Read_2000}; however, its form is rather involved, and we omit it here.

We notice that all three two-component states are Abelian, and hence can be described by $K$ matrix \cite{Wen1995_K_matrix,topo_shift_wen_2, topo_shift3}, as illustrated in Appendix~\ref{app:topological_properties}.

The remaining two candidates are the single-component Pfaffian and
anti-Pfaffian states. The Pfaffian trial wavefunction is
\begin{equation}
    \Psi_{\mathrm{Pf}}
    =
    \operatorname{Pf}\!\left(\frac{1}{z_i-z_j}\right)
    \prod_{i<j}(z_i-z_j)^2,
    \label{eq:psiPf}
\end{equation}
while the anti-Pfaffian is its particle-hole conjugate,
\begin{equation}
    \Psi_{\mathrm{aPf}}
    =
    \mathcal C_{\mathrm{PH}}\Psi_{\mathrm{Pf}}.
\end{equation}
The Pfaffian and anti-Pfaffian are distinct non-Abelian topological
phases, despite being related by the particle-hole conjugation operator  $ \mathcal C_{\mathrm{PH}}$ \cite{PHPfaffian1,PHpfaffian2}.

The low-lying entanglement spectrum provides an important way to
distinguish these candidate states. The expected counting depends on the sector across the entanglement cut, specified in our two-component problem by the particle number $N$ and pseudospin $S^z$. In the main text, we show the counting for the different particle-number sectors with all $S^z$ sectors combined, in which case the counting depends only on the parity of $N$. The corresponding $S^z$-resolved counting is presented in the Appendix~\ref{app:topological_properties}.
For the Halperin $(331)$ state, 
\begin{equation}
    N\ {\rm even}: \qquad 1,2,7,14
\end{equation}
For the JSS/weak-pairing state,
\begin{equation}
     N\ {\rm even}: \qquad
    1,5,15, \dots
\end{equation}
For the strong-pairing state,
\begin{equation}
     N\ {\rm even}: \qquad
    1,1,2,3,5
\end{equation}
For the polarised Pfaffian and anti-Pfaffian states the $S^z$
resolution is absent, and the corresponding even- and odd-sector
countings are
\begin{equation}
    \begin{split}
    N\ {\rm even} &: \qquad
    1,1,3,5,10\\
    N\ {\rm odd} &: \qquad
    1,2,4,7,13.
    \end{split}
\end{equation}
with right- and left-moving chiralities for the Pfaffian and anti-Pfaffian, respectively.

The remaining properties that we use to distinguish the candidate
states are summarised in Table~\ref{tab:candidate_states}. In
particular, the first three states are Abelian and have an eightfold
ground-state degeneracy on the torus. 
Their total quantum dimension is
therefore $ \mathcal D=\sqrt{8}$,
so they also have the same topological entanglement entropy,
$\gamma=\log\sqrt{8}$. Consequently, the ground-state degeneracy and
topological entropy alone cannot distinguish the three two-component
candidates. The shift and sector-resolved entanglement spectrum are
therefore particularly useful. The Pfaffian and anti-Pfaffian, on the
other hand, can already be separated from these states by their complete
pseudospin polarization.

\begin{table}[t]
\centering
\small
\vspace{0.5cm}
\setlength{\tabcolsep}{4pt}
\begin{tabular}{lccccc}
\hline\hline
FQH phase  & SU(2) & GSD & $\mathcal S$ & $\mathcal D$ \\
\hline
Halperin $(331)$    & No  & 0,8 & 3  & $\sqrt{8}$ \\
JSS / weak pairing  & Singlet & 0, 8 & 4  & $\sqrt{8}$ \\
Strong pairing      & Singlet & 0, 8 & 4  & $\sqrt{8}$ \\
Pfaffian           & Polarised & 2,6 & 3  & $\sqrt{8}$ \\
anti-Pfaffian     & Polarised & 2,6 & $-1$ & $\sqrt{8}$  \\
\hline\hline
\end{tabular}
\caption{Topological properties of the candidate states at $\nu=1/2$. The SU(2) column indicates whether the state is an eigenstate of the total pseudospin operator $\hat{S}^2$, and, when applicable, its pseudospin character. The ground-state degeneracy (GSD) is the degeneracy on the torus, with the two entries corresponding to odd and even total particle number $N$, respectively. $\mathcal{S}$ denotes the shift on the sphere, and $\mathcal{D}$ the total quantum dimension.}
\label{tab:candidate_states}
\end{table}

\section{Phase Diagram of the Toy Model}
\label{sec:toy_model}

In this section, we study the Hamiltonian in Eq.~\eqref{eq:hamiltonian_main_ref}, taking $V_I(q)$ to be the pure Coulomb interaction. We find that introducing a finite gate distance $d$ does not modify the results discussed below significantly. We therefore focus on the dependence of the phase diagram on the dimensionless interaction parameter $\Delta V_0/V_0$, where $\Delta V_0$ is the ultra-short-range interaction correction and $V_0$ is the zeroth Coulomb pseudopotential.
 For reference, we estimate in Section~\ref{sec:realistic_interaction} that the value relevant to the physical system is approximately $\Delta V_0/V_0 \simeq -0.08$.

At $\Delta V_0=0$, the ground state is either an unpolarised or a partially polarised composite Fermi liquid (CFL), depending on the number of electrons. The precise polarization of the CFL in the thermodynamic limit remains unknown \cite{sarma_CFL_Monte_carlo,CFL_RezayiRead1994}. On the torus, a characteristic signature of the CFL is the structure and degeneracy of the low-energy manifold. These degeneracies can be understood by filling an effective Fermi sea of composite fermions. Consequently, the ground-state degeneracy depends on the number of electrons and follows the corresponding shell-filling structure, analogous to Hund's rules \cite{Fremling_2018,CFL_N_8,sarma_CFL_Monte_carlo,SharmaPuJain2021_CFL,CFL_Onoda2001}.

We therefore believe that our model in Eq.~\eqref{eq:hamiltonian_main_ref} should include all plausible microscopic perturbations, except for LL mixing, which we study below in section~\ref{sec:realistic_phase_diagram}. Importantly, since this is an SU(2)-symmetric problem, the Pfaffian/anti-Pfaffian state should arise as a fully polarised state. Using Coulomb interaction, we do not numerically observe a fully polarised state for any value of $\Delta V_0<0.1$.

\begin{figure}
 
\includegraphics[width=\linewidth]{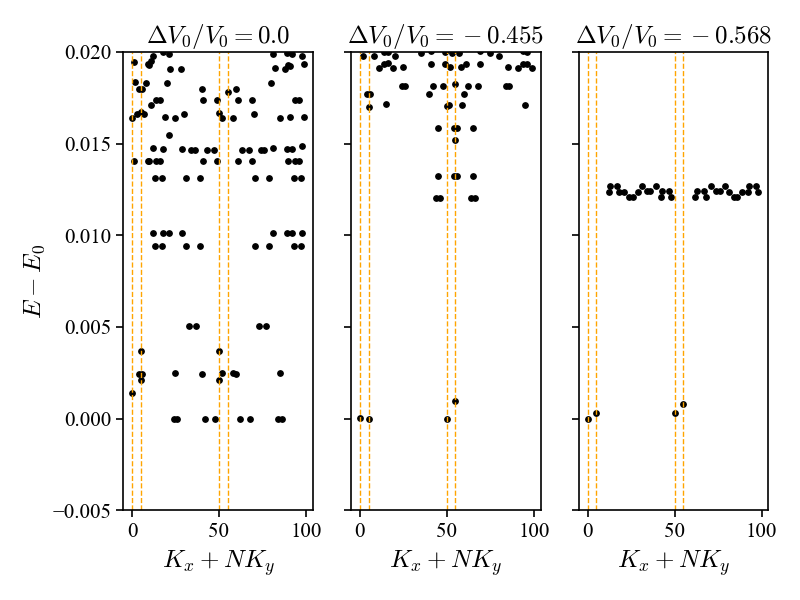}
\includegraphics[width=\linewidth]{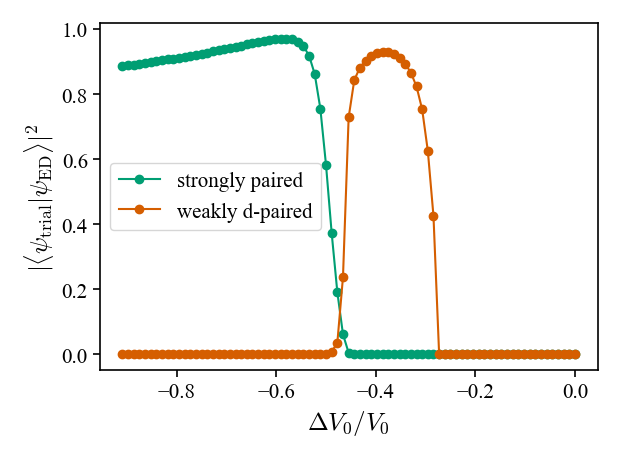}
  \caption{\textit{Evolution of the low-energy exact diagonalization spectrum and the ground-state wavefunction overlap as the interaction strength is tuned through $\Delta V_0/V_0$.} \textit{Top:} Low-energy spectra for three representative interaction strengths: $\Delta V_0/V_0 = 0.0$ (left), $-0.45$ (center), and $-0.57$ (right). The vertical dashed lines indicate the momentum sectors expected for the ground state in the weak- and strong-pairing phases. \textit{Bottom:} Squared overlap of the exact ground state with representative variational wavefunctions for the strong-pairing and weak $d$-paired phases. The overlap reveals two phase transitions as the interaction is varied: first from the composite Fermi liquid (CFL) to the weak $d$-paired phase, and subsequently from the weak $d$-paired phase to the strong-pairing phase. The calculations are performed for $N=10$ particles in the momentum sector $K=(5,5)$.}
    \label{fig:Spectrum_evolution}
\end{figure}

As $\Delta V_0$ is made increasingly negative, the overlap with the weak pseudospin-singlet $d$-paired state wave function increases. In particular, the overlap in the $K=(5,5)$ sector becomes large already around $\Delta V_0/V_0\simeq-0.34$. This alone, however, does not locate the transition between the CFL and the weak $d$-paired state, since the corresponding state in the $K=(5,5)$ sector is not yet the global ground state. The low-energy spectrum provides the complementary information: upon further decreasing $\Delta V_0$, we find a fourfold set of ground states in momentum sectors $K=(0,0),(N_e/2,0),(0,N_e/2),(N_e/2,N_e/2)$, as expected for the weak $d$-paired state. After including the center-of-mass degeneracy, the total degeneracy is eight, again consistent with the expected topological degeneracy. The results are shown in Fig.~\ref{fig:Spectrum_evolution}.

Taking into account the spectrum together with the overlaps in all momentum sectors, we locate the CFL-to-weak-pairing transition at $\Delta V_0/V_0\approx-0.44$. Tracking the wave-function overlaps further indicates a subsequent transition from the weak- to strong-pairing phase at $\Delta V_0/V_0\approx-0.49$. We note that iDMRG and ED predict both transitions at quantitatively similar values of $\Delta V_0$.

Throughout this evolution, the ground state remains unpolarised or partially polarised. This provides evidence that the system retains its two-component character across the phase diagram and, in particular, disfavors an interpretation in terms of a fully polarised Pfaffian state. We additionally find that the overlap with the Halperin $(331)$ state remains small throughout the parameter range studied. A natural explanation is that the $(331)$ state is not an eigenstate of the total pseudospin operator $S^2$, whereas the Hamiltonian considered here commutes with $S^2$.

\begin{figure}[H]
  \includegraphics[width=\linewidth]{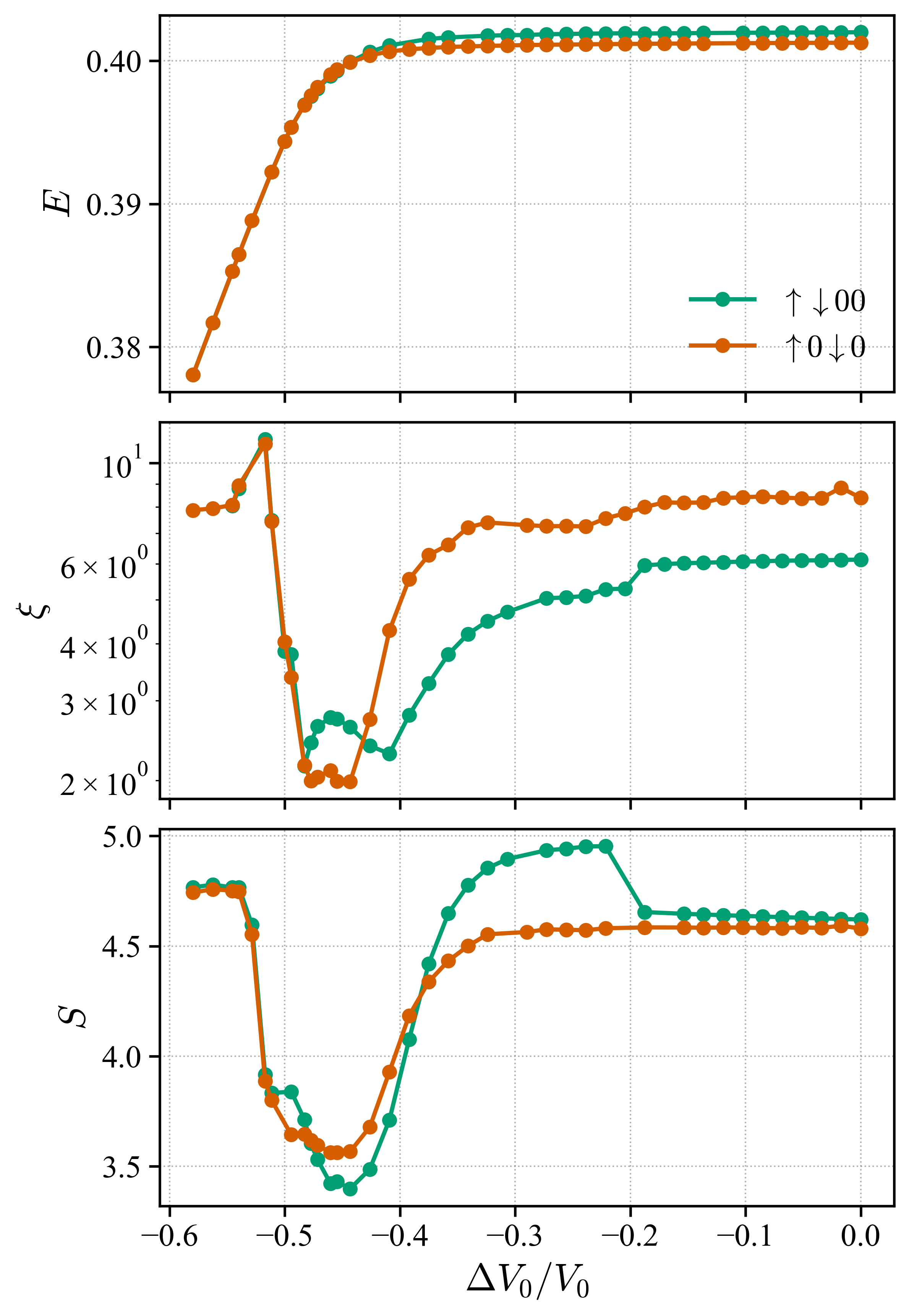}
    \caption{ \textit{Diagnostics of the CFL-to-weak-pairing and weak-to-strong-pairing transitions.} Top panel shows the energies of two root configurations as a function of $\Delta V_0$. At small $\Delta V_0$, the energies are non-degenerate, while they become degenerate around $\Delta V_0/V_0 \approx -0.4$. This indicates a transition from the CFL to the weak-pairing phase, since the different sectors are expected to be degenerate in both the weak- and strong-pairing phases. Middle shows the correlation length as a function of $\Delta V_0$. Initially, the correlation length is relatively large, consistent with the CFL phase. Around $\Delta V_0/V_0 \approx -0.34$, it begins to decrease and eventually saturates, providing further evidence for a transition from the CFL to the weak-pairing phase. The correlation length increases again around $\Delta V_0/V_0 \approx -0.52$, indicating a transition to the strong-pairing phase. We note that the strong-pairing phase is generally difficult to converge in DMRG because it competes with charge-density-wave (CDW) order. Bottom panel shows the entanglement entropy $S$ across the same cut. $S$ decreases in the weak-pairing region and increases again in the strong-pairing region. Data was taken at $L_y=14\ell_B$ and $\chi=8000$.}
        \label{fig:DMRG_V0}
\end{figure}

We next compare these results with DMRG calculations. Figure~\ref{fig:DMRG_V0} shows the evolution of the correlation length and entanglement entropy as a function of $\Delta V_0$. The data indicate a transition from the CFL to a weak-pairing phase near $\Delta V_0/V_0=-0.44$ (see eigenvalue spectrum in Appendix~\ref{app:spectra_data}). The correlation length evolves relatively smoothly across this transition, which is suggestive of a continuous transition, although the available data do not allow us to conclusively rule out a weakly first-order transition. Upon decreasing $\Delta V_0$ further, we observe a second transition, from the weak- to the strong-pairing regime, near $\Delta V_0/V_0 \simeq -0.51$.

The weak-to-strong pairing transition in Ref.~\cite{kun_yang_weak_to_strong} was found to be continuous, proceeding through the gapless Haldane--Rezayi (HR) state. At the transition, an additional state descends from the excited-state spectrum to form the 10-fold ground-state manifold characteristic of the HR state. We observe the same effect in our system. Within a very narrow range of $\Delta V_0/V_0$, between approximately $-0.49$ and $-0.52$, a state in the $K=(0,0)$ sector drops to the ground-state energy and subsequently rises again. The location of this level crossing is consistent with the weak-to-strong pairing transition identified from the overlaps, suggesting a continuous phase transition. Furthermore, in iDMRG we find an apparent divergence of the correlation length near the transition, providing additional evidence for continuous behaviour. Taken together, these observations suggest that the weak-to-strong pairing transition is continuous, although a more detailed analysis is required to establish the nature of the transition conclusively. The detailed eigenvalue spectrum data is provided in the Appendix~\ref{app:spectra_data}.

The discrepancy between the transition points obtained from ED and DMRG is relatively small and can be attributed, at least in part, to finite-size effects in ED. Indeed, the value of $\Delta V_0$ at which the transition occurs fluctuates with system size. In addition, the DMRG calculations require the use of a screened Coulomb interaction in Eq.~\eqref{eq:screening_DMRG} rather than the pure Coulomb interaction. 
Ideally, to determine the transition point more precisely, one would compute the phase diagram for several values of screening lengths $\zeta$ and extrapolate the transition point to the limit $\zeta\to\infty$, corresponding to the unscreened Coulomb interaction. In practice, however, this procedure is challenging for two reasons. First, for each iDMRG calculation, there is some uncertainty in locating the transition precisely, as none of the observables considered exhibits an obvious cusp or other sharp signature at the transition. Second, performing such calculations for multiple values of $\zeta$ would be too computationally expensive.
To at least partially address the discrepancy between the ED and DMRG results, we examine how the transition point obtained from ED scales with $\zeta$. The results are presented in the Appendix~\ref{app:toy_numerics}.

\begin{figure}
  \includegraphics[width=0.49\linewidth]{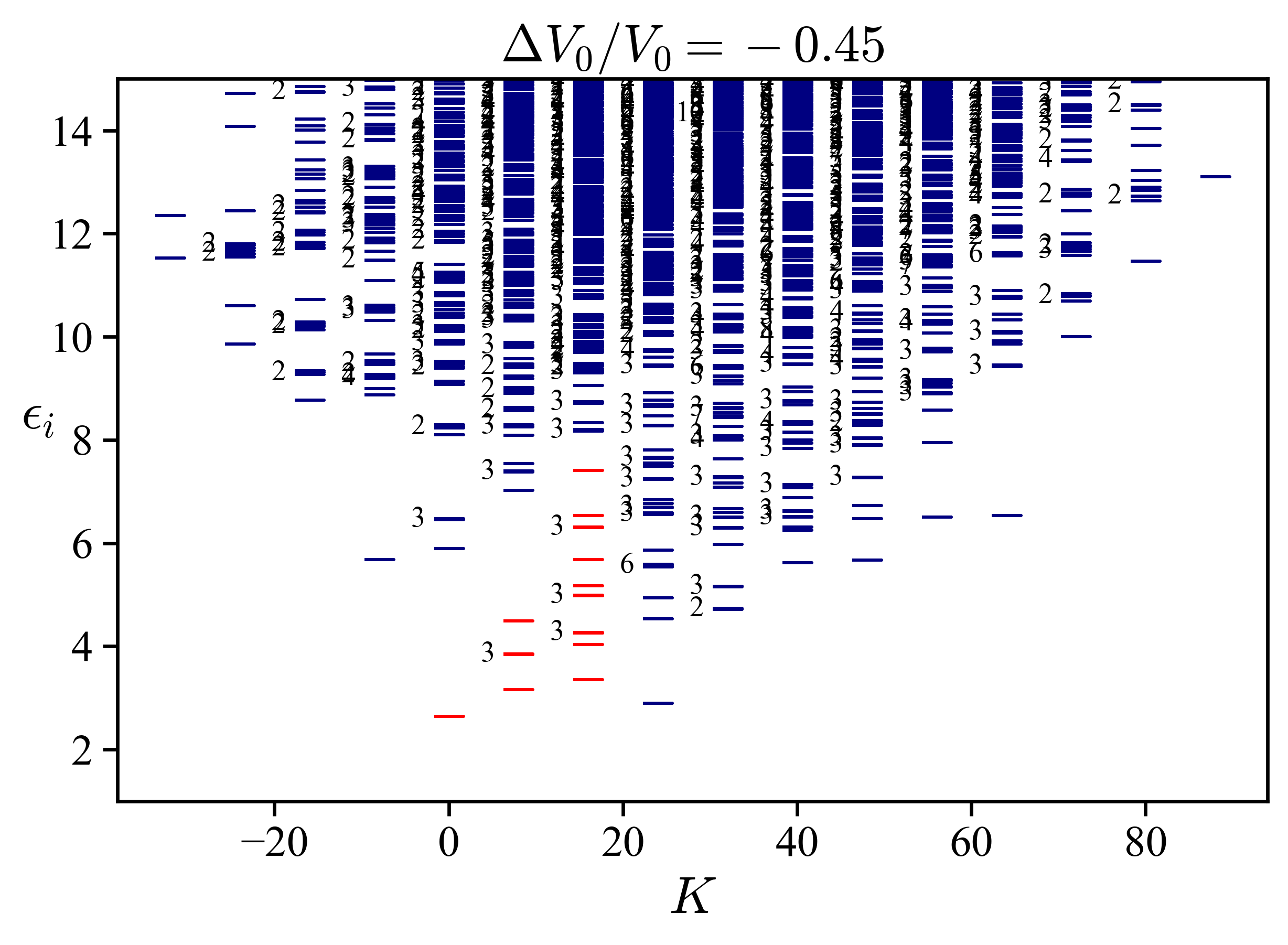}
  \includegraphics[width=0.49\linewidth]{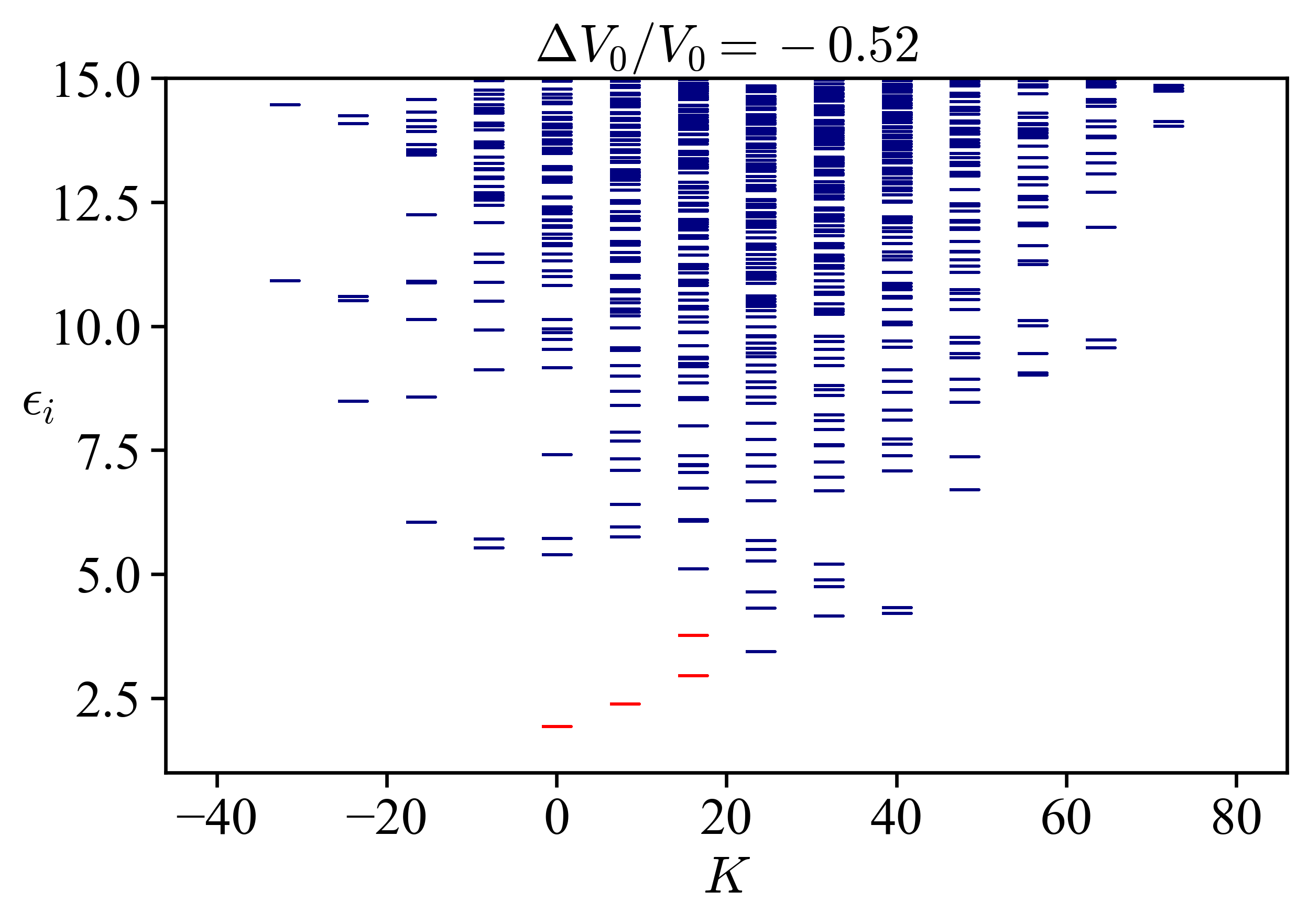}
\includegraphics[width=\linewidth]{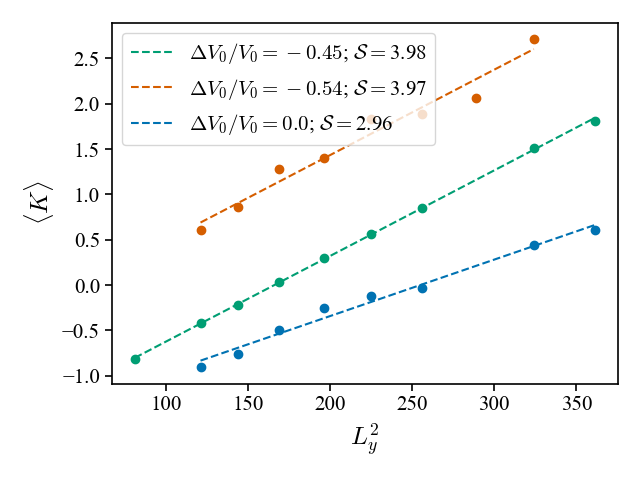}
    \caption{\textit{Entanglement spectra and the topological shift.} \emph{Top:} Entanglement spectra for $\Delta V_0/V_0=-0.45$ and $\Delta V_0/V_0=-0.52$, respectively. We find that the low-lying counting, marked in red, for $\Delta V_0/V_0=-0.45$ is consistent with the weak-pairing state, while that for $\Delta V_0/V_0=-0.52$ is consistent with the strong-pairing state. \emph{Bottom:} Topological shift for the same two interaction strengths. Both $\Delta V_0/V_0=-0.45$ and $\Delta V_0/V_0=-0.54$ yield a shift $\mathcal{S}=4$, consistent with the weak- and strong-pairing phases, respectively. In contrast, $V_0=0$ yields $\mathcal{S}=3$, although the data are considerably less well resolved, as one might expect for a gapless state. Further entanglement spectrum evidence is presented in the Appendix~\ref{app:topological_properties}.
    }
        \label{fig:DMRG_V0_ES}
\end{figure}

Several observations support the identification of the two gapped phases as weak- and strong-pairing states. First, two root-configuration sectors become degenerate in energy upon entering the paired regime from the CFL. This is consistent with an eightfold total degeneracy of the weak- and strong-pairing phases, since each root configuration carries an additional fourfold degeneracy associated with translations. Second, the shift approaches the expected value $\mathcal{S}=4$. Finally, the counting of levels in the entanglement spectrum agrees with the counting expected for the corresponding paired phases. This evidence is shown in Fig.~\ref{fig:DMRG_V0_ES}.

In summary, we find that tuning the short-range attractive interaction $\Delta V_0$ drives the system from a composite Fermi liquid into a weakly paired state and, upon further decreasing $\Delta V_0$, into a strongly paired state. The resulting phase diagram therefore contains two distinct transitions: a CFL-to-weak-pairing transition and the more extensively studied weak-to-strong pairing transition.

In previous works \cite{kharitonov,parton}, an effective attractive $\Delta V_0$ has been argued to arise from finite trilayer separation and from lattice-scale corrections, including electron--phonon coupling. However, the magnitude of $V_0$ required to reach the paired regime in our calculations is substantially larger than the estimated lattice-scale corrections. Such corrections alone therefore appear insufficient to account for the experimentally observed gapped phase. This motivates the inclusion of Landau-level-mixing effects, which we treat within the random-phase approximation (RPA) in Sec.~\ref{sec:realistic_interaction}.

\FloatBarrier

\section{Realistic Interaction Effects}
\label{sec:realistic_interaction}
\subsection{Ultra-short-range interaction corrections}
In this section, we estimate $V_0$ by accounting for corrections beyond the isotropic gated Coulomb interaction. We consider the finite interlayer separation in TLG, the finite lattice-scale extent of the electronic wavefunctions, and electron--phonon coupling. Detailed derivations of these contributions are presented in Appendix~\ref{app:short_range}. To our knowledge, these constitute the dominant ultra-short-range corrections relevant to the present problem.

We subsequently project the resulting interaction onto the two bands retained in our low-energy model. 
After this projection, all lattice-scale corrections reduce to the short-range interaction parametrised by $V_0$. 
In particular, the intervalley mixing terms vanish upon projection. Nevertheless, it is important to retain the complete interaction structure prior to projection. As we discuss in Subsection~\ref{subsec:RPA_effect}, the different interaction channels are renormalised differently by RPA screening, so terms that vanish under direct projection can still contribute to the effective interaction after the application of RPA diagrams.

From here on, we use letters $n,m$ to denote layer number $1,2,3$, letters $i,j$ denote sublattice $A,B$, $s,s'$ denote spin, and $v_1,v_2$ denote valleys.

\subsubsection{Trilayer distance}

We first consider the interlayer distance in trilayer graphene (TLG), incorporating the vertical layer positions $z_n$ ($n=1,2,3$) directly into the interaction. Let $d_{\rm TLG}$ denote the distance between the top ($n=3$) and bottom ($n=1$) layers, and define $\Delta z_{nm}=z_n-z_m$. With gates at $z=\pm d_{g}/2$, and middle layer at $z=0$, the Coulomb interaction becomes to the first order
\begin{equation}\label{eq:modified_interaction}
V_{nm}(q)
\approx V_C(q,d_g)-
 \frac{e^2}{2\epsilon_0\epsilon_r}
|\Delta z_{nm}|,
\end{equation}
where $V_C(q,d_g)$ is the isotropic gated interaction that ignores layer separation. The full derivation is provided in the Appendix~\ref{app:short_range}.

To account for the permittivity mismatch between TLG and hBN, we perform a full electrostatic calculation, effectively equivalent to rescaling $d_g\to d^{\rm eff}_g$ and $\Delta z_{nm}\to\Delta z_{nm}^{\rm eff}$. For typical experimental parameters \cite{chanda,laturia2018dielectric,santos2013electric}, $\Delta z_{nm}^{\rm eff}\approx(2\text{--}2.5)\,\Delta z_{nm}, d^{\rm eff}_g\approx1.4\,d_g$.
The details of the electrostatic calculation, including the derivation of Eq.~\eqref{eq:modified_interaction}, are provided in the Appendix~\ref{app:short_range}.

We introduce a momentum cutoff $\Lambda>K\gg 1/\ell_B$, where $K=4\pi/(3a)$ is the valley momentum. Separating intra- and intervalley scattering yields the perturbation:
\begin{align}
\Delta \hat V
={}&-\frac{e^2}{4\epsilon_0\epsilon_r}
\sum_{\substack{n,m,s,s'\\ i,j,v_1,v_2}}
\Delta z^{\rm eff}_{nm}
\int\frac{d^2\vec q}{(2\pi)^2}\,
\rho^{i,n,s,v_1}_{\vec q}
\rho^{j,m,s',v_2}_{-\vec q}
\nonumber\\
&+\frac{e^2}{4\epsilon_0\epsilon_r}
\sum_{\substack{n,m,s,s'\\ i,j,v_1,v_2}}
\Delta z^{\rm eff}_{nm}
\int\frac{d^2\vec q}{(2\pi)^2}\,
J^{i,n,s,v_1}_{\vec q}
J^{j,m,s',v_2}_{-\vec q}.
\end{align}
where 
\begin{equation}
    J^{insv}_{\vec q}=\sum_{\vec k}c^{\dagger}_{insv}(\vec k+\vec q)\,c_{ins\bar v}(\vec k),
\end{equation}
$\bar v$ denoting the valley opposite $v$.
The first term is regular density-density term,  and the second term is the inter-valley mixing term. Such a term arises due to interactions at wavevectors $\vec q= \vec q_0+\vec Q$, where $|\vec q_0|\ll |\vec Q|$. Projected directly onto the relevant Landau levels intervalley mixing term vanishes.

From the band structure, we observe that the $0LL-$ states near the crossing are localised on sublattice $A$ of the bottom layer, whereas the $0LL+$ states are equally delocalised between the bottom and top layers on sublattice $B$. After projecting onto these states, we obtain the contribution:
\begin{equation}
    \Delta V_0/V_0\approx -0.07
\end{equation}

\subsubsection{Lattice-scale corrections}
We next consider two additional corrections, following \cite{Alicea_2006}. The first arises from the finite spatial extent of the electron wavefunctions. We account for this effect by introducing an on-site Hubbard interaction $U$, while for longer-range interactions we continue to approximate the electron wavefunctions as point-like at each lattice site. The second correction arises from translating these lattice interactions into the continuum description.

We start from
\begin{multline}
  \hat H_{\rm int}=U\sum_{\vec x,i,n}\left[\frac{1}{4}n^2_{\vec x,i,n}-\frac{1}{3}\mathbf S(\vec x,i,n)^2\right]\\
+\frac{1}{2}\sum_{\substack{\vec x\ne\vec x'+\vec a(1-\delta_{ij})\\i,j,n}}
V_C\big(\vec x-\vec x'-\vec a(1-\delta_{ij})\big)\,n_{\vec x,i,n,s}\,n_{\vec x',j,n,s'},
\end{multline}
where $U$ is the Hubbard on-site interaction, $V_C(\vec x)$ the isotropic gated Coulomb interaction in real space.
The sum over $\vec x$ is the sum over triangular lattice sites, $n_{\vec x i n}=\sum_s n_{\vec x ins}$ is the occupancy at lattice site  $\vec x$ in layer $n$.  Here $\vec S^2$ denotes the relevant spin operator.

There is no interlayer interaction term, since the relevant interlayer $A$–$B$ distance is $\sqrt{d_{\rm TLG}^2+a^2}\approx d_{\rm TLG}$ and is already captured by $d$.

Since the interaction is formulated on the lattice, we first transform to momentum space to obtain its continuum representation. We introduce a momentum cutoff $\Lambda>K\gg 1/\ell_B$ and separate each interaction into intra- and intervalley scattering processes. Resolving the operators by valley index ($\pm$) then yields the general expression:
\begin{multline}
    \hat H_{\rm int} =
    -\,\alpha E^{\rm u}_C \frac{a_0}{\ell_B} \frac{1}{2\pi\ell_B^2}
    \int_{q<\Lambda}
    \frac{1}{2}\,
    \rho^{i,n,s}_{\mathbf q}\,
    \rho^{j,n,s'}_{-\mathbf q}
    \frac{d^2\mathbf q}{(2\pi)^2}
    \\
    +\,
    \beta E^{\rm u}_C \frac{a_0}{\ell_B} \frac{1}{2\pi\ell_B^2}
    \int_{q<\Lambda}
    \rho^{A,n,s}_{\mathbf q}\,
    \rho^{B,n,s'}_{-\mathbf q}
    \frac{d^2\mathbf q}{(2\pi)^2}
    \\
    +\,
    \frac{U a_0^2\sqrt{3}}{2}
    \int_q
    \left[
    \frac{1}{4}
    \rho^{i,n,s,v_1}_{\mathbf q}
    \rho^{i,n,s',v_2}_{-\mathbf q}
    \right.
    \\
    \left.
    -
    \frac{1}{3}
    \mathbf S^{i,n,v_1}_{\mathbf q}\cdot
    \mathbf S^{i,n,v_2}_{-\mathbf q}
    \right]
    \frac{d^2\mathbf q}{(2\pi)^2}
    +\hat{H}_V
    \label{eq:term_replaced}
\end{multline}
where $ \alpha\approx 0.6,\beta\approx 0.35$ are obtained numerically, and $\hat{H}_V$ collects all intervalley mixing terms. These terms vanish upon projection onto the two-band subspace, and their explicit form is provided in the Appendix~\ref{app:short_range}.

We assume that ultra-short-range interactions of this type are not appreciably screened by hBN. While screening by the graphene $\sigma$ electrons is in principle possible, we assume this effect to be negligible, such that the relevant energy scale is set by the bare Coulomb interaction, $E^{\rm u}_C \simeq 200~\mathrm{meV}$ (corresponding to $U \approx 12$--$20~\mathrm{eV}$). These short-range interactions therefore retain a stronger repulsive character than their long-range counterparts, which are substantially screened.

In the specific geometry and parameter regime considered here, this repulsion is offset by lattice-scale effects upon passing to the continuum description, so that the overall coupling $V_0$ is negative and takes the value
\begin{equation}
    \Delta V_0/V_0 \approx -0.012
\end{equation}
Importantly, the RPA screening of these short-range interactions differs from that of the long-range Coulomb interaction, and can even lead to a sign change of the effective couplings. In particular, the $U$ term is screened more strongly than the other contributions, resulting in a more repulsive $V_0$ after RPA screening.

\subsubsection{Electron-phonon coupling}
We consider the effective electron-electron interaction generated by coupling to in-plane phonons describing vibrations of nearest-neighbour sites, i.e., sites on opposite sublattices within the same layer. The dominant electron-phonon coupling (EPC) in graphene originates from modes near the $\Gamma$ and $K$ points of the Brillouin zone, which mediate intravalley and intervalley scattering, respectively \cite{zaletel_phonons,Sasaki_phonons,Mariani_phonons}. We approximate the EPC in each layer by that of an isolated monolayer, neglecting modifications of the coupling due to interlayer hybridization. We also neglect out-of-plane phonon modes, whose EPC is expected to be substantially weaker than that of the relevant in-plane modes \cite{bernevig_phonons,more_phonon_values,trilayer_modes_phonons}.

The effective Hamiltonian from $K$-point phonons between nearest neighbours is
\begin{equation}
    \hat H_{K\text{-ph}}=-\frac{1}{2}\frac{N_\Phi}{N}\frac{|g|^2}{\hbar\omega}\sum_{ss'}\int\frac{d^2\vec q}{(2\pi)^2}\,K^{invs}_{\vec q}K^{jnv's'}_{-\vec q},
\end{equation}
with $i$ the sublattice, $n$ the layer, $v$ the valley, $s,s'$ spins, and
\begin{equation}
    K^{insv}_{\vec q}=\sum_{\vec k}c^{\dagger}_{insv}(\vec k+\vec q)\,c_{\bar i n s\bar v}(\vec k).
\end{equation}
Here $|g|$ is the electron-phonon coupling strength, $\omega$ the phonon frequency, $N_\Phi$ the number of magnetic unit cells, and $N$ the number of graphene unit cells. Using $|g|=1~\mathrm{eV}$ and $\hbar\omega=0.166~\mathrm{eV}$ \cite{monolayer_values}, this energy scale is $E_{\rm phonon}=\frac{N_\Phi}{N}\frac{|g|^2}{\hbar\omega}\approx0.68~\mathrm{meV}$ at $B=10~\mathrm T$. 

Similarly, $\Gamma$-point phonons between nearest neighbours yield
\begin{equation}
     \hat H_{\Gamma\text{-ph}}=-\frac{1}{2}\frac{N_\Phi}{N}\frac{|g|^2}{\hbar\omega}\int\frac{d^2\vec q}{(2\pi)^2}\,J^{i,n,s',v}_{\vec q}J^{j,n,s,v'}_{-\vec q},
\end{equation}
with $i,j$ sublattices, $n$ the layer, $s,s'$ spins, $v,v'$ valleys, and
\begin{equation}
    J^{insv}_{\vec q}=\sum_{\vec k}c^{\dagger}_{insv}(\vec k+\vec q)\,c_{\bar i n s v}(\vec k).
\end{equation}
Using $|g|=0.6~\mathrm{eV}$ and $\hbar\omega=0.166~\mathrm{eV}$, \cite{monolayer_values}  we obtain $E_{\rm phonon}\approx0.4~\mathrm{meV}$ at $B=16~\mathrm{T}$. We also estimate the contribution from next-nearest-neighbour vibrations to be approximately $0.02~\mathrm{meV}$, and therefore neglect it in the following. 

Due to valley-lattice locking, the contribution from the $\Gamma$ phonons vanishes upon projection onto the two bands, leaving only the $K$-phonon contribution. We obtain $\Delta V_0/V_0\approx 0.001 $ if both bands are the same spin and $\Delta V_0 = 0$ if they are the opposite spin.

\subsection{Random-Phase Approximation}
\label{subsec:RPA_effect}
As discussed in the previous section, the corrections arising from the
lattice-orbital structure and electron--phonon coupling are not sufficiently
large, by themselves, to drive the composite Fermi liquid (CFL)
into an incompressible phase. It is therefore important to account for
Landau-level mixing. Here we incorporate these effects approximately through
the random-phase approximation (RPA), which captures the screening of the
effective interaction by virtual particle--hole excitations outside the
low-energy subspace.

\begin{figure}
\includegraphics[width=\linewidth]{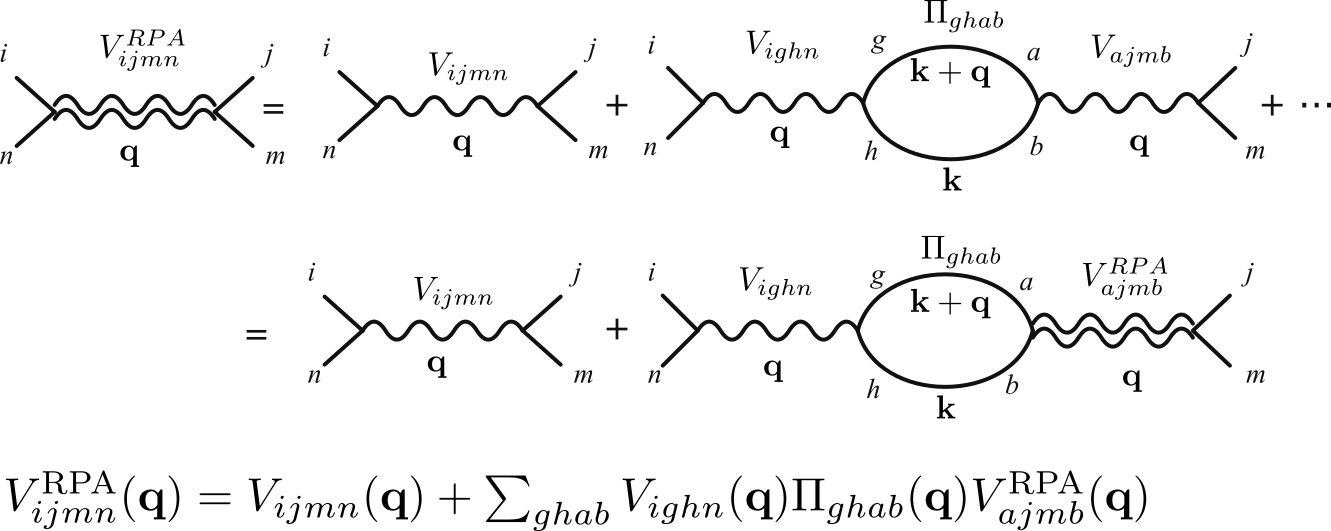}
    \caption{\textit{RPA screening of the Coulomb interaction.}
Diagrammatic representation of the Dyson equation for the RPA-screened interaction $V^{\rm RPA}$. Single wavy lines denote the bare Coulomb interaction $V$, while double wavy lines denote the RPA-screened interaction $V^{\rm RPA}$. Solid lines represent single-particle propagators, and $\Pi$ denotes the polarization four-tensor. The indices $i,j,m,n,g,h,a,b$ collectively represent the suppressed layer, valley, sublattice, Landau-level, and spin degrees of freedom. The bottom equation gives the corresponding self-consistent Dyson equation for $V^{\rm RPA}$.}
        \label{fig:RPA}
\end{figure}

Our procedure consists of three steps:
\begin{enumerate}
    \item We first calculate the perturbative corrections to the interaction,
    which we denote by $\Delta \hat V$. These include the ultra-short-range
    (point-like) interaction terms generated by the microscopic perturbations
    discussed above.

    \item We then apply the RPA diagrams to the full interaction, using the bare single-particle dispersion. In particular, we solve the four-tensor Dyson equation \begin{equation}
\begin{aligned}
V_{ijmn}^{\rm RPA}(\vec{q})
&= V_{ijmn}(\vec{q}) \\
&\quad
+ \sum_{g,h,a,b}
V_{ighn}(\vec{q})
\Pi_{ghab}(\vec{q})
V_{ajmb}^{\rm RPA}(\vec{q}).
\end{aligned}
\end{equation}
Here, the indices $i,j,m,n,g,h,a,b$ collectively represent the suppressed layer, valley, sublattice, Landau-level, and spin degrees of freedom, while $\Pi_{ghab}(\vec{q})$ is the polarization four-tensor and $V_{ijmn}(\vec{q})$ is the bare interaction four-tensor. The corresponding diagrams are shown in Fig.~\ref{fig:RPA}.

Furthermore, there is a subtlety in applying RPA to the short-range interaction. Namely, we have
$\Delta V_0 \int \frac{d^2q}{(2\pi)^2}\, :\!\rho^{\uparrow}_{\mathbf q}\,
    \rho^{\uparrow}_{-\mathbf q}\!: \;=\; \frac{\Delta V_0}{2\pi}\int d^2x\,
    :\!\rho^{\uparrow}(\mathbf{x})\,\rho^{\uparrow}(\mathbf{x})\!:\; = 0,$ by the Pauli principle, since two same-spin fermions cannot occupy the
same point. RPA keeps only the Hartree (density) bubble and does not
generate exchange, so a naive RPA treatment of this term is ambiguous:
parts of the Hartree contribution are expected to be cancelled by the
Fock term, which is absent from the RPA expansion. We discuss this
point in detail in Appendix~\ref{app:RPA_short_range}.
    
    \item Finally, after performing the RPA resummation as depicted in Fig.~\ref{fig:RPA}, we project the
    resulting interaction onto the two Landau levels relevant for the
    many-body calculation.
\end{enumerate}

After summing the RPA diagrams, we expand the resulting interaction
to first order in $\Delta\hat V$. The zeroth-order term gives the
RPA-screened Coulomb interaction, while the term linear in $\Delta\hat V$
gives the RPA-renormalised short-range interaction. After projection onto the two low-energy bands, we recover the effective Hamiltonian introduced in Eq.~\eqref{eq:hamiltonian_main_ref}:
\begin{equation}
\label{eq:RPA_effective_H}
\begin{split}
\hat{H} \approx \frac{1}{2}\sum_{s,s'}
\int \frac{d^2\vec q}{(2\pi)^2}V^{\rm RPA}_C(\vec q)\, \rho^{0}_s(\vec q)\, \rho^{0}_{s'}(-\vec q)
\\+ \Delta V^{\rm RPA}_0 \int \frac{d^2\vec q}{(2\pi)^2} \rho^0_1(\vec{q})\, \rho^0_2(-\vec{q}) 
\end{split}
\end{equation}

The first term in Eq.~\eqref{eq:RPA_effective_H} is the RPA screened Coulomb
interaction,
\begin{equation}
    V_C^{\rm RPA}(q)
    =
    \frac{V_C(q)}
    {1+\Pi_0(q)V_C(q)},
\end{equation}
where $\Pi_0(q)$ is the static polarization function.

The second term in Eq.~\eqref{eq:RPA_effective_H}
contains the
RPA-renormalised short-range perturbations. In contrast to the Coulomb
interaction, its screened form depends on the full polarization tensor
$\Pi_{ijkl}$; the explicit expression is given in
Appendix~\ref{app:RPA}. The RPA diagrams generally introduce a
momentum dependence into this interaction. Upon projection onto the LLL,
however, we find that the zeroth pseudopotential $\Delta V_0^{\rm RPA}$
dominates, exceeding $\Delta V_1^{\rm RPA}$ by one to two orders of magnitude.
We therefore retain only $\Delta V_0^{\rm RPA}$ in
Eq.~\eqref{eq:RPA_effective_H}. This is consistent with findings in the literature \cite{chunli}.

The effect of the RPA resummation is thus two-fold. First, the bare Coulomb
interaction $V_C$ is replaced by the screened interaction
$V_C^{\rm RPA}$. Second, the microscopic ultra-short-range perturbations are
renormalised into an effective interaction $V_0^{\rm RPA}$.

There is an additional complication arising from the single-particle dispersion. In the regime of interest, the characteristic interaction scale is comparable to the energy separation between the relevant bands. Consequently, the dispersion itself can receive appreciable Hartree--Fock corrections. At the same time, the RPA-screened interaction depends on the polarization function and hence on the dispersion used to calculate it. A fully self-consistent treatment would therefore require iterating the Hartree--Fock and RPA calculations until both the dispersion and the screened interaction converge \cite{Hedin1965, HolmVonBarth1998}. Such a self-consistent Hartree--Fock--RPA scheme is beyond the scope of the present work. Instead, we evaluate the RPA screening using either the bare dispersion, and discuss the Hartree--Fock-renormalised dispersion in Appendix~\ref{app:dispersion_renormalisation}.

Because of this sensitivity and also the microscopic modelling uncertainty (discussed in more detail in Sec.~\ref{sec:realistic_phase_diagram}), the polarization function cannot be determined with sufficient precision to justify treating its microscopic value as a fixed parameter. Following Ref.~\cite{abanin_papic}, we therefore parametrise its momentum dependence as
\begin{equation}
    \Pi_0(q)=\frac{A_\Pi}{E_C\ell_B^2}\tanh\!\left[b(q\ell_B)^2\right],
    \label{eq:polarization_parametrization}
\end{equation}
where $b$ controls the momentum scale of the crossover and $A_\Pi$ sets the overall magnitude of the polarization. Factors of $\ell_B$ and $E_C$ are fixed so that both $A_{\Pi}$ and $b$ are dimensionless. We keep $A_\Pi$ fixed, since it is set primarily by remote bands and is therefore expected to be relatively robust, while varying $b$. Furthermore, as noted in Ref.~\cite{abanin_papic}, the polarization function is mostly controlled by the product $A_\Pi b$, rather than by $A_\Pi$ and $b$ separately.

\section{ Phase Diagram of the Realistic Model}
\label{sec:realistic_phase_diagram}

We now present the main results of our analysis of the realistic, RPA-screened interaction. Our goal in this section is first to establish the phase diagram of the screened model and then to determine where the microscopic system lies within this phase diagram. The central result is that RPA screening substantially enhances the tendency toward strong pairing. In the parameter regime accessible to our calculations, we find evidence for a direct transition from a CFL-like regime to the strong-pairing phase, with no intervening weak-pairing regime. For representative RPA parameters, this transition occurs already near $\Delta V_0/V_0\simeq -0.2$, substantially closer to the unperturbed interaction than in the absence of screening. Here, $V_0$ denotes the zeroth pseudopotential of the relevant RPA-screened interaction.

\begin{figure}
\includegraphics[width=\linewidth]{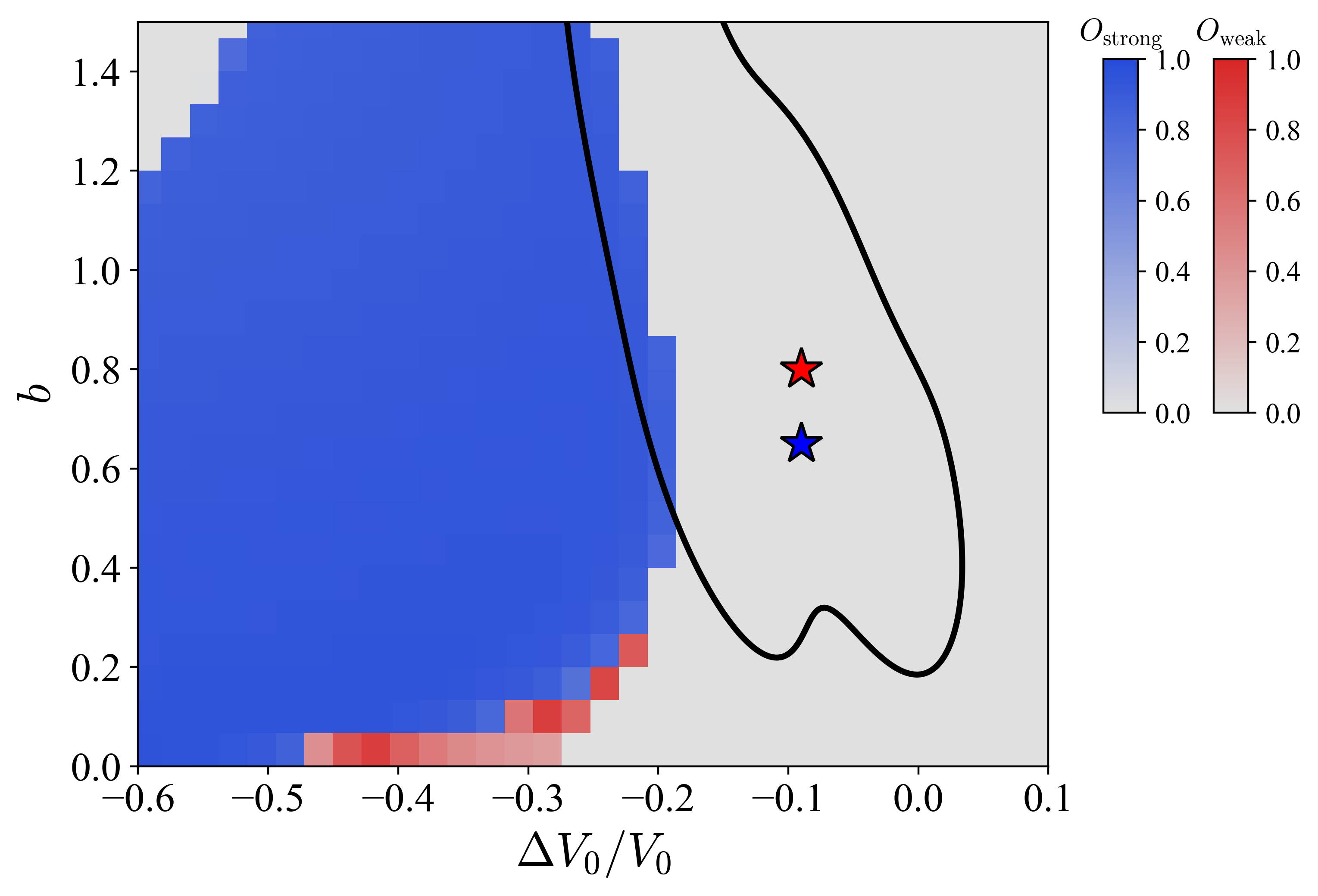}
  \caption{\textit{Overlap with the weak- and strong-pairing phases.}
  Squared overlap of the exact-diagonalization ground state with the weak-pairing (red) and strong-pairing (blue) trial states as a function of $\Delta V_0/V_0$ and $b$, for RPA parameter $A_{\Pi}=1.95 $ and $N=8$ electrons. At each point, the colour indicates the trial state with the larger overlap, while the colour intensity gives the magnitude of the corresponding squared overlap, $|\langle \psi_{\rm trial} | \psi_{\rm ED} \rangle|^2$.
    The grey elliptical contours indicate the range of parameters consistent with the experimentally reported values.
The blue and red stars mark the values obtained using the bare dispersion reported in Ref.~\cite{chanda} at $\nu=\frac{7}{2}$ and $\nu=\frac{5}{2}$, respectively. RPA screening strongly suppresses the weak-pairing region and places the physical parameter estimates close to the phase boundary, indicating that the system lies near the transition between the two competing pairing phases. We believe that grey region on the right corresponds to the CFL, as we discuss in the Appendix~\ref{app:realistic_numerics}.
}
        \label{fig:RPA_overlaps}
\end{figure}

Our analysis proceeds in two stages. We first construct the phase diagram by treating $b$ and $V_0^{\rm RPA}$ as independent parameters. This allows us to determine which phases are favoured over a broad range of screening strengths and short-range interactions without relying on a precise microscopic estimate of either quantity. We subsequently evaluate the polarization and the RPA-renormalised short-range interaction using the dispersion associated with the microscopic Hamiltonian. The resulting values of $b$ and $V_0^{\rm RPA}$ then identify the points in the phase diagram corresponding to the physical system.

Since it is computationally prohibitive to map out the full two-dimensional phase diagram using DMRG, we first explore it using exact diagonalization (ED). The resulting overlaps with representative weak- and strong-pairing trial states are shown in Fig.~\ref{fig:RPA_overlaps}. Within the parameter regime accessible to ED, the inclusion of RPA screening strongly suppresses the weak-pairing regime. Instead, the system evolves directly from the gapless state realised near $\Delta V_0=0$ into a region with substantial overlap with the strong-pairing state. RPA screening also shifts this transition substantially toward weaker short-range correction: for representative RPA parameters, the transition occurs around $\Delta V_0/V_0\simeq -0.2$, compared with approximately $\Delta V_0/V_0\simeq -0.5$ without RPA screening. Thus, although screening removes the intermediate weak-pairing regime found previously, its overall effect is to bring the unperturbed system significantly closer to an incompressible paired phase.

The overlap with the Halperin 331 state remains small throughout the relevant parameter regime.  Moreover, the ground state remains unpolarised or partially polarised, disfavouring a description in terms of a fully polarised single-component state such as Pfaffian or anti-Pfaffian. This two-component interpretation is also consistent with the experimentally observed widths of the incompressible regions. In Appendix~\ref{app:incompressible_widths}, we vary the Landau-level splitting and determine the value at which the incompressible unpolarised state is destroyed. This allows us to extract the corresponding width $\Delta D$ of the incompressible region, which we find to be consistent with experiment.

\begin{figure}

\includegraphics[width=\linewidth]{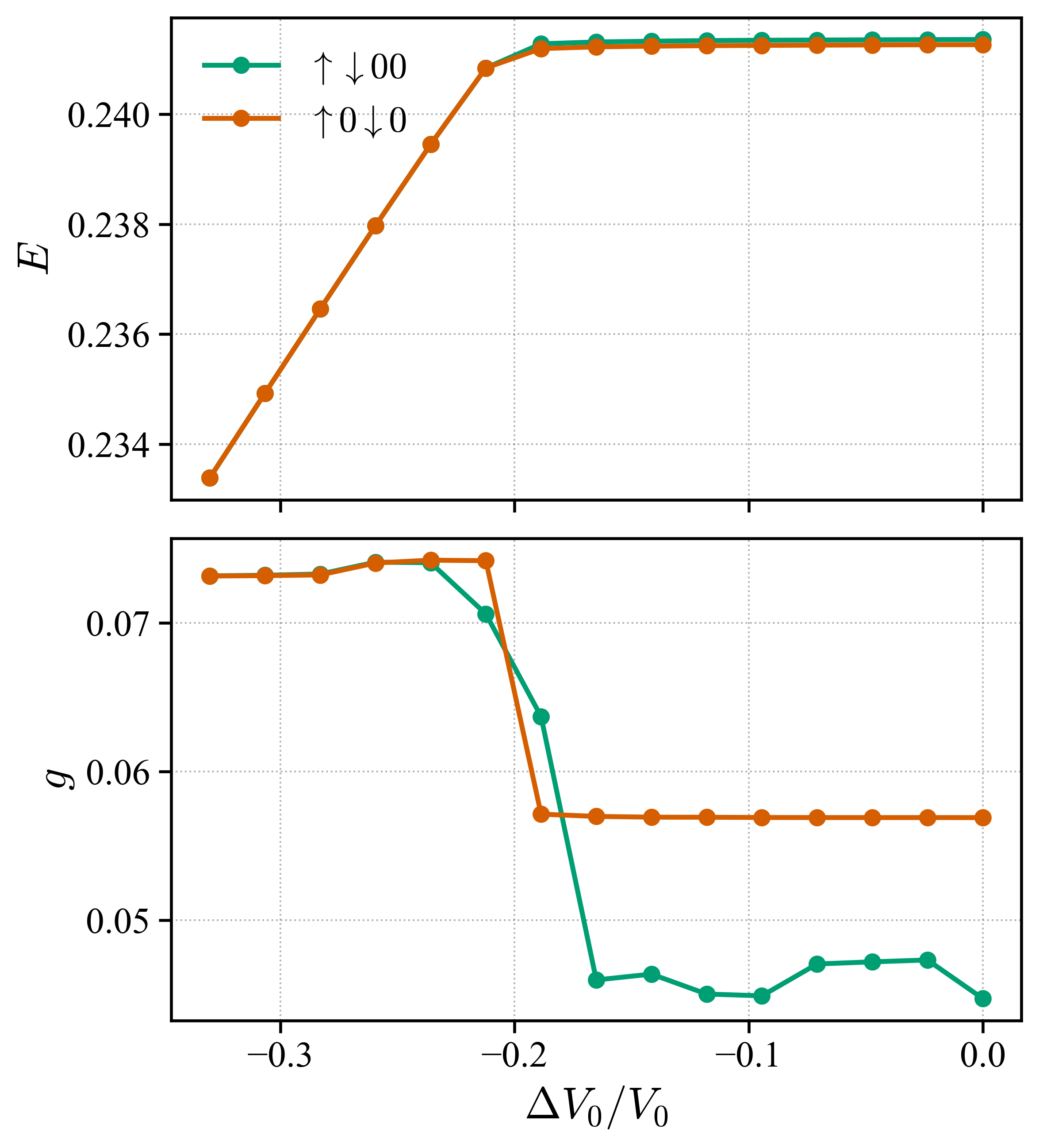}
    \caption{\textit{Evidence for a first-order CFL-to-strong-pairing transition.}
Energy $E$ (top) and correlation function $\langle \rho_{1}(0)\rho_{2}(0)\rangle$ (bottom) as a function of $\Delta V_0/V_0$ for fixed RPA values $A_{\Pi}=1.95$, $b=0.6$. A sharp change near $\Delta V_0/V_0 \simeq -0.2$ is visible in both quantities: the energy has a cusp and starts to decrease rapidly after that point.
    Similarly, we have discontinuous jump in $g=\langle \rho_{1}(0)\rho_{2}(0)\rangle$ at $\Delta V_0/V_0=-0.19$. 
    We interpret these coincident features as evidence for a first order phase transition at $\Delta V_0/V_0 \simeq -0.2$, consistent with a direct transition from the composite Fermi liquid (CFL) to a strongly paired state. The data was taken at $L_y=14\ell_B$ and $\chi=8000$.}
\label{fig:RPA_DMRG_correlation}
\end{figure}

To further characterise this transition beyond the system sizes accessible with ED, we perform DMRG calculations along a cut through the RPA phase diagram, varying $\Delta V_0$ while fixing the RPA parameters to $b=0.6 $ and $A_{\Pi}=1.95$. Figure~\ref{fig:RPA_DMRG_correlation} shows the ground-state energy and the correlator $g=\langle \rho_1(0)\rho_2(0)\rangle$ along this cut. 
Both quantities display a pronounced feature near $\Delta V_0/V_0\simeq-0.2$. Upon entering the more attractive side of the transition, the energy exhibits a cusp followed by a rapid decrease, while the correlator $g$ exhibits a discontinuous jump at approximately the same value of $\Delta V_0/V_0$. The coincidence of these features provides direct evidence for a phase transition near $\Delta V_0/V_0\simeq-0.2$. Since the jump in $g$ is clearly discontinuous and the energy exhibits a cusp, this is a first-order phase transition.

\begin{figure}
\vspace{0.45cm}
\includegraphics[width=0.49\linewidth]{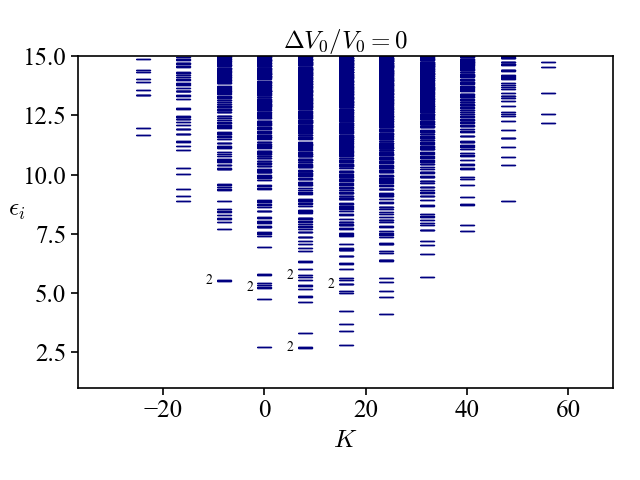}
\includegraphics[width=0.49\linewidth]{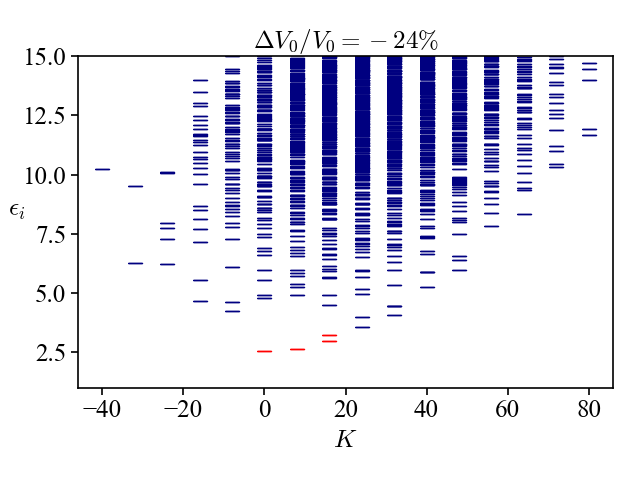}
\includegraphics[width=\linewidth]{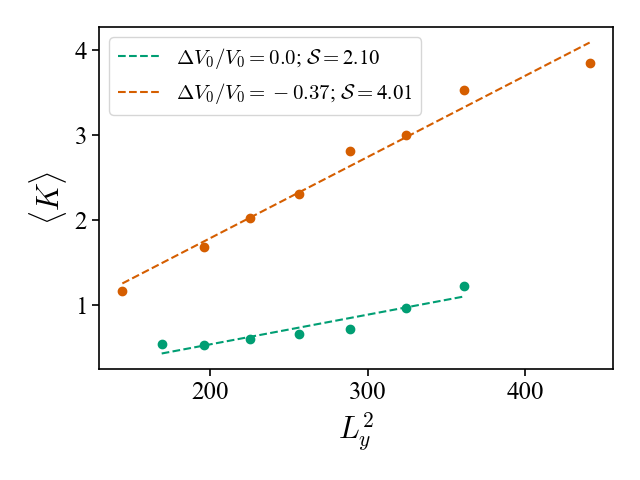}
\caption{\textit{Entanglement spectra and the topological shift.}  \emph{Top:} ES at $\Delta V_0/V_0=0$ (left) and $\Delta V_0/V_0=-0.24$ (right). For $\Delta V_0/V_0<-0.2$, the low-lying ES clearly exhibits the characteristic $1,1,2,\ldots$ counting expected for the strongly paired state, providing further evidence for the identification of this phase. At $\Delta V_0/V_0=0$, the low-lying entanglement spectrum is inconsistent with any of the trial states and varies with $\chi$, indicating a gapless phase. The data is taken at $\chi=8000$ and $L_y=15\ell_B$.
 \emph{Bottom:} Momentum polarization $\langle K\rangle$ as a function of $L_y^2$ for $\Delta V_0/V_0=0.0$ and $\Delta V_0/V_0=-0.37$, with linear fits used to determine the topological shift $\mathcal{S}$. For $\Delta V_0/V_0=-0.37$ we obtain $\mathcal{S}=4.01$, consistent with strong pairing. For $\Delta V_0/V_0=0.0$ the data are considerably less well resolved, as one might expect for a gapless state.}

\label{fig:RPA_DMRG_ES}
\end{figure}

The nature of the phase on the attractive side can be identified more directly from its entanglement spectrum and topological shift. As shown in Fig.~\ref{fig:RPA_DMRG_ES}, for $\Delta V_0/V_0<-0.2$ the low-lying entanglement spectrum develops the $1,1,2$ counting, while also exhibiting a topological shift $\mathcal{S}\approx 4$ — both signatures of the strong-pairing phase. Together with the ED overlaps, this provides strong evidence that the phase reached upon crossing the transition is the strong-pairing state. We therefore identify the feature near $\Delta V_0/V_0\simeq-0.2$ as the onset of strong pairing.

Our results identify the phase on the other side of the transition, including the unperturbed point $\Delta V_0=0$, as a gapless phase (most likely a CFL). Importantly, although the two momentum sectors are nearly degenerate in energy in this phase, the correlator $g$ takes different values in each sector. This indicates that the two sectors are not topologically degenerate, since they can be distinguished by a local correlator.
At $\Delta V_0=0$, we further examine the scaling of the correlation length with $\chi$ and $L_y$. It grows without saturation as either $\chi$ or $L_y$ is increased, indicating gaplessness (see App.~\ref{app:realistic_numerics}). The entanglement spectrum likewise varies with system size and does not exhibit the structure characteristic of a gapped phase.

The torus spectrum provides further evidence for the CFL. Rather than a well-defined manifold of topologically degenerate ground states, we find a larger number of low-energy states whose number and structure vary with system size. Such size-dependent low-energy structure is naturally associated with a CFL, where the finite-size spectrum reflects the shell structure of the composite Fermi surface, rather than with a gapped topological phase possessing a fixed topological ground-state degeneracy. Taken together, these results consistently identify the phase as a CFL. The corresponding correlation-length and torus-energy-spectrum data, with additional supporting evidence for the CFL, are presented in App.~\ref{app:realistic_numerics}.

Our data therefore suggest a direct first order transition from gapless CFL regime to the gapped strong-pairing phase near $\Delta V_0/V_0\simeq-0.2$. RPA screening thus substantially shifts the strong-pairing instability toward the microscopic interaction while suppressing the intermediate weak-pairing regime.

Finally, we identify the regions of the phase diagram corresponding to parameters obtained from the microscopic model. The precise tight-binding parameters are not known and may themselves be renormalised by various interaction effects. We therefore consider a reasonably broad range of parameters consistent with experimentally reported values and map the corresponding values onto the phase diagram. We find that a substantial portion of this parameter range lies within the strong-pairing phase. The range of parameters used is provided in the Appendix~\ref{app:tight_binding}. The stars in Fig.~\ref{fig:RPA_overlaps} correspond to the specific tight-binding parameters used in Ref.~\cite{chanda}. Although these points remain in the compressible phase, they lie substantially closer to the strong-pairing transition than in the model without RPA screening.
These results suggest that the realistic system may lie near a strong-pairing instability. However, given the approximate nature of the RPA treatment and the uncertainties in the microscopic parameters, this conclusion should not be regarded as definitive. Further numerical and analytical studies are needed to establish whether the strong-pairing phase is realised in the microscopic model.

Finally, we note that although the single-particle spectrum contains four symmetry-related Landau-level crossings, incompressible states are observed experimentally at only three of them. This suggests that the stability of the incompressible phase is sensitive to Landau-level mixing with more remote bands, rather than being determined solely by the four bands participating directly in the crossing. It also indicates that the incompressible phase is relatively fragile, consistent with the system lying close to a competing compressible state.

\section{Extension to Monolayer Graphene}
\label{sec:extension_monolayer}

Incompressible states have also been observed at half-integer filling of the zeroth Landau
level in monolayer graphene on hBN substrate, in narrow windows of magnetic field at $\nu=\pm1/2$ but
not at $\nu=\pm3/2$~\cite{Zibrov2018}. It is natural to ask whether the mechanism
identified above for trilayer graphene --- short-range interaction corrections
enhanced by RPA screening --- can also account for these observations.

The two systems differ in an important respect. In trilayer graphene the relevant Landau-level crossing is already present at the single-particle level and is tuned explicitly by the displacement field $D$ through the Slonczewski--Weiss--McClure spectrum (Sec.~\ref{sec:model}), with the incompressible state appearing in a narrow window of $D$. 
In monolayer graphene, the bare single-particle spectrum of the four flavours $0LL+\uparrow$, $0LL+\downarrow$, $0LL-\uparrow$, $0LL-\downarrow$ is set by the sublattice splitting $\Delta_{AB}$ (approximately $B$-independent) and the Zeeman energy. The incompressible state appears in a narrow window of magnetic field $B$~\cite{Zibrov2018}, but this is not generated by Landau-level crossing in the bare dispersion.
Instead, the effective crossing used to define a two-band problem only emerges once interactions are included.
In particular, interactions generate a valley/sublattice anisotropy, captured phenomenologically by the constants $g_z, g_\perp$ of Refs.~\cite{kharitonov,Zibrov2018,hegde_sodemann,anisotropy_2,anisotropy_3,anisotropy_4,anisotropy_5,anisotropy_6}, through the anisotropic Hamiltonian:
\begin{multline}
    \hat H_{\rm aniso} =2\pi \ell^2_B\frac{a_0}{\ell_B}E_C\int d^2 \vec x
\Big[\frac{g_z}{2}\big(\hat\rho_+(\mathbf x)-\hat\rho_-(\mathbf x)\big)^2
\;\\
+\;2g_\perp\hat J_+(\mathbf x)\hat J_-(\mathbf x)\Big],
\end{multline}
where $a_0$ is the graphene lattice constant and
$\hat J_+ = \sum_{s}\hat \psi^{\dagger}_{+,s}(\vec x)\hat \psi_{-,s}(\vec x) = \hat J_-^\dagger$
is the inter-valley coherence operator, with $s$ labelling spin and $\hat\psi$ the electron annihilation operator. There has been a substantial body of work on estimating and renormalising the short-range anisotropic couplings $g_z, g_\perp$ (or their generalisations beyond the ultra-short-range ansatz) at integer filling $\nu =0$~\cite{chunli,anisotropy_1,anisotropy_2,anisotropy_3,Coissard2022}, whereas comparatively little attention has been paid to the fractionally
filled case considered here~\cite{hegde_sodemann,anisotropy_6}.

The anisotropic Hamiltonian scales with $B$ (through $1/\ell_B$)  while $\Delta_{AB}$ does not. At $\nu=-1/2$, taking $0\mathrm{LL}_{+\uparrow}$ to be fully filled, and calculating the Hartree--Fock renormalised dispersion, it is the remaining two bands, $0\mathrm{LL}_{+\downarrow}$ and
$0\mathrm{LL}_{-\downarrow}$, that cross as $B$ is tuned. In
particular, denoting by $\varepsilon_{0\mathrm{LL}\pm\downarrow}(B)$ the
Hartree--Fock single-particle energy of the corresponding band, one finds
\begin{equation}\label{eq:monolayer_crossing}
\varepsilon_{0\mathrm{LL}+\downarrow}(B) - \varepsilon_{0\mathrm{LL}-\downarrow}(B)
\;=\; 2\Delta_{AB} - 2 g_z\,\frac{a_0}{\ell_B}E_C,
\end{equation}
while the fourth band, $0\mathrm{LL}_{-\uparrow}$, remains higher in energy throughout and can be neglected.
It is therefore the sweep of $B$ that tunes the two bands through this interaction-generated crossing, in contrast to trilayer graphene where $D$ tunes a single-particle crossing. $\nu=+1/2$ follows analogously by particle--hole symmetry.
The most obvious issue with projecting the system onto two bands is that the band separation, as well as the interactions, are generated by $g_z,g_{\perp}$. A potential way to resolve this would be to use a self-consistent Hartree--Fock--RPA treatment, which we do not attempt here. Instead, we use the two-band projection as a crude model to assess the plausibility of obtaining the weak $d$-paired or strong-paired state, using both first-principles theoretical calculations and experimental data.

\begin{figure}
    \centering
    \includegraphics[width=\linewidth]{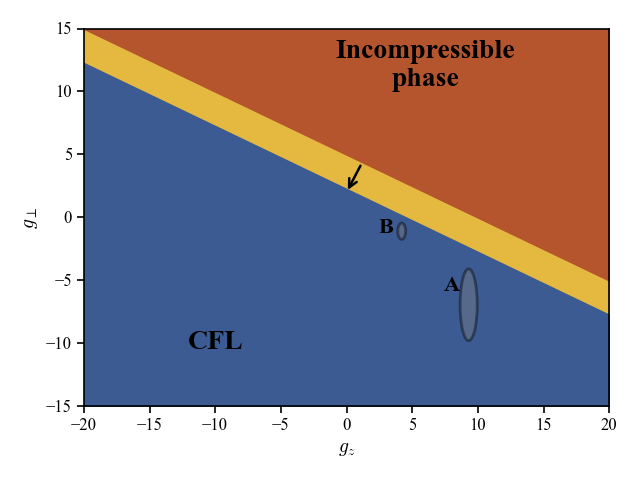}
        \caption{\textit{Monolayer phase diagram.} Estimated phase diagram as a function of the anisotropy constants $g_z,g_\perp$ at $\nu=-1/2$. The incompressible-to-CFL transition is evaluated using the RPA-screened interaction with $A_\Pi=0.5,\ b=0.4$. The yellow band shows how the transition line shifts across the field range $B=8$--$30\,$T typically accessed in experiment (arrow indicates the direction of increasing $B$); axes are restricted to experimentally relevant~\cite{Zibrov2018,kharitonov} range of $g_z,g_\perp$ values. A large fraction of this parameter space lies within the incompressible phase, corresponding to the strongly-paired state. The grey ellipse $A$ indicates the range of parameters obtained by combining estimates of $g_{\perp}$ from Ref.~\cite{Young2014} with values of $g_z$ extracted by fitting the sample A data of Ref.~\cite{Zibrov2018}, whereas ellipse $B$ indicates the parameter range inferred from Ref.~\cite{magnon}. For both estimates, we assume $\epsilon_{\rm hBN}=4.3$.
        }
    \label{fig:phase_diagram_monolayer}
\end{figure}

Projecting our microscopic short-range corrections (lattice-scale and electron--phonon, Appendix~\ref{app:short_range}) directly onto this two-band picture, as we did for trilayer graphene, gives $\Delta V_0/V_0 \approx -0.03$ before RPA screening and about $-0.01$ after screening using the naive bare dispersion at $B=30~\rm T$.
This is far short of the strong-pairing transition, which occurs near $\Delta V_0/V_0\simeq-0.2$ in our model. In the trilayer case, the dominant contribution arises from the interlayer separation, which is absent in the present case, thus rendering $\Delta V_0/V_0$ much smaller than in the trilayer case.

We turn to experimental data for $g_z,g_{\perp}$ instead. 
Several experiments have constrained $g_z,g_\perp$ from the $\nu=0$ phase diagram~\cite{Zibrov2018,Young2014,magnon,Coissard2022,yazdani}, but the resulting estimates vary substantially from sample to sample.
For example, the estimate based on sample A in Ref.~\cite{Zibrov2018} yields $g_z \approx 10$, whereas Ref.~\cite{magnon} gives $g_z \approx 4$, indicating a substantial discrepancy between the two estimates. A similar variation is found for $g_{\perp}$: Ref.~\cite{magnon} gives $g_{\perp} \approx -1$, whereas the estimate based on Ref.~\cite{Young2014} yields $g_{\perp} \approx -5$.
Moreover, most of these experiments constrain only one of $g_z,g_\perp$ at a time, rather than providing a joint estimate from a single device. Compounding this, the extracted values are also fairly sensitive to the choice of permittivity $\epsilon_{\rm hBN}$ used in the fit, adding a further source of ambiguity to any quoted $g_z,g_\perp$. We discuss these various experiments and the resulting choices of $g$ in Appendix~\ref{app:monolayer}.

We present the phase diagram in Fig.~\ref{fig:phase_diagram_monolayer}, which shows, as a function of $g_z,g_\perp$ at half-integer filling, whether the system achieves an incompressible (strong-paired) state or a compressible CFL. The phase diagram is restricted to the range of $g_z,g_\perp$ and $B=8$--$30\,$T considered experimentally realisable. As can be seen from the figure, the incompressible phase occupies a large part of the phase diagram. However, the parameters extracted from experiments (ellipses $A$ and $B$) fall on the compressible, CFL side of the transition, albeit relatively close to the transition line. There is large uncertainty in $g_z,g_\perp$: the extracted values vary substantially from sample to sample, are strongly sensitive to the interaction details, and rely on a somewhat unjustified Hartree--Fock approximation. Hence, it remains plausible that the physical parameters actually lie on the incompressible side of the transition. 

Separately, the estimate omits an additional isotropic short-range term, $g_0\frac{a_0}{\ell_B}E_C\int d^2\vec x\,\hat\rho(\vec x)^2$, for which no experimental estimate currently exists. Since most of the lattice-scale corrections we compute are attractive, it is plausible that $g_0<0$, which would push the system  into the strong-pairing regime.

We note that $g_z,g_\perp$ here are experimentally extracted values, so any RPA screening of the anisotropic interaction is already built into them; we are not separately screening $g_z,g_\perp$ with RPA. The role of RPA in this calculation is therefore distinct: it is used only to determine where the transition in $\Delta V_0/V_0$ occurs, not to renormalise $g_z,g_\perp$ themselves.

Importantly, it has been proposed that $g_z,g_\perp$ are substantially enlarged relative to their bare (unrenormalised) values, by interaction renormalisation~\cite{kharitonov}, which the RPA with bare dispersion does not capture well. It is possible that using RPA with a self-consistently renormalised dispersion resolves this by scaling $g_z,g_\perp$ up to their observed values, but it is also possible that diagrams beyond RPA are needed to do so. Indeed, Ref.~\cite{chunli}, working to second order in exchange diagrams, finds that such diagrams significantly enhance $g_z,g_\perp$.
It is furthermore possible that no diagrammatic expansion is sufficient here: at the fields relevant to experiment, Landau-level mixing may be strong enough that the system sits deep in a regime where a perturbative diagrammatic treatment cannot accurately capture the physics.

We conclude this section by discussing the overall plausibility of the monolayer scenario.
We concede that the monolayer estimate is less compelling than the trilayer result, as it relies on experimentally-fitted parameters, which we are unable to reconstruct microscopically, and the projection onto the two-band model is less justified. Nevertheless, taken at face value, our calculation suggests that the same qualitative picture --- short-range interactions reshaped by screening, favouring a strongly paired state near a Landau-level crossing --- naturally extends
from trilayer to monolayer graphene. Future theoretical work with a more rigorous treatment of the interaction effects in such a multi-band strongly-coupled problem would be required to reach a more solid understanding of the problem.

\section{Conclusion}
\label{sec:outlook}

The problem considered in this work arises when two Landau levels with different orbital character are tuned close to degeneracy, forming an effective two-component system. At half filling of such a system, experiments observe robust incompressible states. The central question is which topological phase is realised and what microscopic interactions stabilise it. Despite considerable experimental interest and a number of proposed candidate states, the problem has received remarkably little quantitative theoretical attention. In particular, there has been no systematic comparison of the candidate phases or quantitative assessment of the mechanisms proposed to stabilise them. In this work, we provide such an analysis by directly comparing the energetics of the candidate states and evaluating the strength of the proposed stabilisation mechanisms. We identify short-range interactions and Landau-level (LL) mixing as the dominant effects and show that, together, they can stabilise the strong-pairing phase.

A variety of interpretations of the incompressible states observed near Landau-level degeneracy have been put forward in the literature. Most works have argued for two-component phases \cite{Zibrov2018,parton,trilayer_expt}, including the Halperin 331 state \cite{trilayer_expt} and a weakly paired $d$-wave state \cite{parton}. The latter work, however, neither considers the strong-pairing regime nor accounts for RPA screening, which we find to be essential in determining the energetic ordering of the competing phases.  
By contrast, Ref.~\cite{chanda} identifies the experimentally observed states with the Pfaffian and anti-Pfaffian phases based, in part, on the reported daughter states. We do not regard the present experimental evidence as sufficient to unambiguously establish this identification. We remind the reader that the presence of a daughter state does not necessarily imply the presence of a parent.  Furthermore, the daughter states have not been independently confirmed, for example through thermal Hall measurements, and their reported filling factors carry non-negligible experimental uncertainty. The expected daughter states occur at $\nu=8/15\approx 0.533$ and $\nu=8/17$ in the strong-pairing regime \cite{daughter_states}. In Ref.~\cite{chanda}, the relevant daughter state is observed at $\nu=7/13\simeq0.538$, which lies close to $\nu=8/15$: indeed, $|8/15-7/13|\simeq 0.005$, comparable to the experimental uncertainty reported in that work. The reported daughter states therefore do not, by themselves, rule out the strong-pairing interpretation. This leaves considerable scope for further experimental work to establish the topological order of the incompressible phases. An additional useful experimental measurement would be a finer sampling of the gap near the transition, which could help determine whether the transition from the CFL to the incompressible state is continuous or first order.

We have treated LL mixing in its simplest form, by incorporating screening from nearby Landau levels within RPA and subsequently projecting the resulting effective interaction into the active Landau levels. Despite the simplicity of this approximation, we find that LL mixing can modify the interaction sufficiently to make the stabilisation of the strong-pairing phase plausible. An important feature of our projected Hamiltonian is that it is SU(2) symmetric in pseudospin. Explicit admixture of nearby Landau levels may break this symmetry and generate effective interactions absent from the RPA-projected description, potentially altering the competition between candidate phases.

A natural direction for future work is therefore to retain a restricted set of nearby Landau levels and study the resulting enlarged Hilbert space using restricted exact diagonalisation and DMRG. Such calculations could determine whether the strong-pairing phase found here survives explicit LL mixing, is destabilised in favour of Pfaffian or anti-Pfaffian order, or gives way to a qualitatively different phase. More generally, such calculations would establish whether the stabilisation mechanism identified here survives beyond the projected RPA description.

There remains considerable scope for further experimental work on monolayer graphene.  First, no dataset currently provides $g_z$ and $g_\perp$ jointly from the same sample; obtaining both from a single device would remove one obvious source of the inconsistencies discussed above. Second, $g_z,g_\perp$ are typically extracted at $\nu=0$, not $\nu=1/2$, and the two values can generally disagree significantly. This is because the experimentally extracted $g_z,g_\perp$ already incorporate renormalisation effects such as RPA screening, and the corresponding diagrams, and hence the renormalisation, depend on the filling.
An appropriate experiment would therefore determine $g_z,g_\perp$ directly at $\nu=1/2$. Measuring the sublattice splitting $\Delta_{AB}$ together with the crossing field $B$ already gives one estimate of $g_z$ at this filling, through Eq.~\eqref{eq:monolayer_crossing}. An independent, compatible estimate of $g_\perp$ at $\nu=1/2$ does not currently exist. 

Although there were some experimental tests at $\nu=0$ \cite{Young2012}, the $B$-dependence assumed for $g_z,g_\perp$ terms has itself never been directly tested at $\nu=1/2$. The commonly assumed $a_0/\ell_B$ scaling follows from treating the underlying corrections as lattice-scale in origin, but this scaling has not been directly tested and need not hold in general. For example, electron--phonon contributions to $g_\perp$ can be sizeable and carry no intrinsic magnetic-field dependence. RPA screening can further introduce its own, non-trivial dependence on $B$.

\vspace{0.5cm}
\section*{Acknowledgements}
We thank Konstantinos Vasiliou, Chunli Huang, and Biao Lian for valuable discussions. 
We are especially grateful to M. Zaletel, R. Mong, and F. Pollmann for granting us permission to use their code for compressing matrix product operators into MPO graph representations \cite{Zaletel_2013,Zaletel_2015}. We further acknowledge the use of the TeNPy library for tensor-network simulations and DiagHam for exact-diagonalisation calculations.
D.P. and Z.W. acknowledge funding from Leverhulme Trust International Professorship (Grant No. LIP-202-014).  S.H.S. acknowledges support from UK Engineering and Physical Sciences Research Council EPSRC (EP/X030881/1).

\bibliography{manuscript}

@article{Bolotin2009,
    author  = {Bolotin, Kirill I. and Ghahari, Fereshte and Shulman, Michael D. and Stormer, Horst L. and Kim, Philip},
    title   = {Observation of the fractional quantum Hall effect in graphene},
    journal = {Nature},
    volume  = {462},
    pages   = {196--199},
    year    = {2009},
    doi     = {10.1038/nature08582}
}

@article{TokeJain2007,
    author  = {T{\H{o}}ke, Csaba and Jain, J. K.},
    title   = {{SU(4)} composite fermions in graphene: Fractional quantum Hall states without analog in {GaAs}},
    journal = {Physical Review B},
    volume  = {75},
    number  = {24},
    pages   = {245440},
    year    = {2007},
    doi     = {10.1103/PhysRevB.75.245440}
}

@article{SodemannMacDonald2014,
    author  = {Sodemann, I. and MacDonald, A. H.},
    title   = {Broken {SU(4)} symmetry and the fractional quantum Hall effect in graphene},
    journal = {Physical Review Letters},
    volume  = {112},
    number  = {12},
    pages   = {126804},
    year    = {2014},
    doi     = {10.1103/PhysRevLett.112.126804}
}

@article{Zibrov2018,
    author  = {Zibrov, A. A. and Spanton, E. M. and Zhou, H. and Kometter, C. and Taniguchi, T. and Watanabe, K. and Young, A. F.},
    title   = {Even-denominator fractional quantum Hall states at an isospin transition in monolayer graphene},
    journal = {Nature Physics},
    volume  = {14},
    pages   = {930--935},
    year    = {2018},
    doi     = {10.1038/s41567-018-0190-0}
}

@article{Huang2022,
    author  = {Huang, Ke and Fu, Hailong and Hickey, Danielle Reifsnyder and Alem, Nasim and Lin, Xi and Watanabe, Kenji and Taniguchi, Takashi and Zhu, Jun},
    title   = {Valley Isospin Controlled Fractional Quantum Hall States in Bilayer Graphene},
    journal = {Physical Review X},
    volume  = {12},
    number  = {3},
    pages   = {031019},
    year    = {2022},
    doi     = {10.1103/PhysRevX.12.031019}
}

@article{PetersonNayak2014,
    author  = {Peterson, Michael R. and Nayak, Chetan},
    title   = {Effects of Landau level mixing on the fractional quantum Hall effect in monolayer graphene},
    journal = {Physical Review Letters},
    volume  = {113},
    number  = {8},
    pages   = {086401},
    year    = {2014},
    doi     = {10.1103/PhysRevLett.113.086401}
}

@article{Sokolik2017,
    author  = {Sokolik, A. A. and Zabolotskiy, A. D. and Lozovik, Yu. E.},
    title   = {Many-body effects of Coulomb interaction on Landau levels in graphene},
    journal = {Physical Review B},
    volume  = {95},
    pages   = {125402},
    year    = {2017},
    doi     = {10.1103/PhysRevB.95.125402}
}

@article{Kim2019,
    author  = {Kim, Youngwook and Balram, Ajit C. and Taniguchi, Takashi and Watanabe, Kenji and Jain, Jainendra K. and Smet, Jurgen H.},
    title   = {Even denominator fractional quantum Hall states in higher Landau levels of graphene},
    journal = {Nature Physics},
    volume  = {15},
    pages   = {154--158},
    year    = {2019},
    doi     = {10.1038/s41567-018-0355-x}
}

@article{trilayer_expt,
    author  = {Chen, Yiwei and Huang, Yan and Li, Qingxin and Tong, Bingbing and Kuang, Guangli and Xi, Chuanying and Watanabe, Kenji and Taniguchi, Takashi and Liu, Guangtong and Zhu, Zheng and Lu, Li and Zhang, Fu-Chun and Wu, Ying-Hai and Wang, Lei},
    title   = {Tunable even- and odd-denominator fractional quantum Hall states in trilayer graphene},
    journal = {Nature Communications},
    volume  = {15},
    pages   = {6236},
    year    = {2024},
    doi     = {10.1038/s41467-024-50589-2}
}

@misc{chanda,
      title={Even denominator fractional quantum Hall states in the zeroth Landau level of monolayer-like band of ABA trilayer graphene}, 
      author={Tanima Chanda and Simrandeep Kaur and Harsimran Singh and Kenji Watanabe and Takashi Taniguchi and Manish Jain and Udit Khanna and Ajit C. Balram and Aveek Bid},
      year={2025},
      eprint={2502.06245},
      archivePrefix={arXiv},
      primaryClass={cond-mat.mes-hall},
      url={https://arxiv.org/abs/2502.06245}, 
}

@article{Bilayer,
    author  = {Zibrov, A. A. and Kometter, C. and Zhou, H. and Spanton, E. M. and Taniguchi, T. and Watanabe, K. and Zaletel, M. P. and Young, A. F.},
    title   = {Tunable interacting composite fermion phases in a half-filled bilayer-graphene Landau level},
    journal = {Nature},
    volume  = {549},
    number  = {7672},
    pages   = {360--364},
    year    = {2017},
    doi     = {10.1038/nature23893}
}

@article{HalperinLeeRead1993,
  author  = {Halperin, B. I. and Lee, P. A. and Read, N.},
  title   = {Theory of the Half-Filled Landau Level},
  journal = {Physical Review B},
  volume  = {47},
  number  = {12},
  pages   = {7312--7343},
  year    = {1993},
  doi     = {10.1103/PhysRevB.47.7312}
}

@article{Morf1998,
  author  = {Morf, R. H.},
  title   = {Transition from Quantum Hall to Compressible States in the Second Landau Level: New Light on the Enigma},
  journal = {Physical Review Letters},
  volume  = {80},
  number  = {7},
  pages   = {1505--1508},
  year    = {1998},
  doi     = {10.1103/PhysRevLett.80.1505}
}

@article{Geraedts2016,
  author  = {Geraedts, Scott D. and Zaletel, Michael P. and Mong, Roger S. K. and Metlitski, Max A. and Vishwanath, Ashvin and Motrunich, Olexei I.},
  title   = {The Half-Filled Landau Level: The Case for Dirac Composite Fermions},
  journal = {Science},
  volume  = {352},
  number  = {6282},
  pages   = {197--201},
  year    = {2016},
  doi     = {10.1126/science.aad4302}
}

@article{sarma_CFL_Monte_carlo,
  author  = {Biddle, J. and Peterson, Michael R. and Das Sarma, S.},
  title   = {Variational Monte Carlo study of spin-polarization stability of fractional quantum Hall states against realistic effects in half-filled Landau levels},
  journal = {Physical Review B},
  volume  = {87},
  number  = {23},
  pages   = {235134},
  year    = {2013},
  doi     = {10.1103/PhysRevB.87.235134}
}

@article{parton,
  title = {Two-component parton fractional quantum Hall state in graphene},
  author = {Wu, Ying-Hai},
  journal = {Phys. Rev. B},
  volume = {106},
  issue = {15},
  pages = {155132},
  numpages = {8},
  year = {2022},
  month = {Oct},
  publisher = {American Physical Society},
  doi = {10.1103/PhysRevB.106.155132},
  url = {https://link.aps.org/doi/10.1103/PhysRevB.106.155132}
}

@article{Alicea_2006,
  author  = {Alicea, Jason and Fisher, Matthew P. A.},
  title   = {Graphene integer quantum {H}all effect in the ferromagnetic and paramagnetic regimes},
  journal = {Physical Review B},
  volume  = {74},
  pages   = {075422},
  year    = {2006},
  doi     = {10.1103/PhysRevB.74.075422}
}

@article{kharitonov,
  title = {Phase diagram for the $\ensuremath{\nu}=0$ quantum Hall state in monolayer graphene},
  author = {Kharitonov, Maxim},
  journal = {Phys. Rev. B},
  volume = {85},
  issue = {15},
  pages = {155439},
  numpages = {23},
  year = {2012},
  month = {Apr},
  publisher = {American Physical Society},
  doi = {10.1103/PhysRevB.85.155439},
  url = {https://link.aps.org/doi/10.1103/PhysRevB.85.155439}
}

@article{zaletel_phonons,
  author  = {Chatterjee, Shubhayu and Bultinck, Nick and Zaletel, Michael P.},
  title   = {Symmetry breaking and skyrmionic transport in twisted bilayer graphene},
  journal = {Physical Review B},
  volume  = {101},
  pages   = {165141},
  year    = {2020},
  doi     = {10.1103/PhysRevB.101.165141}
}

@misc{snizhko2015,
      title={Fractional quantum Hall effect in bilayer graphene beyond the single Landau level approximation}, 
      author={Kyrylo Snizhko and Vadim Cheianov and Steven H. Simon},
      year={2015},
      eprint={1505.03787},
      archivePrefix={arXiv},
      primaryClass={cond-mat.str-el},
      url={https://arxiv.org/abs/1505.03787}, 
}

@article{abanin_papic,
  title = {Topological Phases in the Zeroth Landau Level of Bilayer Graphene},
  author = {Papic, Z. and Abanin, D. A.},
  journal = {Phys. Rev. Lett.},
  volume = {112},
  issue = {4},
  pages = {046602},
  numpages = {5},
  year = {2014},
  month = {Jan},
  publisher = {American Physical Society},
  doi = {10.1103/PhysRevLett.112.046602},
  url = {https://link.aps.org/doi/10.1103/PhysRevLett.112.046602}
}

@article{Zaletel_2013,
  author  = {Zaletel, Michael P. and Mong, Roger S. K. and Pollmann, Frank},
  title   = {Topological characterization of fractional quantum {H}all ground states from microscopic {H}amiltonians},
  journal = {Physical Review Letters},
  volume  = {110},
  pages   = {236801},
  year    = {2013},
  doi     = {10.1103/PhysRevLett.110.236801}
}

@article{Zaletel_2015,
  author  = {Zaletel, Michael P. and Mong, Roger S. K. and Pollmann, Frank and Rezayi, Edward H.},
  title   = {Infinite density matrix renormalization group for multicomponent quantum {H}all systems},
  journal = {Physical Review B},
  volume  = {91},
  pages   = {045115},
  year    = {2015},
  doi     = {10.1103/PhysRevB.91.045115}
}

@misc{tenpy,
  author = {Johannes Hauschild and Frank Pollmann},
  title = {TeNPy: Tensor Network Python Library},
  year = {2018},
  note = {https://github.com/tenpy/tenpy}
}

@article{milovanovic,
  author  = {Moran, N. and Sterdyniak, A. and Vidanovi{\'c}, I. and Regnault, N. and Milovanovi{\'c}, M. V.},
  title   = {Topological $d$-wave pairing structures in {J}ain states},
  journal = {Physical Review B},
  volume  = {85},
  pages   = {245307},
  year    = {2012},
  doi     = {10.1103/PhysRevB.85.245307}
}

@article{Read_2000,
   title={Paired states of fermions in two dimensions with breaking of parity and time-reversal symmetries and the fractional quantum Hall effect},
   volume={61},
   ISSN={1095-3795},
   url={http://dx.doi.org/10.1103/PhysRevB.61.10267},
   DOI={10.1103/physrevb.61.10267},
   number={15},
   journal={Physical Review B},
   publisher={American Physical Society (APS)},
   author={Read, N. and Green, Dmitry},
   year={2000},
   month=Apr, pages={10267–10297} }

@article{Fremling_2018,
  author  = {Fremling, M. and Moran, N. and Slingerland, J. K. and Simon, S. H.},
  title   = {Trial wave functions for a composite {F}ermi liquid on a torus},
  journal = {Physical Review B},
  volume  = {97},
  pages   = {035149},
  year    = {2018},
  doi     = {10.1103/PhysRevB.97.035149}
}

@article{kun_yang_weak_to_strong,
  author  = {Zhang, Yuhui and Rezayi, E. H. and Yang, Kun},
  title   = {Existence of strong-pairing quantum {H}all phase in bilayer cold-atom systems with dipolar interactions},
  journal = {Physical Review B},
  volume  = {90},
  pages   = {165102},
  year    = {2014},
  doi     = {10.1103/PhysRevB.90.165102}
}

@article{Yuan_RPA,
   title={Polarization of graphene in a strong magnetic field beyond the Dirac cone approximation},
   volume={152},
   ISSN={0038-1098},
   url={http://dx.doi.org/10.1016/j.ssc.2012.04.041},
   DOI={10.1016/j.ssc.2012.04.041},
   number={15},
   journal={Solid State Communications},
   publisher={Elsevier BV},
   author={Yuan, Shengjun and Roldán, Rafael and Katsnelson, Mikhail I.},
   year={2012},
   month=Aug, pages={1446–1455} }

@article{CFL_Haldane_rezayi,
  author  = {Rezayi, E. H. and Haldane, F. D. M.},
  title   = {Incompressible Paired Hall State, Stripe Order, and the Composite Fermion Liquid Phase in Half-Filled Landau Levels},
  journal = {Physical Review Letters},
  volume  = {84},
  number  = {20},
  pages   = {4685--4688},
  year    = {2000},
  doi     = {10.1103/PhysRevLett.84.4685}
}

@article{SharmaPuJain2021_CFL,
  author  = {Sharma, Anirban and Pu, Songyang and Jain, J. K.},
  title   = {Bardeen-Cooper-Schrieffer pairing of composite fermions},
  journal = {Physical Review B},
  volume  = {104},
  number  = {20},
  pages   = {205303},
  year    = {2021},
  doi     = {10.1103/PhysRevB.104.205303}
}

@article{Ando2006Screening_RPA,
  author  = {Ando, Tsuneya},
  title   = {Screening Effect and Impurity Scattering in Monolayer Graphene},
  journal = {Journal of the Physical Society of Japan},
  volume  = {75},
  number  = {7},
  pages   = {074716},
  year    = {2006},
  doi     = {10.1143/JPSJ.75.074716}
}

@article{NomuraMacDonald2006_RPA,
  author  = {Nomura, Kentaro and MacDonald, Allan H.},
  title   = {Quantum Hall Ferromagnetism in Graphene},
  journal = {Physical Review Letters},
  volume  = {96},
  number  = {25},
  pages   = {256602},
  year    = {2006},
  doi     = {10.1103/PhysRevLett.96.256602}
}

@article{PyatkovskiyGusynin2011_RPA,
  author  = {Pyatkovskiy, P. K. and Gusynin, V. P.},
  title   = {Dynamical polarization of graphene in a magnetic field},
  journal = {Physical Review B},
  volume  = {83},
  number  = {7},
  pages   = {075422},
  year    = {2011},
  doi     = {10.1103/PhysRevB.83.075422}
}

@article{PHPfaffian1,
  author  = {Levin, Michael and Halperin, Bertrand I. and Rosenow, Bernd},
  title   = {Particle-Hole Symmetry and the Pfaffian State},
  journal = {Physical Review Letters},
  volume  = {99},
  number  = {23},
  pages   = {236806},
  year    = {2007},
  doi     = {10.1103/PhysRevLett.99.236806}
}

@article{PHpfaffian2,
  author  = {Lee, Sung-Sik and Ryu, Shinsei and Nayak, Chetan
             and Fisher, Matthew P. A.},
  title   = {Particle-Hole Symmetry and the {$\nu=5/2$} Quantum Hall State},
  journal = {Physical Review Letters},
  volume  = {99},
  number  = {23},
  pages   = {236807},
  year    = {2007},
  doi     = {10.1103/PhysRevLett.99.236807}
}

@article{daughter_states,
  author  = {Yutushui, Misha and Hermanns, Maria and Mross, David F.},
  title   = {Paired fermions in strong magnetic fields and daughters of even-denominator Hall plateaus},
  journal = {Physical Review B},
  volume  = {110},
  number  = {16},
  pages   = {165402},
  year    = {2024},
  doi     = {10.1103/PhysRevB.110.165402}
}

@book{monolayer_values,
author = {Egger, Reinhold and Matrasulov, Davron and Rakhimov, Khamdam},
year = {2013},
month = {01},
pages = {},
title = {Low-Dimensional Functional Materials},
isbn = {978-94-007-6617-4},
doi = {10.1007/978-94-007-6618-1}
}

@article{Sasaki_phonons,
  author  = {Sasaki, Ken-ichi and Kato, Keiko and Tokura, Yasuhiro and Suzuki, Satoru and Sogawa, Tetsuomi},
  title   = {Decay and frequency shift of both intervalley and intravalley phonons in graphene: {D}irac-cone migration},
  journal = {Physical Review B},
  volume  = {86},
  pages   = {201403},
  year    = {2012},
  doi     = {10.1103/PhysRevB.86.201403}
}

@article{Mariani_phonons,
  author  = {Mariani, Eros and von Oppen, Felix},
  title   = {Flexural phonons in free-standing graphene},
  journal = {Physical Review Letters},
  volume  = {100},
  pages   = {076801},
  year    = {2008},
  doi     = {10.1103/PhysRevLett.100.076801}
}

@article{more_phonon_values,
  title = {Electron-phonon interactions for optical-phonon modes in few-layer graphene: First-principles calculations},
  author = {Yan, Jia-An and Ruan, W. Y. and Chou, M. Y.},
  journal = {Phys. Rev. B},
  volume = {79},
  issue = {11},
  pages = {115443},
  numpages = {6},
  year = {2009},
  month = {Mar},
  publisher = {American Physical Society},
  doi = {10.1103/PhysRevB.79.115443},
  url = {https://link.aps.org/doi/10.1103/PhysRevB.79.115443}
}

@article{bernevig_phonons,
  title = {Electron--$K$-phonon interaction in twisted bilayer graphene},
  author = {Liu, Chao-Xing and Chen, Yulin and Yazdani, Ali and Bernevig, B. Andrei},
  journal = {Phys. Rev. B},
  volume = {110},
  issue = {4},
  pages = {045133},
  numpages = {7},
  year = {2024},
  month = {Jul},
  publisher = {American Physical Society},
  doi = {10.1103/PhysRevB.110.045133},
  url = {https://link.aps.org/doi/10.1103/PhysRevB.110.045133}
}

@article{trilayer_modes_phonons,
  title = {Phonon dispersions and vibrational properties of monolayer, bilayer, and trilayer graphene: Density-functional perturbation theory},
  author = {Yan, Jia-An and Ruan, W. Y. and Chou, M. Y.},
  journal = {Phys. Rev. B},
  volume = {77},
  issue = {12},
  pages = {125401},
  numpages = {7},
  year = {2008},
  month = {Mar},
  publisher = {American Physical Society},
  doi = {10.1103/PhysRevB.77.125401},
  url = {https://link.aps.org/doi/10.1103/PhysRevB.77.125401}
}

@article{Slonczewski1958,
  title = {Band Structure of Graphite},
  author = {Slonczewski, J. C. and Weiss, P. R.},
  journal = {Phys. Rev.},
  volume = {109},
  pages = {272--279},
  year = {1958},
  doi = {10.1103/PhysRev.109.272}
}

@article{McClure1957,
  title = {Band Structure of Graphite and de Haas-van Alphen Effect},
  author = {McClure, J. W.},
  journal = {Phys. Rev.},
  volume = {108},
  pages = {612--618},
  year = {1957},
  doi = {10.1103/PhysRev.108.612}
}

@article{Li_2008,
  author  = {Li, Hui and Haldane, F. D. M.},
  title   = {Entanglement spectrum as a generalization of entanglement entropy: Identification of topological order in non-{A}belian fractional quantum {H}all effect states},
  journal = {Physical Review Letters},
  volume  = {101},
  pages   = {010504},
  year    = {2008},
  doi     = {10.1103/PhysRevLett.101.010504}
}

@article{Hedin1965,
  author  = {Hedin, Lars},
  title   = {New Method for Calculating the One-Particle Green's Function with Application to the Electron-Gas Problem},
  journal = {Phys. Rev.},
  volume  = {139},
  number  = {3A},
  pages   = {A796--A823},
  year    = {1965},
  doi     = {10.1103/PhysRev.139.A796}
}

@article{HolmVonBarth1998,
  author  = {Holm, B. and von Barth, U.},
  title   = {Fully self-consistent {$GW$} self-energy of the electron gas},
  journal = {Phys. Rev. B},
  volume  = {57},
  number  = {4},
  pages   = {2108--2117},
  year    = {1998},
  doi     = {10.1103/PhysRevB.57.2108}
}

@article{Luo_2022,
  author  = {Luo, Wenchen and Zhang, Wei and Hu, Yutao and Wang, Hao},
  title   = {Phase transitions and topological properties of the $5/2$ quantum {H}all states with strong {L}andau-level mixing},
  journal = {Physical Review B},
  volume  = {106},
  pages   = {245142},
  year    = {2022},
  doi     = {10.1103/PhysRevB.106.245142}
}

@article{JSS,
  title = {Theory of spin-singlet fractional quantum Hall effect at \ensuremath{\nu}=1/2},
  author = {Belkhir, L. and Jain, J. K.},
  journal = {Phys. Rev. Lett.},
  volume = {70},
  issue = {5},
  pages = {643--646},
  numpages = {0},
  year = {1993},
  month = {Feb},
  publisher = {American Physical Society},
  doi = {10.1103/PhysRevLett.70.643},
  url = {https://link.aps.org/doi/10.1103/PhysRevLett.70.643}
}

@article{MooreRead1991,
  author  = {Moore, Gregory and Read, Nicholas},
  title   = {Nonabelions in the Fractional Quantum Hall Effect},
  journal = {Nuclear Physics B},
  volume  = {360},
  number  = {2-3},
  pages   = {362--396},
  year    = {1991},
  doi     = {10.1016/0550-3213(91)90407-O}
}

@article{Halperin_states,
  author  = {Halperin, Bertrand I.},
  title   = {Theory of the Quantized Hall Conductance},
  journal = {Helvetica Physica Acta},
  volume  = {56},
  pages   = {75--102},
  year    = {1983},
  doi     = {10.5169/seals-115362}
}

@article{topo_shift_wen_1,
  author  = {Wen, X. G. and Zee, A.},
  title   = {Shift and Spin Vector: New Topological Quantum Numbers for the Hall Fluids},
  journal = {Physical Review Letters},
  volume  = {69},
  number  = {6},
  pages   = {953--956},
  year    = {1992},
  doi     = {10.1103/PhysRevLett.69.953}
}

@article{topo_shift_wen_2,
  author  = {Wen, X. G. and Zee, A.},
  title   = {Classification of Abelian Quantum Hall States and Matrix Formulation of Topological Fluids},
  journal = {Physical Review B},
  volume  = {46},
  number  = {4},
  pages   = {2290--2301},
  year    = {1992},
  doi     = {10.1103/PhysRevB.46.2290}
}

@article{topo_shift3,
  author  = {Hansson, T. H. and Hermanns, M. and Simon, S. H. and Viefers, S. F.},
  title   = {Quantum Hall Physics: Hierarchies and Conformal Field Theory Techniques},
  journal = {Reviews of Modern Physics},
  volume  = {89},
  number  = {2},
  pages   = {025005},
  year    = {2017},
  doi     = {10.1103/RevModPhys.89.025005}
}

@article{KitaevPreskill2006_topo_entropy,
  author  = {Kitaev, Alexei and Preskill, John},
  title   = {Topological Entanglement Entropy},
  journal = {Physical Review Letters},
  volume  = {96},
  number  = {11},
  pages   = {110404},
  year    = {2006},
  doi     = {10.1103/PhysRevLett.96.110404}
}

@article{LevinWen2006_topo_entropy,
  author  = {Levin, Michael and Wen, Xiao-Gang},
  title   = {Detecting Topological Order in a Ground State Wave Function},
  journal = {Physical Review Letters},
  volume  = {96},
  number  = {11},
  pages   = {110405},
  year    = {2006},
  doi     = {10.1103/PhysRevLett.96.110405}
}

@article{WenNiu1990,
  author  = {Wen, X. G. and Niu, Q.},
  title   = {Ground-State Degeneracy of the Fractional Quantum Hall States
             in the Presence of a Random Potential and on High-Genus
             Riemann Surfaces},
  journal = {Physical Review B},
  volume  = {41},
  number  = {13},
  pages   = {9377--9396},
  year    = {1990},
  doi     = {10.1103/PhysRevB.41.9377}
}

@article{Kitaev2003,
  author  = {Kitaev, A. Yu.},
  title   = {Fault-Tolerant Quantum Computation by Anyons},
  journal = {Annals of Physics},
  volume  = {303},
  number  = {1},
  pages   = {2--30},
  year    = {2003},
  doi     = {10.1016/S0003-4916(02)00018-0}
}

@book{Simon2023_book,
  author    = {Simon, Steven H.},
  title     = {Topological Quantum},
  publisher = {Oxford University Press},
  year      = {2023},
  isbn      = {9780198886723},
  doi       = {10.1093/oso/9780198886723.001.0001}
}

@article{Wen1995_K_matrix,
  author  = {Wen, Xiao-Gang},
  title   = {Topological Orders and Edge Excitations in Fractional
             Quantum Hall States},
  journal = {Advances in Physics},
  volume  = {44},
  number  = {5},
  pages   = {405--473},
  year    = {1995},
  doi     = {10.1080/00018739500101566}
}

@article{laturia2018dielectric,
  title   = {Dielectric properties of hexagonal boron nitride and
             transition metal dichalcogenides: from monolayer to bulk},
  author  = {Laturia, Akash and Van de Put, Maarten L. and
             Vandenberghe, William G.},
  journal = {npj 2D Materials and Applications},
  volume  = {2},
  pages   = {6},
  year    = {2018},
  doi     = {10.1038/s41699-018-0050-x}
}

@article{santos2013electric,
  author  = {Santos, Elton J. G. and Kaxiras, Efthimios},
  title   = {Electric-field dependence of the effective dielectric constant in graphene},
  journal = {Nano Letters},
  volume  = {13},
  pages   = {898--902},
  year    = {2013},
  doi     = {10.1021/nl303611v}
}

@article{WuJain2013,
  author  = {Wu, Ying-Hai and Jain, Jainendra K.},
  title   = {Quantum Hall Effect of Two-Component Bosons at Fractional and Integral Fillings},
  journal = {Physical Review B},
  volume  = {87},
  pages   = {245123},
  year    = {2013},
  doi     = {10.1103/PhysRevB.87.245123}
}

@article{CFL_N_8,
  title = {Valley Stoner instability of the composite Fermi sea},
  author = {Zhu, Zheng and Sheng, D. N. and Fu, Liang and Sodemann, Inti},
  journal = {Phys. Rev. B},
  volume = {98},
  issue = {15},
  pages = {155104},
  numpages = {7},
  year = {2018},
  month = {Oct},
  publisher = {American Physical Society},
  doi = {10.1103/PhysRevB.98.155104},
  url = {https://link.aps.org/doi/10.1103/PhysRevB.98.155104}
}

@article{CFL_RezayiRead1994,
    author  = {Rezayi, E. and Read, N.},
    title   = {Fermi-Liquid-Like State in a Half-Filled Landau Level},
    journal = {Physical Review Letters},
    volume  = {72},
    number  = {6},
    pages   = {900--903},
    year    = {1994},
    doi     = {10.1103/PhysRevLett.72.900}
}

@article{CFL_Onoda2001,
    author  = {Onoda, M. and Mizusaki, T. and Otsuka, T. and Aoki, H.},
    title   = {Composite fermion picture and the spin states in the fractional quantum Hall system---a numerical study},
    journal = {Physica B: Condensed Matter},
    volume  = {298},
    pages   = {173--176},
    year    = {2001},
    doi     = {10.1016/S0921-4526(01)00295-2}
}

@article{BalramJain2017,
    author  = {Balram, Ajit C. and Jain, J. K.},
    title   = {Fermi wave vector for the partially spin-polarized composite-fermion Fermi sea},
    journal = {Physical Review B},
    volume  = {96},
    number  = {23},
    pages   = {235102},
    year    = {2017},
    doi     = {10.1103/PhysRevB.96.235102}
}

@article{magnon,
  title   = {Strong-Magnetic-Field Magnon Transport in Monolayer Graphene},
  author  = {Zhou, Haoxin and Huang, Chunli and Wei, Nemin and Taniguchi, Takashi and Watanabe, Kenji and Zaletel, Michael P. and Papi\'{c}, Zlatko and MacDonald, Allan H. and Young, Andrea F.},
  journal = {Physical Review X},
  volume  = {12},
  pages   = {021060},
  year    = {2022},
  doi     = {10.1103/PhysRevX.12.021060}
}

@article{Young2014,
  title   = {Tunable symmetry breaking and helical edge transport in a graphene quantum spin Hall state},
  author  = {Young, A. F. and Sanchez-Yamagishi, J. D. and Hunt, B. and Choi, S. H. and Watanabe, K. and Taniguchi, T. and Ashoori, R. C. and Jarillo-Herrero, P.},
  journal = {Nature},
  volume  = {505},
  number  = {7484},
  pages   = {528--532},
  year    = {2014},
  month   = jan,
  doi     = {10.1038/nature12800}
}

@article{Ardonne2011SpinfulQH,
  title   = {Structure of spinful quantum {Hall} states: A squeezing perspective},
  author  = {Ardonne, E. and Regnault, N.},
  journal = {Physical Review B},
  volume  = {84},
  pages   = {205134},
  year    = {2011},
  doi     = {10.1103/PhysRevB.84.205134}
}

@article{Regnault2009JainHierarchy,
  title   = {Topological Entanglement and Clustering of {Jain} Hierarchy States},
  author  = {Regnault, N. and Bernevig, B. A. and Haldane, F. D. M.},
  journal = {Physical Review Letters},
  volume  = {103},
  pages   = {016801},
  year    = {2009},
  doi     = {10.1103/PhysRevLett.103.016801}
}

@article{Seidel2011HaldaneRezayi,
  title   = {Gapless excitations in the {Haldane}-{Rezayi} state: The thin-torus limit},
  author  = {Seidel, Alexander and Yang, Kun},
  journal = {Physical Review B},
  volume  = {84},
  pages   = {085122},
  year    = {2011},
  doi     = {10.1103/PhysRevB.84.085122}
}

@article{papic_dmrg,
  title = {Competing Abelian and non-Abelian topological orders in $\nu = 1/3 + 1/3$ quantum Hall bilayers},
  author = {Geraedts, Scott and Zaletel, Michael P. and Papi\'c, Zlatko and Mong, Roger S. K.},
  journal = {Physical Review B},
  volume = {91},
  number = {20},
  pages = {205139},
  year = {2015},
  publisher = {American Physical Society},
  doi = {10.1103/PhysRevB.91.205139}
}

@article{hegde_sodemann,
  author  = {Hegde, Suraj S. and Sodemann Villadiego, Inti},
  title   = {Theory of competing charge density wave, {K}ekul\'e, and antiferromagnetically ordered fractional quantum {H}all states in graphene aligned with boron nitride},
  journal = {Physical Review B},
  volume  = {105},
  number  = {19},
  pages   = {195417},
  year    = {2022},
  doi     = {10.1103/PhysRevB.105.195417}
}

@article{chunli,
  author  = {Wei, Nemin and Xu, Guopeng and Sodemann Villadiego, Inti and Huang, Chunli},
  title   = {Landau-Level Mixing and Symmetry Breaking in Graphene},
  journal = {Physical Review Letters},
  volume  = {134},
  number  = {4},
  pages   = {046501},
  year    = {2025},
  doi     = {10.1103/PhysRevLett.134.046501}
}

@article{anisotropy_1,
  author  = {Xu, Guopeng and Huang, Chunli},
  title   = {Influence of the {D}irac sea on phase transitions in monolayer graphene under strong magnetic fields},
  journal = {Physical Review B},
  volume  = {111},
  pages   = {125127},
  year    = {2025},
  doi     = {10.1103/PhysRevB.111.125127}
}

@article{anisotropy_2,
  author  = {De, Suman Jyoti and Das, Anirban and Rao, Sumathi and Kaul, Ribhu K. and Murthy, Ganpathy},
  title   = {Global phase diagram of charge-neutral graphene in the quantum {H}all regime for generic interactions},
  journal = {Physical Review B},
  volume  = {107},
  number  = {12},
  pages   = {125422},
  year    = {2023},
  doi     = {10.1103/PhysRevB.107.125422}
}

@article{anisotropy_3,
  author  = {Das, Anirban and Kaul, Ribhu K. and Murthy, Ganpathy},
  title   = {Coexistence of Canted Antiferromagnetism and Bond Order in $\nu=0$ Graphene},
  journal = {Physical Review Letters},
  volume  = {128},
  pages   = {106803},
  year    = {2022},
  doi     = {10.1103/PhysRevLett.128.106803}
}

@article{anisotropy_4,
  author  = {Stefanidis, Nikolaos and Sodemann Villadiego, Inti},
  title   = {Spin-valley entangled quantum {H}all states in graphene},
  journal = {Physical Review B},
  volume  = {108},
  pages   = {235137},
  year    = {2023},
  doi     = {10.1103/PhysRevB.108.235137}
}

@article{anisotropy_5,
  author  = {Stefanidis, Nikolaos and Sodemann Villadiego, Inti},
  title   = {Competing spin-valley entangled and broken symmetry states in the $N=1$ {L}andau level of graphene},
  journal = {Physical Review B},
  volume  = {107},
  pages   = {045132},
  year    = {2023},
  doi     = {10.1103/PhysRevB.107.045132}
}

@article{anisotropy_6,
  author  = {An, Junkai and Balram, Ajit C. and Murthy, Ganpathy},
  title   = {Magnetic and lattice ordered fractional quantum {H}all phases in graphene},
  journal = {Physical Review B},
  volume  = {110},
  pages   = {L081103},
  year    = {2024},
  doi     = {10.1103/PhysRevB.110.L081103}
}

@article{Coissard2022,
  author = {Coissard, Alexis and Wander, David and Vignaud, Hadrien
            and Grushin, Adolfo G. and Repellin, C{\'e}cile
            and Watanabe, Kenji and Taniguchi, Takashi
            and Gay, Fr{\'e}d{\'e}ric and Winkelmann, Clemens B.
            and Courtois, Herv{\'e} and Sellier, Hermann
            and Sac{\'e}p{\'e}, Benjamin},
  title = {Imaging tunable quantum {H}all broken-symmetry orders in graphene},
  journal = {Nature},
  volume = {605},
  pages = {51--56},
  year = {2022},
  doi = {10.1038/s41586-022-04513-7}
}

@article{yazdani,
  title   = {Visualizing broken symmetry and topological defects in a quantum Hall ferromagnet},
  author  = {Liu, Xiaomeng and Farahi, Gelareh and Chiu, Cheng-Li and Papi\'c, Zlatko and Watanabe, Kenji and Taniguchi, Takashi and Zaletel, Michael P. and Yazdani, Ali},
  journal = {Science},
  volume  = {375},
  number  = {6578},
  pages   = {321--326},
  year    = {2022},
  doi     = {10.1126/science.abm3770}
}

@article{Young2012,
  author  = {Young, A. F. and Dean, C. R. and Wang, L. and Ren, H. and Cadden-Zimansky, P. and Watanabe, K. and Taniguchi, T. and Hone, J. and Shepard, K. L. and Kim, P.},
  title   = {Spin and valley quantum Hall ferromagnetism in graphene},
  journal = {Nature Physics},
  volume  = {8},
  pages   = {550--556},
  year    = {2012},
  doi     = {10.1038/nphys2307}
}

@article{Zhu2016BilayerTunneling,
  author  = {Zhu, W. and Liu, Zhao and Haldane, F. D. M. and Sheng, D. N.},
  title   = {Fractional quantum {H}all bilayers at half filling: Tunneling-driven non-{A}belian phase},
  journal = {Physical Review B},
  volume  = {94},
  pages   = {245147},
  year    = {2016},
  doi     = {10.1103/PhysRevB.94.245147}
}

@article{Gmitra2009SOC,
  author  = {Gmitra, M. and Konschuh, S. and Ertler, C. and Ambrosch-Draxl, C. and Fabian, J.},
  title   = {Band-structure topologies of graphene: Spin-orbit coupling effects from first principles},
  journal = {Physical Review B},
  volume  = {80},
  pages   = {235431},
  year    = {2009}
}

@article{Konschuh2012BLG,
  author  = {Konschuh, S. and Gmitra, M. and Kochan, D. and Fabian, J.},
  title   = {Theory of spin-orbit coupling in bilayer graphene},
  journal = {Physical Review B},
  volume  = {85},
  pages   = {115423},
  year    = {2012}
}
\clearpage

\onecolumngrid

\begin{center}
    {\large\bfseries Supplemental Material}
\end{center}
\appendix

\section{Tight-Binding Hamiltonian and Parameters}
\label{app:tight_binding}

\subsection{Model and Parameters}

We describe Bernal-stacked trilayer graphene (TLG) using the
Slonczewski--Weiss--McClure (SWMcC) tight-binding model. Each graphene
layer contains two sublattices, $A$ and $B$, and we denote the layer
index by $n=1,2,3$.  The SWMcC parameters used throughout this work are
taken from Ref.~\cite{chanda} and are summarized in Table~\ref{tab:swmcc_params}. In the basis
\begin{equation}
    \left\{
    A_1,B_1,A_2,B_2,A_3,B_3
    \right\},
\end{equation}
the single-particle Hamiltonian takes the form
\begin{equation}
H(\mathbf{k}) =
\begin{pmatrix}
\Delta_1+\Delta_2
    & \gamma_0 t_{\mathbf{k}}^{*}
    & \gamma_4 t_{\mathbf{k}}^{*}
    & \gamma_3 t_{\mathbf{k}}
    & \gamma_2/2
    & 0
\\
\gamma_0 t_{\mathbf{k}}
    & \delta+\Delta_1+\Delta_2
    & \gamma_1
    & \gamma_4 t_{\mathbf{k}}^{*}
    & 0
    & \gamma_5/2
\\
\gamma_4 t_{\mathbf{k}}
    & \gamma_1
    & \delta-2\Delta_2
    & \gamma_0 t_{\mathbf{k}}^{*}
    & \gamma_4 t_{\mathbf{k}}
    & \gamma_1
\\
\gamma_3 t_{\mathbf{k}}^{*}
    & \gamma_4 t_{\mathbf{k}}
    & \gamma_0 t_{\mathbf{k}}
    & -2\Delta_2
    & \gamma_3 t_{\mathbf{k}}^{*}
    & \gamma_4 t_{\mathbf{k}}
\\
\gamma_2/2
    & 0
    & \gamma_4 t_{\mathbf{k}}^{*}
    & \gamma_3 t_{\mathbf{k}}
    & -\Delta_1+\Delta_2
    & \gamma_0 t_{\mathbf{k}}^{*}
\\
0
    & \gamma_5/2
    & \gamma_1
    & \gamma_4 t_{\mathbf{k}}^{*}
    & \gamma_0 t_{\mathbf{k}}
    & \delta-\Delta_1+\Delta_2
\end{pmatrix}.
\label{eq:app_tlg_hamiltonian}
\end{equation}
Here $\gamma_0,\ldots,\gamma_5$ are the SWMcC hopping parameters and
$\delta$ accounts for the difference in onsite energy between dimer
and non-dimer sites. The parameters $\Delta_1$ and $\Delta_2$
describe the electrostatic asymmetry between the three layers.
In terms of the layer potentials $V_1,V_2,V_3$, we define
\begin{equation}
    \Delta_1
    =
    -e\,\frac{V_1-V_3}{2},
    \qquad
    \Delta_2
    =
    -e\,\frac{V_1+V_3-2V_2}{6}.
\end{equation}
Thus, $\Delta_1$ parametrizes the potential difference between the
outer layers, while $\Delta_2$ describes the deviation of the middle
layer potential from the average potential of the outer layers. 

In particular, the potential difference $\Delta_1$ is tunable by the displacement field $D$ and is related to it through $ \Delta_1=-\frac{e d_{\perp}}{2\epsilon_{\rm TLG}}D,$ where $d_{\perp}=0.67\,{\rm nm}$ is the separation between the outer layers and $\epsilon_{\rm TLG}\approx 2$ is the dielectric constant of trilayer graphene. In contrast, $\Delta_2$ is not directly controlled and is primarily intended to capture the intrinsic electrostatic field of the trilayer. It is expected to vanish only when the net electric field across the three layers---external and intrinsic combined---is uniform.

The momentum dependence in Eq.~\eqref{eq:app_tlg_hamiltonian} is
contained in the nearest-neighbour structure factor
\begin{equation}
    t_{\mathbf{k}}
    =
    \sum_{j=1}^{3}
    e^{i\mathbf{k}\cdot\boldsymbol{\delta}_j},
\end{equation}
where $\boldsymbol{\delta}_j$ are the three nearest-neighbour vectors
of the honeycomb lattice.

For the low-energy physics relevant to the Landau-level crossings
considered in the main text, we expand the tight-binding Hamiltonian
about the two inequivalent valleys $K_{\pm}$. Introducing
\begin{equation}
    \pi_{\pm}
    =
    \hbar(k_x \pm i k_y),
    \qquad
    \hbar v_i = \frac{\sqrt{3}}{2}a\gamma_i ,
\end{equation}
where $a$ is the graphene lattice constant, the lattice hopping
terms reduce to terms linear in $\pi_{\pm}$. The resulting continuum
Hamiltonians at the two valleys follow directly from
Eq.~\eqref{eq:app_tlg_hamiltonian}.

It is useful to make the mirror symmetry of ABA trilayer graphene
explicit. We therefore transform to the basis
\begin{equation}
\left\{
    \frac{A_1-A_3}{\sqrt{2}},
    \frac{B_1-B_3}{\sqrt{2}},
    \frac{A_1+A_3}{\sqrt{2}},
    \frac{B_1+B_3}{\sqrt{2}},
    A_2,
    B_2
\right\}.
\label{eq:app_mirror_basis}
\end{equation}
In this basis the continuum Hamiltonian can be written as
\begin{equation}
    H =
    \begin{pmatrix}
        H_{\rm slg} & H_{\Delta_1} \\
        H_{\Delta_1}^{\dagger} & H_{\rm blg}
    \end{pmatrix},
\label{eq:app_block_hamiltonian}
\end{equation}
where $H_{\rm slg}$ and $H_{\rm blg}$ describe the monolayer-like
(mirror-odd) and bilayer-like (mirror-even) sectors, respectively,
and
\begin{equation}
H_{\Delta_1}
=
\begin{pmatrix}
    \Delta_1 & 0 & 0 & 0 \\
    0 & \Delta_1 & 0 & 0
\end{pmatrix}.
\end{equation}
Consequently, at $\Delta_1=0$ the mirror-odd and mirror-even sectors
decouple exactly, whereas a finite displacement field hybridizes
them.

For the $K_+$ valley, the monolayer-like block is
\begin{equation}
H_{\rm slg}^{+}
=
\begin{pmatrix}
-\gamma_2/2+\Delta_2
    & v_0\pi_{-}
\\
v_0\pi_{+}
    & -\gamma_5/2+\delta+\Delta_2
\end{pmatrix},
\label{eq:app_slg}
\end{equation}
while the bilayer-like block is
\begin{equation}
H_{\rm blg}^{+}
=
\begin{pmatrix}
\gamma_2/2+\Delta_2
    & v_0\pi_{-}
    & -\sqrt{2}v_4\pi_{-}
    & \sqrt{2}v_3\pi_{+}
\\
v_0\pi_{+}
    & \gamma_5/2+\delta+\Delta_2
    & \sqrt{2}\gamma_1
    & -\sqrt{2}v_4\pi_{-}
\\
-\sqrt{2}v_4\pi_{+}
    & \sqrt{2}\gamma_1
    & \delta-2\Delta_2
    & v_0\pi_{-}
\\
\sqrt{2}v_3\pi_{-}
    & -\sqrt{2}v_4\pi_{+}
    & v_0\pi_{+}
    & -2\Delta_2
\end{pmatrix}.
\label{eq:app_blg}
\end{equation}
The $K_-$ Hamiltonian is obtained by the corresponding valley
transformation $\pi_{+}\leftrightarrow-\pi_{-}$.
A perpendicular magnetic field $B$ is incorporated through minimal
coupling,
\begin{equation}
    \mathbf{p}\rightarrow\mathbf{p}+e\mathbf{A},
\end{equation}
where we take $e>0$ to denote the elementary charge. Equivalently, the
momentum operators entering the continuum Hamiltonian are replaced by
their kinetic counterparts,
\begin{equation}
    \pi_{\pm}
    \rightarrow
    \Pi_{\pm}
    =
    \pi_{\pm}+e(A_x\pm iA_y).
\end{equation}
Introducing the magnetic length $\ell_B=\sqrt{\frac{\hbar}{eB}},$
the operators $\Pi_{\pm}$ can be expressed in terms of the usual
Landau-level raising and lowering operators. We define
\begin{equation}
    a=\frac{\ell_B}{\sqrt{2}\hbar}\Pi_-,
    \qquad
    a^\dagger=\frac{\ell_B}{\sqrt{2}\hbar}\Pi_+,
\end{equation}
such that
\begin{equation}
    a|n\rangle=\sqrt{n}\,|n-1\rangle,
    \qquad
    a^\dagger|n\rangle=\sqrt{n+1}\,|n+1\rangle .
\end{equation}
The continuum Hamiltonian therefore becomes a matrix acting jointly
in sublattice-layer and Landau-level orbital space. We obtain the
single-particle Landau-level spectrum by diagonalizing this Hamiltonian
in a truncated Landau-level basis,
\begin{equation}
    |n\rangle,\qquad n=0,\ldots,n_{\rm max},
\end{equation}
with $n_{\rm max}$ chosen sufficiently large such that the low-energy
Landau levels relevant to this work are well captured.

\begin{table}[t]
    \centering
    \caption{Slonczewski--Weiss--McClure parameters used throughout this work, taken from Ref.~\cite{chanda}.}
    \label{tab:swmcc_params}
    \begin{tabular}{c c}
        \hline\hline
        Parameter & Value (eV) \\
        \hline
        $\gamma_0$ & $3.10$ \\
        $\gamma_1$ & $0.39$ \\
        $\gamma_2$ & $-0.005$ \\
        $\gamma_3$ & $0.275$ \\
        $\gamma_4$ & $0.040$ \\
        $\gamma_5$ & $0.005$ \\
        $\delta$    & $0.0108$ \\
        $\Delta_2$  & $0.003$ \\
        \hline\hline
    \end{tabular}
\end{table}

\subsection{Parameter Range and Its Effect on the Eigenstates}
\label{app:parameter_range}

The tight-binding parameters entering the simulations in Fig.~\ref{fig:RPA_overlaps} were sampled from the following ranges:

\begin{center}
\begin{tabular}{@{}p{18pt}l@{}}
$\gamma_2$ & $\in [-0.0135,\ 0.0035]~\text{eV},$ \\[4pt]
$\gamma_3$ & $\in [0.225,\ 0.325]~\text{eV},$ \\[4pt]
$\gamma_4$ & $\in [-0.035,\ 0.115]~\text{eV},$ \\[4pt]
$\gamma_5$ & $\in [-0.015,\ 0.025]~\text{eV},$ \\[4pt]
\hspace{2pt}$\delta$ & $\in [-0.0022,\ 0.0238]~\text{eV},$ \\[4pt]
$\Delta_2$ & $\in [-6.0,\ 6.0]~\text{meV}.$
\end{tabular}
\end{center}

Importantly, within this range we are varying not only the energy levels but also the eigenstates, i.e., the crossing bands. Across this parameter space we identify two important types of perturbations to the eigenstates. In the first, instead of one eigenstate being localized between layers $1$ and $3$, we obtain significant hybridization with layer $2$, while both relevant bands remain fully within the $0$LL. In the second, one of the bands acquires a larger admixture ($\sim 5$--$10\%$) of higher Landau levels, while the other does not.

In the first case, admixing more of layer $2$ causes RPA to screen the perturbation less effectively, yielding a larger $|\Delta V_0/V_0|$, hence bringing us closer to the strong paired state.
The second case, by contrast, explicitly breaks $SU(2)$. Interestingly, these states seem to have $\Delta V_0/V_0$ near zero, or even positive. Strictly speaking, our treatment assumes a pure $0$LL, so this admixture can induce an error in our calculation for $\Delta V_0/V_0$, which we ignore presuming it is the same order as the mixing, although it can be amplified by the RPA.
Importantly, when performing ED calculations with these bands at $\Delta V_0=0.0$, setting their single-particle energies to be equal, we always found a CFL ground state, since the admixture with higher LLs is large enough to significantly break $SU(2)$, but not to stabilise a gapped state. Moreover, in this case the dispersion often does not even exhibit a crossing, providing a further reason to distrust these states. We still include them in Fig.~\ref{fig:RPA_overlaps}, but for the reasons above assign them a lower weight when plotting the region of values. Importantly, the aim of our study is to establish the plausibility of RPA stabilising the strongly paired state, rather than to claim that a generic set of tight-binding parameters would necessarily stabilise it. Because our goal is this more modest plausibility argument, rather than a generic claim, we allow ourselves the crude approximation of treating states with up to $10\%$ higher-LL admixture as pure $0$LL. In theory, one should also perform both ED and DMRG calculations to see how such an admixture affects the crossing points. Nonetheless, a separate and more careful treatment of this regime should be pursued in future work.

\section{Renormalisation of the Dispersion}
\label{app:dispersion_renormalisation}

The separation between the relevant Landau levels is comparable to the characteristic interaction scale. Interaction effects can therefore appreciably renormalise the effective single-particle dispersion and, in particular, shift the positions of Landau-level crossings.

The purpose of this appendix is to justify our treatment of interaction effects near Landau-level crossings at non-integer filling. We first introduce the Hartree--Fock approximation used for partially filled Landau levels, in which the exchange energy between different Landau levels is obtained from the corresponding integer-filling result by rescaling with the appropriate filling fractions. We test this approximation directly using iDMRG for a filled $1\mathrm{LL}$ coupled to a half-filled $0\mathrm{LL}$ hosting a composite-Fermi-liquid state. Having established the validity of this approximation, we then use it to compare the Landau-level crossings at $\nu=7/2$ and $\nu=4$. We show that, although the Hartree--Fock correction shifts the crossing relative to the bare single-particle dispersion, the resulting shift is the same at the two fillings.

Second, we investigate whether interaction-induced renormalisation of the single-particle dispersion can produce a band inversion while remaining consistent with the experimentally determined band parameters. To this end, we introduce a Hartree--Fock formalism for the dispersion renormalisation and find that such a scenario is realised only within a highly fine-tuned region of parameter space.

The tight-binding parameters introduced in Appendix~\ref{app:tight_binding} are fitted to the dispersion near charge neutrality. We therefore regard this tight-binding Hamiltonian as already containing the interaction-induced renormalisation of the bands up to \(\nu=0\) at zero magnetic field. When considering fillings above charge neutrality, we consequently keep the fitted tight-binding Hamiltonian fixed and include only the additional Hartree--Fock corrections arising from the Landau levels occupied above \(\nu=0\). This avoids double counting interaction effects already absorbed into the fitted dispersion while accounting for the exchange shifts generated by the newly occupied levels.

\subsection{Hartree--Fock Shifts of Landau-Level Crossings}

\subsubsection{Validity of the Hartree--Fock decomposition at non-integer filling}

 We would like to use a Hartree--Fock description of the interactions between different Landau levels. At integer filling, where each Landau level is either completely filled or empty, the Hartree--Fock exchange energy is well defined and can be evaluated analytically. In the problem considered here, however, some Landau levels are only partially filled. We therefore generalize the integer-filling result by assuming that the exchange contribution between two Landau levels is given by the exchange energy of the corresponding fully filled levels, rescaled by their respective filling fractions:
\begin{equation}
\label{eq:Fock}
E_x^{ab}(\nu_a,\nu_b) \approx \nu_a\nu_b,E_x^{ab,\mathrm{full}},
\end{equation}
where $a$ and $b$ label different Landau levels, $\nu_a$ and $\nu_b$ are their respective filling fractions, $E_x^{ab}(\nu_a,\nu_b)$ is the exchange energy between the two Landau levels at fillings $\nu_a$ and $\nu_b$, and $E_x^{ab,\mathrm{full}}$ is the corresponding exchange energy when both Landau levels are fully filled.
In other words, we neglect additional corrections to the inter-Landau-level exchange arising from correlations within the partially filled levels. In this appendix, we test this approximation directly using iDMRG.

To test this assumption, we consider a minimal two-band problem consisting of a fully filled $1\mathrm{LL}$ and a half-filled $0\mathrm{LL}$ hosting a composite-Fermi-liquid (CFL) state, with a fixed single-particle band splitting $\mu$ between the two levels. Under the approximation introduced above, the total energy can be decomposed as
\begin{equation}
\label{eq:E_decomp}
E \approx E(\text{full }1\mathrm{LL}) + E(\text{half-filled }0\mathrm{LL}) + E_x^{\rm Fock},
\end{equation}
where $E(\text{full }1\mathrm{LL})$ and $E(\text{half-filled }0\mathrm{LL})$ are the energies of the two Landau levels evaluated \emph{in isolation}, and $E_x^{\rm Fock}$ is the exchange energy between them, evaluated using Eq.~\eqref{eq:Fock}. Within Hartree--Fock, $E_x^{\rm Fock}$ can be evaluated analytically in closed form. Since the three energies appearing in Eq.~\eqref{eq:E_decomp} are independently accessible from iDMRG, we can directly compare the exchange contribution inferred from iDMRG with the analytic Hartree--Fock prediction.

We perform iDMRG for two cases: the $1\mathrm{LL}$ and $0\mathrm{LL}$ have either the same or different pseudospin. Hartree--Fock predicts a finite exchange energy between the two Landau levels in the former case and zero exchange energy in the latter.

The results are summarized in Table~\ref{tab:Ex_validation}. We find that Hartree--Fock reproduces the DMRG result to within $0$--$25\%$, with a relatively mild dependence on $\mu$ over the range relevant to the physical problem. This confirms that the inter-Landau-level exchange energy is reasonably well captured by the filling-rescaled Hartree--Fock approximation, despite the strong correlations within the half-filled $0\mathrm{LL}$, and supports the decomposition in Eq.~\eqref{eq:E_decomp}.

\begin{table}[h]
\centering
\begin{tabular}{|c|c|}
\hline
$\mu/E_C$ & $(E_x^{\rm DMRG}-E_x^{\rm HF})/E_x^{\rm HF}$ \\
\hline
0.3 & $0.2558$ \\ \hline
0.4 & $0.2543$ \\ \hline
0.6 & $0.2092$ \\ \hline
1.0 & $0.0170$ \\ \hline
\end{tabular}
\caption{Relative deviation $(E_x^{\rm DMRG}-E_x^{\rm HF})/E_x^{\rm HF}$ of the exchange energy
per Landau-level site between a filled $1\mathrm{LL}$ and a half-filled $0\mathrm{LL}$
hosting a CFL state, obtained from iDMRG and compared with the filling-rescaled
Hartree--Fock prediction $E_x^{\rm HF}/N_{\rm Phi}=-0.6437\,E_C$, as a function of $\mu$. Energies
are given in units of $E_C$. Here we used $L_y=17\ell_B$, and screened Coulomb
interaction for both Fock and DMRG with $\zeta=4\ell_B$, $\chi=5000$. We see fairly
good agreement between iDMRG and the Hartree--Fock prediction.}
\label{tab:Ex_validation}
\end{table}
\subsubsection{Crossing point}

Having established the validity of the Hartree--Fock approximation, we now use it to compare the crossing points at $\nu=7/2$ and $\nu=4$. We show that, although exchange shifts the crossing relative to the bare single-particle prediction, it produces the same shift at both fillings.

At $\nu=7/2$, the relevant crossing is between the two approximate configurations
\begin{align}
\text{(A)}\qquad &
1\mathrm{LL}-\uparrow,\ 
1\mathrm{LL}-\downarrow,\ 
0\mathrm{LL}-\uparrow\ \text{filled},
\qquad
0\mathrm{LL}-\downarrow\ \text{half filled},
\\
\text{(B)}\qquad &
1\mathrm{LL}-\uparrow,\ 
1\mathrm{LL}-\downarrow,\ 
0\mathrm{LL}-\uparrow\ \text{filled},
\qquad
0\mathrm{LL}+\uparrow\ \text{half filled}.
\end{align}
Here ${1\mathrm{LL}}$ denotes the pseudospin-filled $1\mathrm{LL}$ manifold, which is common to both configurations and acts as a spectator background. The configurations specify the approximate filling pattern of the full iDMRG ground state rather than a particular many-body wave function. The crossing point $D^*$ is defined by
\begin{equation}
E_A(D^*)=E_B(D^*).
\end{equation}
where $E_i$ is the energy of configuration $i$.

We can determine this crossing within Hartree--Fock by considering the difference $E_A(D)-E_B(D)$. Contributions that are identical in the two configurations cancel, leaving only the single-particle energies difference and the exchange contribution:
\begin{equation}
E_A(D)-E_B(D)=
\frac{1}{2}\mu_{0\mathrm{LL}-\downarrow}(D)-
\frac{1}{2}\mu_{0\mathrm{LL}+\uparrow}(D)
+\frac{1}{2}E_{0\mathrm{LL}-\downarrow,1\mathrm{LL}-\downarrow}^{\rm Fock}
\end{equation}
where $\mu_i(D)$ is the single-particle energy of Landau level $i$, and $E_{0\mathrm{LL}+\uparrow,1\mathrm{LL}+\uparrow}^{\rm Fock}$ is the exchange energy between the indicated $0\mathrm{LL}$ and $1\mathrm{LL}$. Each term carries a factor of $1/2$ because the Landau level that distinguishes the two configurations is half filled. There is no corresponding exchange contribution between the $-$ and $+$ valleys. Setting the energy difference to zero gives
\begin{equation}
\mu_{0\mathrm{LL}+\uparrow}(D^*)
-\mu_{0\mathrm{LL}-\downarrow}(D^*)=E_{0\mathrm{LL}+\uparrow,1\mathrm{LL}+\uparrow}^{\rm Fock}
\end{equation}
instead of the bare single-particle crossing condition $   \mu_{0\mathrm{LL}+\uparrow}(D^*)
=\mu_{0\mathrm{LL}-\downarrow}(D^*)$. Thus, the exchange contribution generally shifts the crossing point in $D$.

We now repeat the argument at $\nu=4$. The corresponding configurations are
\begin{align}
\text{(A)}\qquad &
1\mathrm{LL}-\uparrow,\ 
1\mathrm{LL}-\downarrow,\ 
0\mathrm{LL}-\uparrow,\ 
0\mathrm{LL}-\downarrow
\qquad \text{filled},
\\
\text{(B)}\qquad &
1\mathrm{LL}-\uparrow,\ 
1\mathrm{LL}-\downarrow,\ 
0\mathrm{LL}-\uparrow,\ 
0\mathrm{LL}+\uparrow
\qquad \text{filled}.
\end{align}
The structure of the two configurations is the same as at $\nu=7/2$, except that the Landau level distinguishing them is now fully filled rather than half filled. Consequently, both its single-particle contribution and its exchange contribution are twice those at $\nu=7/2$. The energy difference is therefore
\begin{equation}
E_A(D^*)-E_B(D^*)=
\mu_{0\mathrm{LL}-\downarrow}(D^*)-
\mu_{0\mathrm{LL}+\uparrow}(D^*)
+E_{0\mathrm{LL}-\downarrow,1\mathrm{LL}-\downarrow}^{\rm Fock}=0
\end{equation}
which is simply twice the $\nu=7/2$ energy difference. Multiplying the entire energy difference by two clearly does not change where it vanishes. The crossing therefore occurs at exactly the same value of $D^*$.

Thus, Hartree--Fock exchange can shift the crossing away from the value predicted by the bare single-particle dispersion, but it shifts the $\nu=7/2$ and $\nu=4$ crossings identically. If the two crossings coincide in the bare dispersion, they therefore remain coincident after including the Hartree--Fock correction.

\subsection{Band Inversion}
\label{app:band_inversion}

We finally discuss the possibility of band inversion. Here we use the same Hartree--Fock prescription described above: the fitted tight-binding Hamiltonian is taken as the starting dispersion at charge neutrality, and we include additional exchange corrections from the Landau levels filled above $\nu=0$.

Within this approach, we do not find a robust parameter regime in which a finite band-inverted region occurs while simultaneously reproducing the experimentally relevant integer quantum Hall gap.

At $\nu=5/2$, we consider four candidate filling configurations. The corresponding fully filled and half-filled Landau levels are
\begin{align}
&\text{fully filled: } 
1\mathrm{LL}-\uparrow,\;1\mathrm{LL}-\downarrow;
\qquad
\text{half-filled: }0\mathrm{LL}-\uparrow,
\\
&\text{fully filled: }
1\mathrm{LL}-\uparrow,\;0\mathrm{LL}-\downarrow;
\qquad
\text{half-filled: }1\mathrm{LL}-\downarrow,
\\
&\text{fully filled: }
0\mathrm{LL}-\downarrow,\;0\mathrm{LL}-\uparrow;
\qquad
\text{half-filled: }1\mathrm{LL}-\uparrow,
\\
&\text{fully filled: }
1\mathrm{LL}-\uparrow,\;1\mathrm{LL}-\downarrow;
\qquad
\text{half-filled: }0\mathrm{LL}+\uparrow.
\end{align}

\begin{figure}[h]
    \centering
    \includegraphics[width=0.49\linewidth]{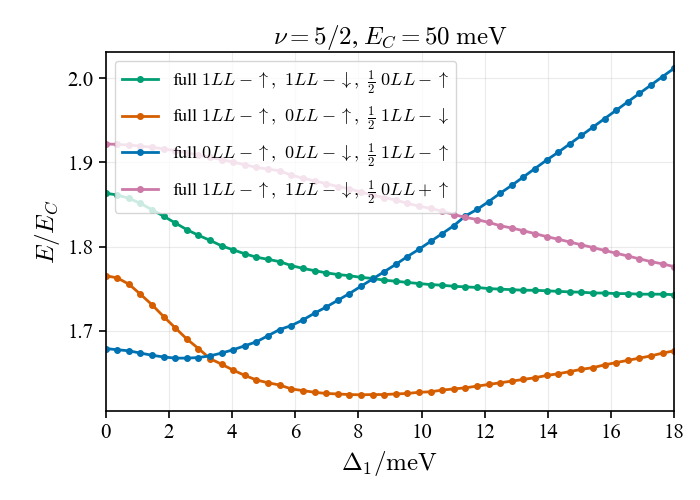}
       \includegraphics[width=0.49\textwidth]{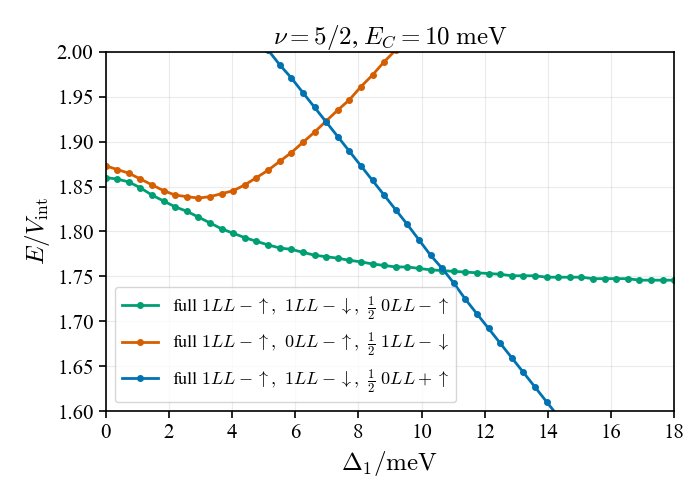}
    \caption{Hartree--Fock energies of the candidate $\nu=5/2$ filling configurations as a function of $\Delta_1$. \emph{Left:} unscreened interaction strength, $E_C=50\,\mathrm{meV}$. \emph{Right:} screened interaction strength, $E_C=10\,\mathrm{meV}$. Screening strongly modifies the relative ordering of the candidate states and suppresses the band-inverted regime, with the relevant $0LL$ levels crossing near $\Delta_1 \simeq 10\,\mathrm{meV}$.}

    \label{fig:5_2_inversion}
\end{figure}

For each configuration, we fix the corresponding occupations and evaluate the Hartree--Fock energy as a function of $\Delta_1$. The resulting energies are shown in Fig.~\ref{fig:5_2_inversion}, with the lowest-energy branch determining the preferred filling configuration. For the half-filled Landau level, we approximate its energy by that of the CFL at half filling, which is sufficient for the level of approximation considered here.

At the unscreened interaction strength, $E_C=50\,\mathrm{meV}$, we find band inversion over a broad range of $\Delta_1$. If instead we use a screened interaction strength, $E_C=10\,\mathrm{meV}$, with the screening expected to arise from RPA, there is no band inversion and the relevant $0LL$ levels cross near $\Delta_1 \simeq 10\,\mathrm{meV}$. We verify that at $\nu=4$ a $0LL$ band crossing also occurs near this value.

These results suggest that tuning $E_C$ can, in principle, restrict band inversion to a finite window of $\Delta_1$. However, obtaining a narrow finite window requires a very fine-tuning of $E_C$. Moreover, the corresponding $\nu=4$ crossing lies far outside the resulting band-inverted region. Thus, bringing the band-inverted region into agreement with the experimentally relevant integer crossing would require additional fine-tuning of the single-particle dispersion.

\begin{figure}[h]
    \centering
    \includegraphics[width=0.49\linewidth]{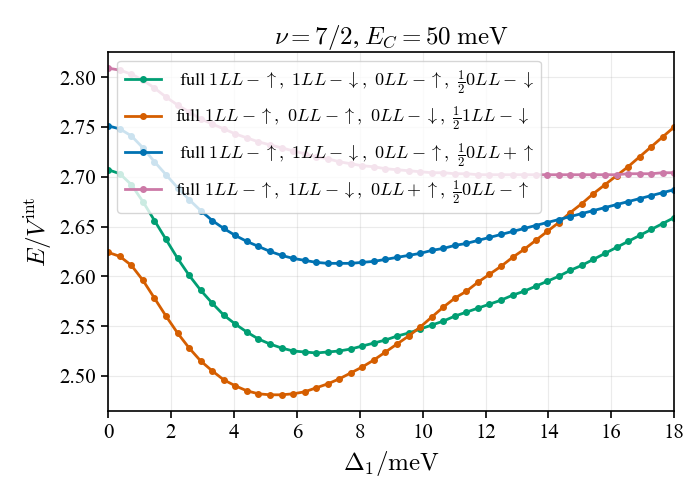}
       \includegraphics[width=0.49\textwidth]{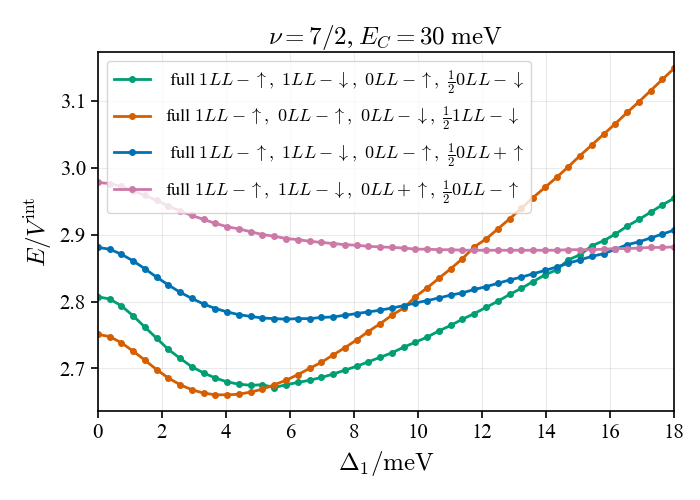}
    \caption{Hartree--Fock energies of the candidate $\nu=7/2$ filling configurations as a function of $\Delta_1$. \emph{Left:} unscreened interaction strength, $E_C=50\,\mathrm{meV}$, for which band inversion persists over a broad range of $\Delta_1 \lesssim 10\,\mathrm{meV}$. \emph{Right:} screened interaction strength, $E_C=30\,\mathrm{meV}$, for which the inverted region is substantially reduced and the relevant $0LL$ levels cross near $\Delta_1 \simeq 15\,\mathrm{meV}$.}

    \label{fig:7_2_inversion}
\end{figure}

Similarly, at $\nu=7/2$, we consider four candidate filling configurations. The corresponding fully filled and half-filled Landau levels are
\begin{align}
&\text{fully filled: } 
1\mathrm{LL}-\downarrow,\;1\mathrm{LL}-\uparrow,\;0\mathrm{LL}-\uparrow;
\qquad
\text{half-filled: }0\mathrm{LL}-\downarrow,
\\
&\text{fully filled: }
1\mathrm{LL}-\uparrow,\;0\mathrm{LL}-\downarrow,\;0\mathrm{LL}-\uparrow;
\qquad
\text{half-filled: }1\mathrm{LL}-\downarrow,
\\
&
\text{fully filled: }
1\mathrm{LL}-\downarrow,\;1\mathrm{LL}-\uparrow,\;0\mathrm{LL}-\uparrow;
\qquad
\text{half-filled: }0\mathrm{LL}+\uparrow,
\\
&\text{fully filled: }
1\mathrm{LL}-\downarrow,\;1\mathrm{LL}-\uparrow,\;0\mathrm{LL}+\uparrow;
\qquad
\text{half-filled: }0\mathrm{LL}-\uparrow.
\end{align}
The results are shown in Fig.~\ref{fig:7_2_inversion}. At the unscreened interaction strength, $E_C=50\,\mathrm{meV}$, we find band inversion over a wide range of $\Delta_1<10\,\mathrm{meV}$. If instead we use a screened interaction strength, $E_C\approx30\,\mathrm{meV}$, this region is suppressed, while the relevant $0LL$ levels cross near $\Delta_1\simeq15\,\mathrm{meV}$.

\section{Topological Properties and Additional Numerical Diagnostics}
\label{app:topological_properties}
\subsection{$K$-Matrix Description of the Two-Component States}
The three two-component candidate states considered in the main text
are Abelian and can therefore be described within the $K$-matrix
formalism. The corresponding Abelian Chern--Simons theory is
\begin{equation}
    \mathcal{L}
    =
    \frac{1}{4\pi}
    K_{IJ}a_I\partial a_J
    +
    \frac{1}{2\pi}
    t_I A\partial a_I ,
\end{equation}
where $K$ is an integer symmetric matrix and $\mathbf{t}$ is the
electromagnetic charge vector. The filling factor and ground-state
degeneracy on the torus are given by
\begin{equation}
    \nu = \mathbf{t}^{T}K^{-1}\mathbf{t},
    \qquad
    \mathrm{GSD}=|\det K|.
\end{equation}

The Halperin $(331)$ state is described by
\begin{equation}
    K_{331}
    =
    \begin{pmatrix}
        3 & 1\\
        1 & 3
    \end{pmatrix},
    \qquad
    \mathbf{t}_{331}
    =
    \begin{pmatrix}
        1\\
        1
    \end{pmatrix}.
\end{equation}
It follows immediately that $\nu=\mathbf{t}_{331}^{T}K_{331}^{-1}\mathbf{t}_{331}=\frac{1}{2},$ and $|\det K_{331}|=8.$ Thus, the $331$ state has an eightfold topological ground-state
degeneracy on the torus.

The weakly paired $d$-wave pseudospin-singlet state is topologically
equivalent to the Jain pseudospin-singlet state. Following
Ref.~\cite{parton}, it can be represented by
\begin{equation}
    K_{\rm weak}
    =
    \begin{pmatrix}
        3 & 2 & -1\\
        2 & 3 & -1\\
        -1 & -1 & 2
    \end{pmatrix},
    \qquad
    \mathbf{t}_{\rm weak}
    =
    \begin{pmatrix}
        1\\
        1\\
        0
    \end{pmatrix}.
\end{equation}
For this representation, $\nu=\mathbf{t}_{\rm weak}^{T}K_{\rm weak}^{-1}\mathbf{t}_{\rm weak}=\frac{1}{2}$ and $|\det K_{\rm weak}|=8$. The $3\times3$ form of $K_{\rm weak}$ corresponds to three bosonic
edge fields. Despite having the same filling factor and torus
degeneracy as the $(331)$ state, the weak-pairing state represents
distinct topological order.

Finally, in the strong-pairing regime the composite fermions form
tightly bound pairs of charge $2e$. The resulting Abelian topological
order is described by the one-component theory \cite{daughter_states}
\begin{equation}
    K_{\rm strong}=(8),
    \qquad
    \mathbf{t}_{\rm strong}=(2).
\end{equation}
Consequently, $\nu=\mathbf{t}_{\rm strong}^{T}K_{\rm strong}^{-1}\mathbf{t}_{\rm strong}=\frac{1}{2}$ and $|\det K_{\rm strong}|=8$. The strong $d$-wave paired state is therefore topologically equivalent
to the conventional strongly paired spin-singlet state. In particular,
all three two-component candidate phases have $\nu=1/2$ and an
eightfold ground-state degeneracy on the torus, despite corresponding to distinct Abelian topological orders.

\subsection{Low-lying part of the Entanglement spectrum}

For many model fractional quantum Hall states, the counting of the entanglement spectrum can be obtained from the root configuration and its associated squeezing rules. In particular, when the state is the exact zero-energy ground state of a suitable parent Hamiltonian, the wave function can often be represented in terms of Jack polynomials, from which generalized Pauli principles and squeezing rules can be obtained \cite{Ardonne2011SpinfulQH,Regnault2009JainHierarchy,Seidel2011HaldaneRezayi}. The weak d-paired and strong paired states considered here do not have such a construction. We therefore determine the expected entanglement-spectrum counting from the edge theory.

The weak-pairing state is more complicated, so we study this one first.
The weak-pairing Abelian phase of the \(d\)-wave paired state was analysed by Read and Green~\cite{Read_2000}. They showed that the edge contains a spin-$1/2$ doublet of chiral Dirac fermions. Upon bosonisation, these give two chiral bosons. In the fractional quantum Hall realization, a third chiral boson describes the charge mode, giving three chiral bosonic degrees of freedom in total~\cite{Read_2000}.

To obtain the entanglement-spectrum counting from the edge theory, we proceed as follows. The edge Hamiltonian decomposes into sectors labelled by the conserved quantum numbers of the entanglement spectrum, such as particle number $N$ and spin $S_z$. We refer to these as zero-mode sectors. For a fixed zero-mode sector, the edge Hamiltonian describes three chiral bosons, and hence gives the three-boson counting. Thus, when all quantum numbers specifying the sector are fixed, the entanglement-spectrum counting is simply that of three chiral bosons.

When considering an entanglement spectrum containing different sectors, one
must also determine where the spectrum associated with each sector begins
in angular momentum $\Delta L_z$. Each sector has the same three-boson
oscillator counting, but its lowest state may occur at a different
$\Delta L_z$. Read and Green classified these sectors and determined their
relative momentum shifts~\cite{Read_2000}, which we use below to obtain
the entanglement-spectrum counting.

For the purpose of counting, the corresponding edge Hamiltonian in some fixed sector may be written as
\begin{equation}
H_{\rm edge}
=
\frac{2\pi}{L_y}
\sum_{a=1}^{3}
v_a \Delta L_z^a
+
\mathrm{const.},
\end{equation}
where $L_y$ is the circumference of the edge, $a=1,2,3$ labels the three chiral boson branches, and $v_a$ is the propagation velocity of branch $a$. The angular-momentum excitation carried by branch $a$ is
\begin{equation}
\Delta L_z^a
=
\sum_{n=1}^{\infty} n N_{n,a},
\end{equation}
where $n=1,2,\ldots$ labels the oscillator modes and $N_{n,a}=0,1,2,\ldots$ is the occupation number of mode $n$ in branch $a$.

The velocities $v_a$ affect energy the positions of the levels but not their counting. The total angular-momentum of the excitation is
\begin{equation}
\Delta L_z
=
\sum_{a=1}^{3} \Delta L_z^a
=
\sum_{a=1}^{3}
\sum_{n=1}^{\infty}
n N_{n,a}.
\end{equation}

Thus, within a fixed zero-mode sector, the three chiral bosons give the counting for weak d-paired state:
\begin{equation}
1,3,9,22,51,\ldots .
\end{equation}

We now turn to the classification of the zero-mode sectors given by
Read and Green~\cite{Read_2000}. They introduce three integers
$r_1,r_2,r_3$ defining the excitation lattice
\begin{equation}
\mathbf{v}
=
\left(
\frac{r_1}{\sqrt{2}},
\frac{r_2}{\sqrt{2}},
\frac{r_3}{2\sqrt{2}}
\right),
\qquad
r_1+r_2+r_3=0\pmod 2,
\end{equation}
and relate them to the quantum numbers of each sector through
\begin{equation}
S_z=\frac{r_1}{2},
\qquad
M=\frac{r_2}{2},
\qquad
Q=\frac{r_3}{4}.
\end{equation}
Here $S_z$ and $Q$ are the physical spin and charge, while $M$ is an
additional neutral quantum number associated with the Landau-level index
in the Jain construction~\cite{Read_2000}. Unlike $S_z$ and $Q$, $M$
is not associated with a microscopic conserved quantity and the
corresponding symmetry can be broken by the dynamics. Here $Q \bmod 1=0$ corresponds to the even-$N$ sector, while
$Q \bmod 1=1/2$ corresponds to the odd-$N$ sector.

The excitation lattice also determines the angular-momentum offset of
each zero-mode sector. For fixed $(Q,S_z,M)$, the angular momentum
relative to the reference ground state is
\begin{equation}
\Delta L_z
=
\frac{\mathbf{v}^2}{2}
+
\sum_{a=1}^{3}
\sum_{n=1}^{\infty}
n N_{n,a},
\end{equation}
where the first term gives the starting angular momentum of the sector
and the second gives the three-boson excitations above it.

\begin{figure}[h]
    \centering
    \includegraphics[width=0.49\linewidth]{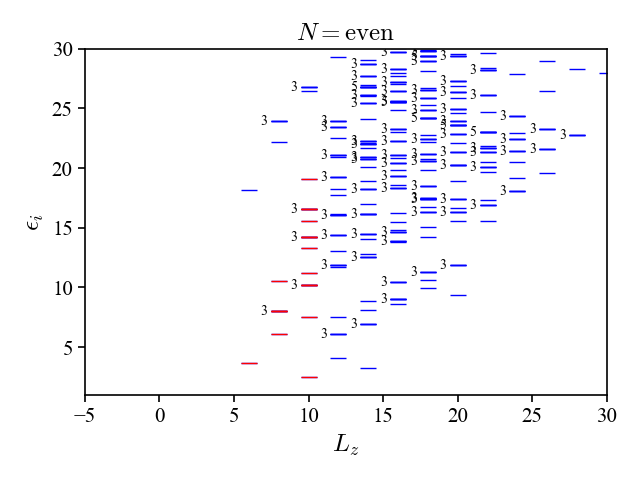}
    \includegraphics[width=0.49\linewidth]{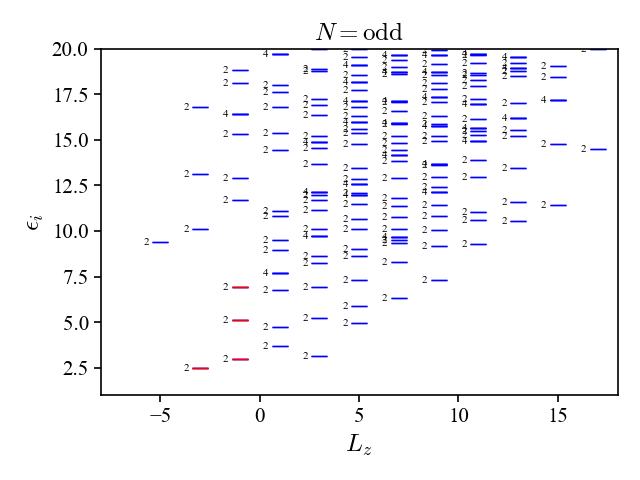}
    \caption{
    Entanglement spectrum obtained from the composite-fermion trial wavefunction for $N=10$. The numbers next to the levels indicate their degeneracies. Since the system has SU(2) symmetry, these degeneracies correspond to different total-spin multiplets. In particular, for even $N$, at $\Delta L_z=1$ we find two singlets and one triplet, as predicted, while at $\Delta L_z=2$ we find six singlets and three triplets. Thus, the low-lying entanglement spectrum agrees with the predicted counting (indicated by red bars). Owing to the small system size, however, the structure of the spectrum cannot be reliably resolved at larger $\Delta L_z$. The additional branch is most likely associated with the presence of the $M$ quantum number and appears at substantially higher entanglement energies.}
    \label{fig:ED_entanglement_spectra}
\end{figure}

This gives us enough information to determine the relevant counting. Since $M$ is not a conserved quantity imposed by the Hamiltonian, we neglect it in the present analysis and take $M=0$. The resulting counting for both even and odd $N$ is summarized in Table~\ref{tab:JSS_counting}. To verify this counting, we also calculate the entanglement spectrum of the weak-pairing trial wave function. The resulting spectra are consistent with the edge-theory counting and are shown in Fig.~\ref{fig:ED_entanglement_spectra}. A similar analysis of entanglement spectra obtained from trial wave functions was carried out by Wu and Jain for bosonic two-component Jain states~\cite{WuJain2013}, which are closely related to the state considered here.

\begin{table}[h]
    \centering
    \renewcommand{\arraystretch}{1.15}
    \begin{tabular}{c@{\qquad\qquad}c}
        $\begin{array}{c|cccc}
            \Delta L_z & 0 & 1 & 2 & 3 \\ \hline
            S_z=0  & 1 & 3 & 9 & 22 \\
            S_z=+1 & 0 & 1 & 3 & 9 \\
            S_z=-1 & 0 & 1 & 3 & 9 \\ \hline
            \text{All }S_z & 1 & 5 & 15 & 40
        \end{array}$
        &
        $\begin{array}{c|cccc}
            \Delta L_z & 0 & 1 & 2 & 3 \\ \hline
            S_z=+\frac{1}{2} & 1 & 3 & 9 & 22 \\
            S_z=-\frac{1}{2} & 1 & 3 & 9 & 22 \\
            S_z=+\frac{3}{2} & 0 & 0 & 1 & 3 \\
            S_z=-\frac{3}{2} & 0 & 0 & 1 & 3 \\ \hline
            \text{All }S_z & 2 & 6 & 20 & 50
        \end{array}$ \\[1ex]
        (a) $N$ even
        &
        (b) $N$ odd
    \end{tabular}
    \caption{Entanglement-spectrum counting predicted by the Read--Green edge theory for the weakly paired JSS state. The counting is shown as a function of $\Delta L_z$ in fixed-$S_z$ sectors for even and odd $N$. The final row in each panel gives the counting after summing over the displayed $S_z$ sectors.}
    \label{tab:JSS_counting}
\end{table}

We can use a similar approach to study the strong-pairing regime. Read and Green showed that in this phase the low-energy degrees of freedom are spin-singlet pairs~\cite{Read_2000}. Consequently, the low-energy edge spectrum contains only even-$N$ sectors with $S_z=0$; odd $N$ or $S_z\neq0$ requires breaking a pair and lies above the pairing gap. The remaining edge degree of freedom is a single chiral charge boson, giving the counting \begin{equation}
1,1,2,3,5,7,\ldots .
\end{equation}

\subsection{Additional Entanglement-Spectrum Data}

\begin{figure}[h]
    \centering
\includegraphics[width=0.43\linewidth]{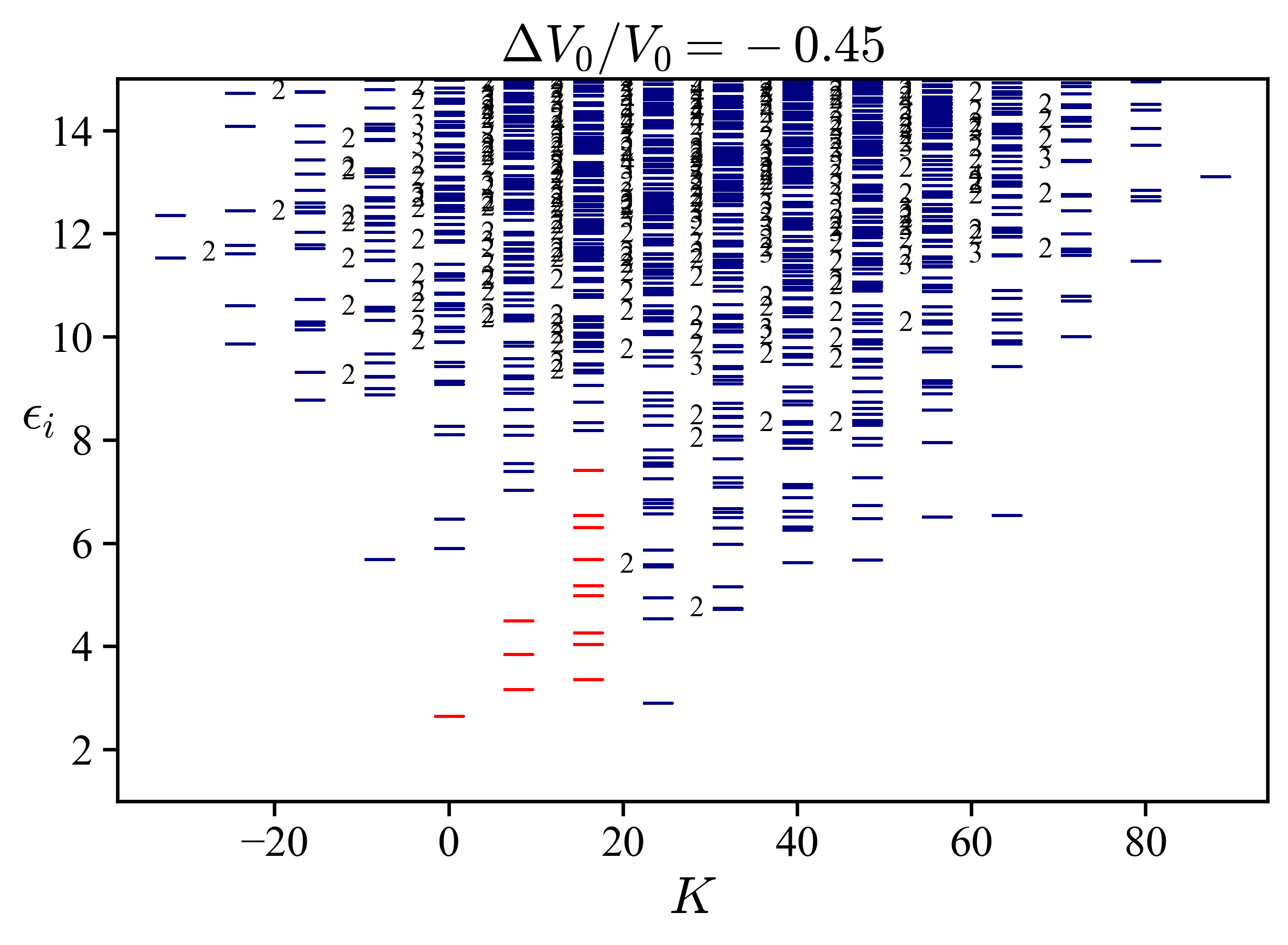}
\includegraphics[width=0.43\linewidth]{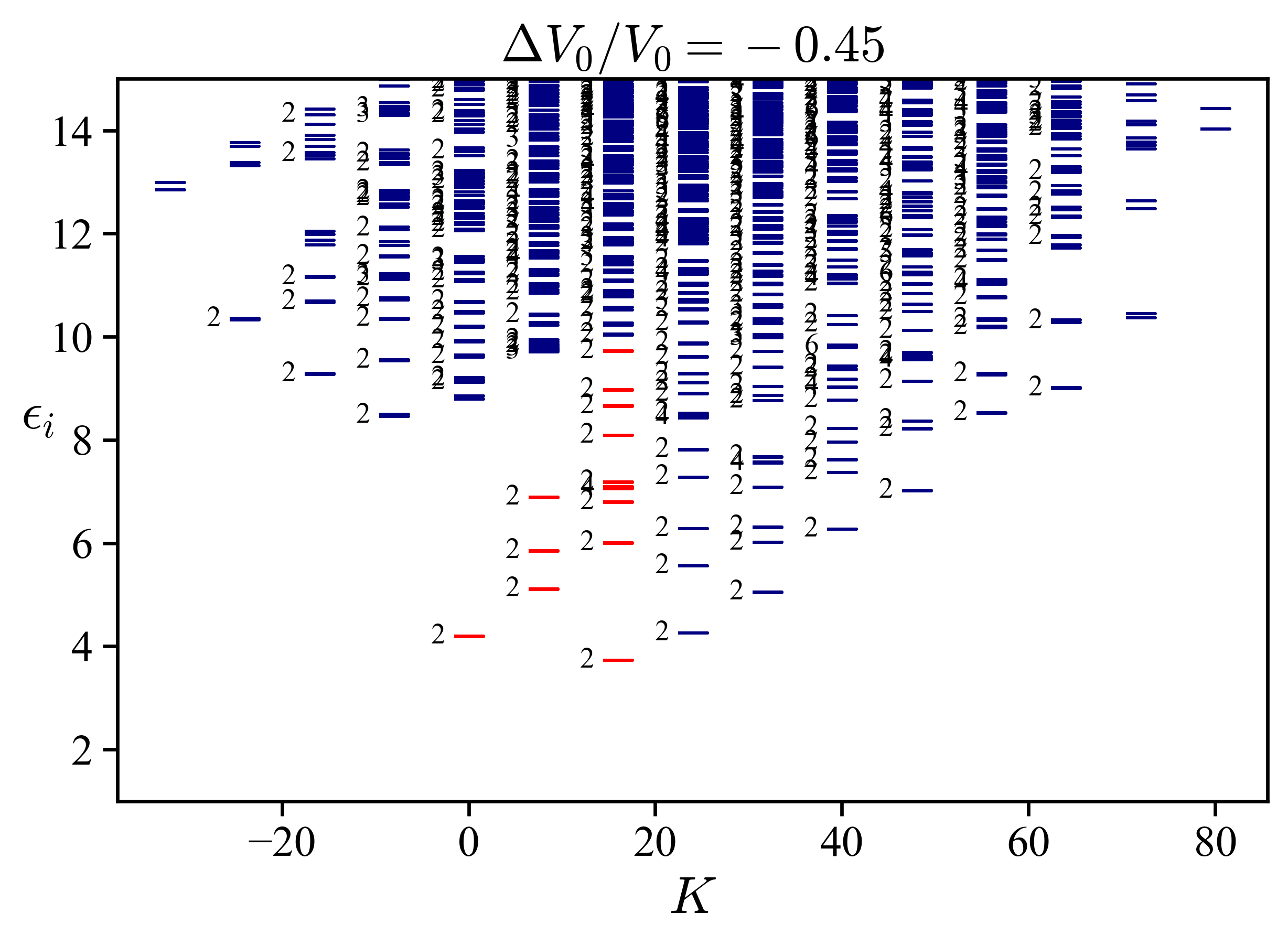}
\includegraphics[width=0.43\linewidth]{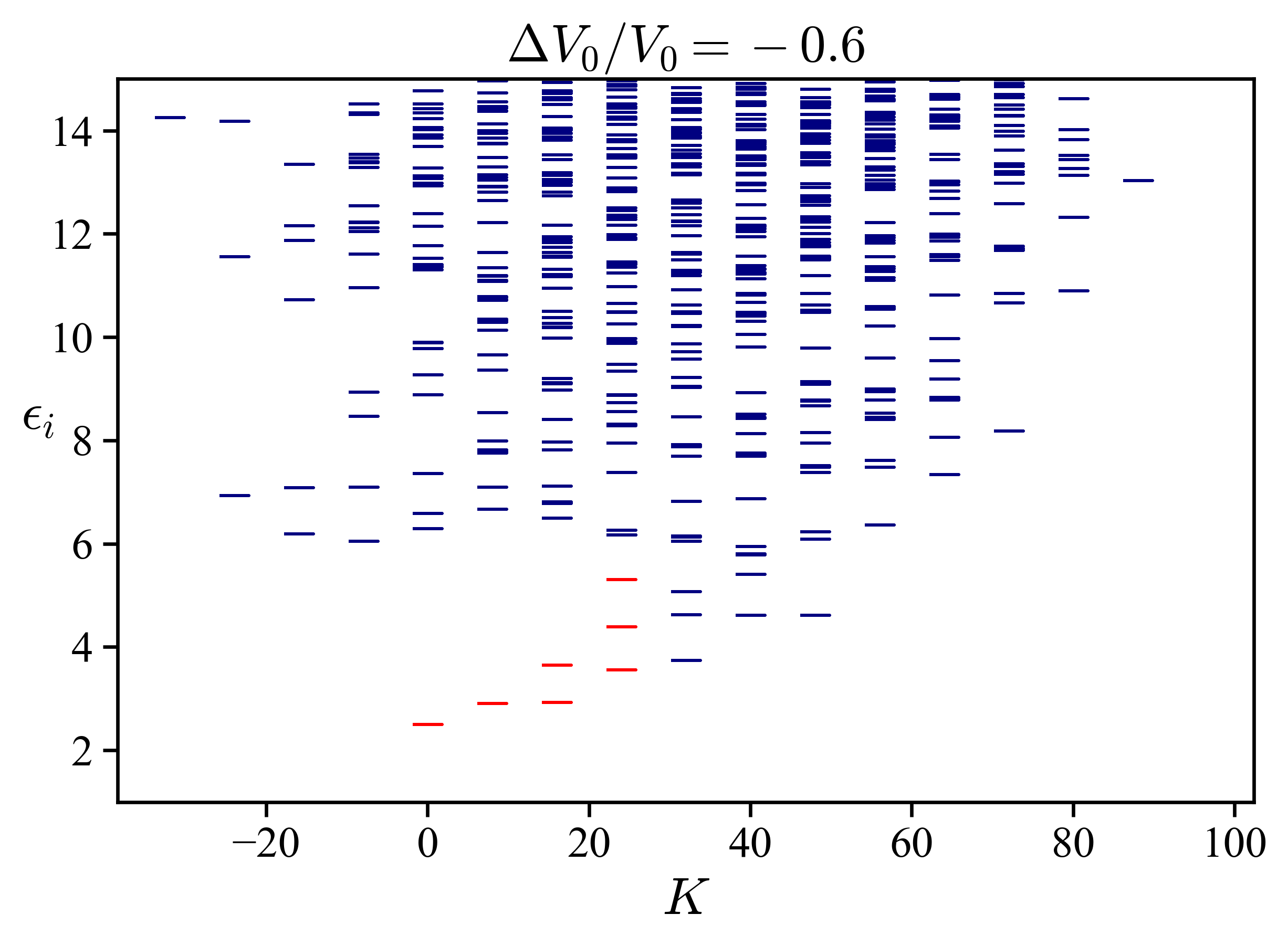}
    \caption{ The top left and top right panels show representative entanglement spectra in the weak-pairing regime at $\Delta V_0/V_0=-0.45$ for the even- and odd-$N$ sectors, respectively. The low-lying levels highlighted in red exhibit the expected counting $1,3,9$ for even $N$ and $2,6$ for odd $N$. The bottom panel shows the entanglement spectrum in the strong-pairing regime at $\Delta V_0/V_0=-0.6$, where the highlighted low-lying levels exhibit the characteristic counting $1,1,2,3$. All spectra were obtained using bond dimension $\chi=8000$, and $L_y=18\ell_B$.}
    \label{fig:additional_ES_data}
\end{figure}

In this appendix, we present additional entanglement-spectrum data that were omitted from the main text section~\ref{sec:toy_model}. The results are shown in Fig.~\ref{fig:additional_ES_data}.

In left panel of Fig.~\ref{fig:additional_ES_data}, we first consider the fixed $S_z=0$ sector in the weak-pairing regime at $\Delta V_0/V_0=-0.45$. As expected, we observe the counting $1,3,9$. For typical parameters in the weak-pairing regime, the $\Delta L_z=2$ counting is difficult to resolve clearly. Nevertheless, the counting in the $\Delta L_z=0,1$ sectors generally agrees with the expected values in both the even- and odd-$N$ sectors for the appropriate spin sectors. In right panel of Fig.~\ref{fig:additional_ES_data}, we show the corresponding data for the odd-$N$ sector at $\Delta V_0/V_0=-0.45$, where we observe the expected counting $2,6$.

Finally, in the Fig.~\ref{fig:additional_ES_data} c), we show the entanglement spectrum in the strong-pairing regime at $\Delta V_0/V_0=-0.6$, where the characteristic counting $1,1,2,3$ is clearly resolved.
The entanglement spectrum for odd $N$ does not seem to have any apparent low-lying structure.

\section{Microscopic Short-Range Interactions}
\label{app:short_range}

In this appendix, we provide the microscopic derivation of the short-range perturbations introduced in the main text. We separate two effects. First, we discuss the correction due to the finite vertical separation between the graphene layers. Second, we consider lattice-scale corrections arising from the fact that the electronic density is not completely point-like on the scale of the graphene lattice.

\subsection{Finite Trilayer Separation}

The first perturbation comes from the finite interlayer separation in trilayer graphene. We incorporate the vertical positions of the graphene layers directly into the Coulomb interaction. Let \(i,j=1,2,3\) label the graphene layers, with vertical coordinates \(z_i,z_j\). We define
\begin{equation}
    \Delta z_{ij}=z_i-z_j ,
\end{equation}
and denote by \(d_{\rm TLG}\) the distance between the top and bottom graphene layers.

For metallic gates located at \(z=\pm d/2\), the gated interaction between charges in layers \(i\) and \(j\) is
\begin{equation}
    V_{ij}(q)
    =\frac{e^2}{4\pi\epsilon_0 \epsilon_{\rm hBN}}
    \frac{2\pi}{q}
    \left[
    -
    \frac{\cosh\!\left((z_i+z_j)q\right)}{\sinh(dq)}
    +
    e^{qd}
    \frac{\cosh\!\left(\Delta z_{ij}q\right)}{\sinh(dq)}
    -
    e^{q|\Delta z_{ij}|}
    \right].
\end{equation}
Expanding this expression for small layer separation gives
\begin{equation}
    V_{ij}(q)
    \approx
    V_C(q,d)-\frac{e^2}{2\epsilon_0 \epsilon_{\rm hBN}} |\Delta z_{ij}|,
\end{equation}
where \(V_C(q,d)\) is the isotropic gated Coulomb interaction obtained by neglecting the vertical separation between the graphene layers in Eq.~\ref{eq:gated_coulomb}. 

The expression above assumes a homogeneous dielectric environment. In the actual device, however, the trilayer graphene is embedded in hBN, and the dielectric constants of hBN and trilayer graphene are different. We therefore perform a separate electrostatic calculation and absorb the effect of the dielectric mismatch into effective parameters,
\begin{equation}
    d\rightarrow d_{\rm eff},
    \qquad
    \Delta z_{ij}\rightarrow \Delta z^{\rm eff}_{ij}.
\end{equation}

\begin{figure}[h]
    \centering
    \includegraphics[width=0.5\linewidth]{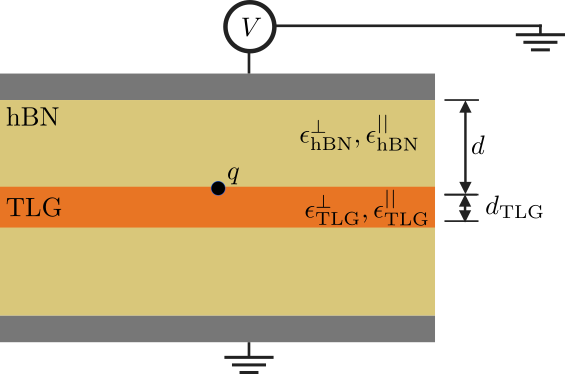}
   \caption{Electrostatic geometry used to determine the effective interaction between the TLG layers. The TLG slab, with dielectric response $\epsilon_{\rm TLG}^{\perp,\parallel}$, is encapsulated by hBN with $\epsilon_{\rm hBN}^{\perp,\parallel}$ and bounded by metallic gates. The bottom gate is grounded and a voltage $V$ is applied to the top gate. A point-like source charge $q$ is placed in the TLG, and the resulting electrostatic potential is evaluated on the graphene layers. Here $d$ denotes the hBN thickness and $d_{\rm TLG}$ the TLG thickness.}

    \label{fig:geometry_trilayer}
\end{figure}

To obtain the effective electrostatic parameters, we solve the Poisson equation for the full hBN--trilayer-graphene--hBN geometry. We place a source charge on one graphene layer and compute the electrostatic potential on another layer. The solution is fixed by the standard electrostatic boundary conditions: the potential vanishes at the metallic gates, the normal component of the displacement field \(D_\perp\) is continuous across dielectric interfaces without free charge, and the electric field is continuous parallel to the interfaces. At the position of the source charge, \(D_\perp\) has the discontinuity required by Gauss's law. Repeating this calculation for each layer pair gives the effective interaction between the graphene layers. The geometry is depicted in the Fig.~\ref{fig:geometry_trilayer}.

The numerical interaction is fitted to the analytic homogeneous-dielectric form
\begin{equation}
    V_{\rm fit}(q)
    =
    \frac{
    V_{ij}\!\left(q;z_{\rm eff},d_{\rm eff}\right)
    }
    {\epsilon_{\rm eff}},
    \label{eq:electrostatic_fit}
\end{equation}
where \(z_{\rm eff}\), \(d_{\rm eff}\), and \(\epsilon_{\rm eff}\) are fitting parameters. Thus \(z_{\rm eff}\) captures the effective vertical separation of the graphene layers, \(d_{\rm eff}\) captures the effective gate distance, and \(\epsilon_{\rm eff}\) captures the overall dielectric screening. This allows us to retain the simple analytic form of the gated Coulomb interaction while incorporating the electrostatics of the full layered dielectric structure. 

For typical experimental parameters, we find
\begin{equation}
    \Delta z_{ij}^{\rm eff}
    \approx
    2.5 \Delta z_{ij},
    \qquad
    d_{\rm eff}
    \approx
    1.4d .
\end{equation}
where we used $\epsilon^{\perp}_{\rm hbN}=3.6$, $\epsilon^{||}_{\rm hbN}=6.6$, $\epsilon^{\perp}_{\rm TLG}=1.5$ and $\epsilon^{||}_{\rm TLG}=2$ \cite{chanda,laturia2018dielectric,santos2013electric}.
We denote the effective top-to-bottom trilayer separation by \(d_{\rm TLG}^{\rm eff}\).

The finite-thickness correction to the bare interaction therefore takes the density--density form
\begin{equation}
    \Delta V_{ij}
    =
    -\frac{e^2}{4\epsilon_0 \epsilon_{\rm hBN}} d_{\rm TLG}^{\rm eff}
    \begin{pmatrix}
        0 & 1 & 2 \\
        1 & 0 & 1 \\
        2 & 1 & 0
    \end{pmatrix}_{ij}.
    \label{eq:finite_thickness_matrix}
\end{equation}
Here \(i,j\) are layer indices. Since this perturbation is independent of spin and valley, it is sufficient at this stage to treat it as a \(3\times 3\) matrix in layer space. 

In addition to the usual intervalley scattering discussed in the next subsection, the finite-thickness correction generates additional intervalley terms with momentum transfer close to the valley separation, $\mathbf q=\mathbf q_0+\mathbf Q$, where $\mathbf Q$ connects the two valleys. These additional terms take the form
\begin{equation}
    H_{\rm hop}
    =
    \frac{1}{2}
    \sum_{ijmn}
    \sum_{ss'v}
    \Delta V_{nm}(\mathbf Q)
    \int
    J^{i,n,s,v}_{\mathbf q}
    J^{j,m,s',\bar v}_{-\mathbf q}
    \frac{d^2\mathbf q}{(2\pi)^2},
\end{equation}
where \(i,j\) label sublattices, \(n,m\) label layers, and \(s,s'\) label spin. Here
\begin{equation}
    J^{i n s v}_{\mathbf q}
    =
    \sum_{\mathbf k}
    c^\dagger_{i n s v}(\mathbf k+\mathbf q)
    c_{i n s \bar v}(\mathbf k),
\end{equation}
with \(\bar v\) denoting the valley opposite to \(v\). Because the low-energy Landau-level wave functions are sublattice--valley locked, this hopping term vanishes after direct projection onto the relevant band subset. We include these terms in the subsequent RPA analysis, as they can, in principle, generate relevant perturbations after the appropriate RPA treatment and projection onto the relevant low-energy subspace. However, we find that their contributions remain negligible even after the RPA treatment.

\subsection{Lattice-Scale Interaction Corrections}

We next consider corrections due to the lattice-scale structure of the electronic density. The treatment follows the approach of Ref.~\cite{Alicea_2006}, adapted to the trilayer geometry. Within each graphene layer, we take the microscopic interaction to be
\begin{equation}
  \hat H_{\rm int}
  =
  U\sum_{\mathbf x,i,n}
  \left[
  \frac{1}{4}n^2_{\mathbf x,i,n}
  -
  \frac{1}{3}\mathbf S(\mathbf x,i,n)^2
  \right]
  +
  \frac{1}{2}
  \sum_{\mathbf{x}\ne\mathbf{x}'+\mathbf a(1-\delta_{ij}),i,j,n}
  V_C\!\left(\mathbf{x}-\mathbf{x}'-\mathbf a(1-\delta_{ij})\right)
  n_{\mathbf{x},i,n,s}
  n_{\mathbf{x}',j,n,s'} .
  \label{eq:lattice_H}
\end{equation}
Here \(U\) is the onsite Hubbard interaction, \(n\) labels the graphene layer, \(i,j=A,B\) are sublattice labels, and \(s,s'\) are spin indices. The operator \(n_{\mathbf x i n}\) denotes the total density on sublattice \(i\) in layer \(n\),
\begin{equation}
    n_{\mathbf x i n}=\sum_s n_{\mathbf x i n s}.
\end{equation}
The valley label does not appear explicitly in this real-space formulation.

We do not include an additional lattice-scale correction between different graphene layers. The reason is that the distance between relevant sites in different layers is
\(\sqrt{d_{\rm TLG}^2+a^2}\), which is very close to \(d_{\rm TLG}\) for the vertical separations considered here. Thus this effect is already accurately captured by the finite interlayer separation discussed above. 

After Fourier transforming and passing to the continuum, the interaction can be written as
\begin{multline}
     \hat H_{\rm int}
     =
     \frac{1}{2}
     \sum_{ins,jms'}
     \int_{q<\Lambda}
     V_C(\mathbf q)
     \rho^{ins}_{\mathbf q}
     \rho^{jms'}_{-\mathbf q}
     \frac{d^2\mathbf q}{(2\pi)^2}
     +
     \frac{1}{2}
     \sum_{ijnss'}
     \int_{q<\Lambda}
     \Delta V_{ij}(\mathbf q)
     \rho^{i,s,n}_{\mathbf q}
     \rho^{j,n,s'}_{-\mathbf q}
     \frac{d^2\mathbf q}{(2\pi)^2}
     \\
     +
     2u_0
     \sum_{i,n}
     \int_{q<\Lambda}
     \left[
     \frac{1}{4}
     \sum_{ss'}
     \rho^{ins}_{\mathbf q}
     \rho^{ins'}_{-\mathbf q}
     -
     \frac{1}{3}
     \mathbf S^{i,n}_{\mathbf q}\cdot
     \mathbf S^{i,n}_{-\mathbf q}
     \right]
     \frac{d^2\mathbf q}{(2\pi)^2}.
     \label{eq:lattice_continuum}
\end{multline}
Here $u_0= U\frac{\sqrt{3}a_0^2}{4}$ is the continuum onsite coupling, with \(a_0\) the graphene lattice constant. The ultraviolet cutoff satisfies \(\Lambda\gg |\mathbf Q|\), so that both small-momentum and intervalley processes can be represented within the same continuum formulation. 

The lattice correction to the Coulomb interaction is
\begin{equation}
    \Delta V_{ij}(q)
    =
    A_{\rm cell}
    \sum_{\mathbf r\neq 0}
    e^{i\mathbf q\cdot\mathbf r}
    V_C(\mathbf r)
    -
    \frac{2\pi}{|q|}
    +
    (1-\delta_{ij})A_{\rm cell}
    \left[
    V_C(\mathbf a)
    +
    \sum_{\mathbf r\neq 0}
    e^{i\mathbf q\cdot\mathbf r}
    \left(
    V_C(\mathbf r-\mathbf a)-V_C(\mathbf r)
    \right)
    \right].
    \label{eq:DeltaV_lattice}
\end{equation}
Expanding this expression at small momentum gives the approximate local form
\begin{multline}
    \frac{1}{2}
    \sum_{ij}
    \int_{q<\Lambda}
    \Delta V_{ij}(\mathbf q)
    \rho^{i,n,s}_{\mathbf q}
    \rho^{j,n,s'}_{-\mathbf q}
    \frac{d^2\mathbf q}{(2\pi)^2}
    \approx
    -
    \alpha E_C
    \frac{a_0}{\ell_B}
    2\pi\ell_B^2
    \int_{q<\Lambda}
    \frac{1}{2}
    \rho^{i,n,s}_{\mathbf q}
    \rho^{j,n,s'}_{-\mathbf q}
    \frac{d^2\mathbf q}{(2\pi)^2}
    \\
    +
    \beta E_C
    \frac{a_0}{\ell_B}
    2\pi\ell_B^2
    \int_{q<\Lambda}
    \rho^{A,n,s}_{\mathbf q}
    \rho^{B,n,s'}_{-\mathbf q}
    \frac{d^2\mathbf q}{(2\pi)^2},
    \label{eq:lattice_alpha_beta}
\end{multline}
with $ \alpha\approx 0.6,\beta\approx 0.35$ which we obtain from numerical calculation. However, since $E_C \propto 1/\ell_B$, the expression is in fact independent of $\ell_B$. It is written in this form for convenience, as it allows us to directly read off the zeroth Haldane pseudopotentials as $-\alpha a_0/(\ell_B)$ from the first term and $-\beta a_0/\ell_B$ from the second.

We now separate the continuum fields into valley components. After this decomposition, the interaction can be organized into terms that preserve valley and terms that hop between valleys. Schematically,
\begin{equation}
    \hat H_{\rm int}
    =
    \hat H_C
    +
    \hat H_{1,V}
    +
    \hat H_{1,U}
    +
    \hat H_{C,\rm hop}
    +
    \hat H_{V,\rm hop}
    +
    \hat H_{U,\rm hop}.
\end{equation}
More explicitly,
\begin{multline}
     \hat H_{\rm int}
     =
     \hat H_C
     +
     \frac{1}{2}
     \int_q
     \Delta V_{ij}(\mathbf q)
     \rho^{i,s,n,v_1}_{\mathbf q}
     \rho^{j,n,s',v_2}_{-\mathbf q}
     \frac{d^2\mathbf q}{(2\pi)^2}
     +
     2u_0
     \int_q
     \left[
     \frac{1}{4}
     \rho^{insv_1}_{\mathbf q}
     \rho^{ins'v_2}_{-\mathbf q}
     -
     \frac{1}{3}
     \mathbf S^{i,n,v_1}_{\mathbf q}\cdot
     \mathbf S^{i,n,v_2}_{-\mathbf q}
     \right]
     \frac{d^2\mathbf q}{(2\pi)^2}
     \\
     +
     \frac{V_C(\mathbf Q)}{2}
     \int
     J^{insv_1}_{\mathbf q}
     J^{jms'v_2}_{-\mathbf q}
     \frac{d^2\mathbf q}{(2\pi)^2}
     +
     \frac{\Delta V_{ij}(\mathbf Q)}{2}
     \int
     J^{i,s,n,v_1}_{\mathbf q}
     J^{j,n,s',v_2}_{-\mathbf q}
     \frac{d^2\mathbf q}{(2\pi)^2}
     \\
     +
     2u_0
     \int
     \left[
     \frac{1}{4}
     J^{insv_1}_{\mathbf q}
     J^{ins'v_2}_{-\mathbf q}
     -
     \frac{1}{3}
     \mathbf K^{i,n,v_1}_{\mathbf q}\cdot
     \mathbf K^{i,n,v_2}_{-\mathbf q}
     \right]
     \frac{d^2\mathbf q}{(2\pi)^2}.
     \label{eq:valley_decomposed_interaction}
\end{multline}
For compactness, the sums over repeated spin, valley, sublattice, and layer indices have been suppressed.

The intervalley operators are defined as
\begin{equation}
    J^{insv}_{\mathbf q}
    =
    \sum_{\mathbf k}
    c^\dagger_{insv}(\mathbf k+\mathbf q)
    c_{ins\bar v}(\mathbf k),
    \qquad
    \mathbf K^{inv}_{\mathbf q}
    =
    \sum_{\mathbf k}
    \boldsymbol{\sigma}_{ss'}
    c^\dagger_{insv}(\mathbf k+\mathbf q)
    c_{ins'\bar v}(\mathbf k),
\end{equation}
where \(\bar v\) denotes the valley opposite to \(v\). The term \(\hat H_C\) is the isotropic Coulomb interaction. The terms denoted by \(\hat H_1\) preserve valley, while the terms containing \(J\) or \(\mathbf K\) are intervalley hopping terms.

Due to sublattice--valley locking, the intervalley hopping terms vanish after direct projection to the relevant Landau levels. However, they must be retained before applying RPA, since screening can mix them into other interaction channels that do not vanish after projection. The valley-preserving short-range terms survive projection directly and are therefore expected to give the dominant microscopic anisotropies. Moreover, \(V_C(\mathbf Q)\), \(\Delta V_{ij}(\mathbf Q)\), and \(u_0\) are of comparable short-distance scale because the latter two do not carry the small-\(q\) Coulomb suppression.

\section{Random-phase approximation}
\label{app:RPA}

In this appendix we describe the random-phase approximation (RPA) used to incorporate screening by Landau levels outside the explicitly retained low-energy subspace. Because the microscopic interactions considered in this work may depend on spin, valley, and layer--sublattice indices, it is useful to formulate the RPA first for a general four-index interaction. Density--density interactions and the usual isotropic Coulomb RPA then follow as special cases.

The central object is the static polarization tensor. After introducing it, we derive the corresponding screened interaction, specialize the result to density--density interactions, and finally obtain the screening of a weak anisotropic perturbation about an isotropic Coulomb interaction. We conclude by describing the phenomenological parametrization of the scalar polarization function used in the numerical calculations.

\subsection{Polarisation tensor}

\begin{figure}
\centering
\includegraphics[width=0.7\linewidth]{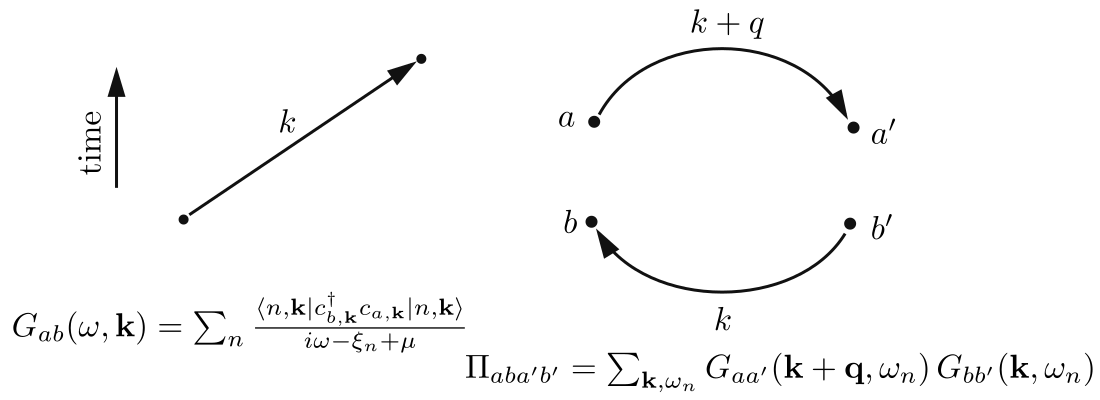}
\caption{Single-particle propagator and polarization bubble used in the tensor formulation of the RPA.}
\label{fig:RPA_bubble}
\end{figure}

We first introduce the single-particle quantities used in the RPA analysis. The single-particle Green's function is
\begin{equation}
G_{ab}(\mathbf{k},i\omega)=\sum_n\frac{\langle n,\mathbf{k}|c^\dagger_{b,\mathbf{k}}c_{a,\mathbf{k}}|n,\mathbf{k}\rangle}{i\omega-\epsilon_n+\mu},
\end{equation}
where $n$ labels the single-particle eigenstates, $\epsilon_n$ are the corresponding single-particle energies, and $\mu$ is the chemical potential. The indices $a$ and $b$ denote the internal degrees of freedom relevant to the interaction, which, depending on the context, may include spin, valley, and layer--sublattice quantum numbers. Throughout this section, we use Latin indices $a,b,c,d,\ldots$ for such internal indices. The propagator is illustrated in Fig.~\ref{fig:RPA_bubble}.

We further introduce the generalized density operator
\begin{equation}
\rho_{ab}(\mathbf q)
=
\sum_{\mathbf k}
c^\dagger_{a,\mathbf{k}+\mathbf q}
c_{b,\mathbf k}.
\end{equation}
In a Landau-level basis this can be written as
\begin{equation}
\rho_{ab}(\mathbf q)
=
\sum_{\alpha,\beta,k_y}
F_{\alpha\beta}(\mathbf q),
a^\dagger_{\alpha,a,k_y+q_y}
a_{\beta,b,k_y},
\end{equation}
where \(F_{\alpha\beta}(\mathbf q)\) is the standard Landau-level form factor, and $\alpha,\beta$ refer to relevant Landau levels.

The static polarization bubble is consequently a rank-four tensor,
\begin{equation}
\Pi_{abcd}(\mathbf q)
\equiv
\frac{1}{\beta A}
\sum_{\mathbf k,\omega}
G_{ac}(\mathbf k+\mathbf q,i\omega)
G_{db}(\mathbf k,i\omega).
\label{eq:RPA_Pi_definition}
\end{equation}
Evaluating the Matsubara sum at zero temperature gives
\begin{equation}
\Pi_{abcd}(\mathbf q)
=
\frac{2}{A}
\sum_{X_i,X_j}
\sum_{\substack{n>n_F, m\leq n_F}}
\frac{
\operatorname{Re}
\left[
\langle m,X_i|\rho_{ab}(\mathbf q)|n,X_j\rangle
\langle n,X_j|\rho_{dc}(-\mathbf q)|m,X_i\rangle
\right]
}
{\epsilon_n-\epsilon_m}.
\end{equation}
Here, we use the Landau gauge, such that $|n,X_i\rangle$ denotes a single-particle eigenstate, with $n$ labelling the eigenstate and $X_i$ its guiding-center coordinate. The index $m$ runs over occupied states and $n$ over unoccupied states, with $n_F$ denoting the highest-energy filled band. This tensor form is useful because the intermediate propagator need not preserve the individual internal indices: such mixing is encoded directly in the off-diagonal Green's functions $G_{ab}$.

\subsection{RPA for a general four-point interaction}

We now formulate the RPA for a general interaction which is not assumed to be diagonal in the internal indices. This is needed when the interaction can scatter spin, valley, layer, or sublattice labels, so that the screened vertex remains a genuine four-index object.

\begin{figure}
\centering
\includegraphics[width=0.7\linewidth]{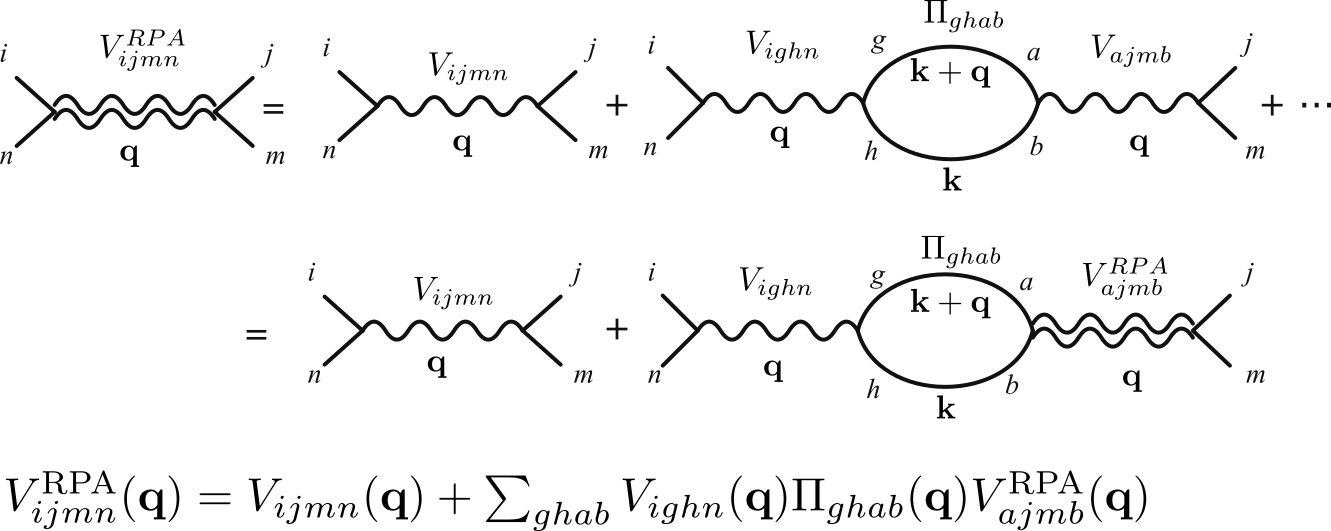}
\caption{RPA series for a general four-point interaction.}
\label{fig:RPA_anisotropic}
\end{figure}

We start from
\begin{equation}
\hat V
=
\frac{1}{2A}
\sum_q
\sum_{ijmn}
V_{ijmn}(q)\,
c^\dagger_{i,k+q}
c^\dagger_{j,k'-q}
c_{m,k'}
c_{n,k}.
\end{equation}
Let \(V^{\rm RPA}_{ijmn}(q)\) denote the RPA-screened vertex and
\(\Pi_{abcd}(q)\) the polarization bubble. Suppressing the common momentum argument, the RPA ladder of bubble insertions gives (see Fig.~\ref{fig:RPA_anisotropic})
\begin{equation}
V^{\rm RPA}_{ijmn}
=
V_{ijmn}
-
\sum_{abgh}
V_{ighn}\,
\Pi_{ghab}\,
V^{\rm RPA}_{ajmb}.
\label{eq:RPA_tensor_Dyson}
\end{equation}
The important point is that the bubble contracts two legs of the interaction vertex, while the remaining two external legs are left open. Thus the RPA equation is a matrix equation only after pairs of indices are grouped into composite indices.

Moving the bubble term to the left, we write
$\sum_{ab}
V^{\rm RPA}_{ajmb}\,
A_{iabn}
=
V_{ijmn},$
where
\begin{equation}
A_{iabn}
\equiv
\delta_{ia}\delta_{nb}
+
\sum_{gh}
\Pi_{ghab}\,
V_{ighn}.
\label{eq:RPA_tensor_A}
\end{equation}
Here \(A\) should be understood as a matrix in the composite index space
\((i,n)\leftrightarrow(a,b)\). Defining \(B=A^{-1}\), with
\begin{equation}
\sum_{in}
A_{iabn}B_{cind}
=
\delta_{ad}\delta_{bc},
\end{equation}
the solution is
\begin{equation}
V^{\rm RPA}_{djmc}
=
\sum_{in}
V_{ijmn}\,
B_{cind}.
\label{eq:RPA_tensor_solution}
\end{equation}

This is the useful form of the generic four-tensor RPA: for each fixed pair of spectator indices \((j,m)\), the screened vertex is obtained by inverting the bubble-renormalization matrix \(A\) in the remaining two-index space. In compact notation,
\begin{equation}
\boxed{
W
=
V(\id+\Pi V)^{-1}}
\label{eq:RPA}
\end{equation}
with the understanding that the inverse is taken in the composite-index space defined above.

\subsection{Density--density interactions}

Suppose the interaction has the form
\begin{equation}
\hat V
=
\frac{1}{2A}
\sum_{\mathbf q}\sum_{ij}
V_{ij}(\mathbf q)
:\rho_i(\mathbf q)\rho_j(-\mathbf q): .
\label{eq:dens_dens_equation}
\end{equation}
In the four-index notation this corresponds to
\begin{equation}
V_{ijmn}(\mathbf q)
=
\delta_{in}\delta_{jm}V_{ij}(\mathbf q).
\end{equation}
Consequently, only the density components of the polarization tensor enter. We define
\begin{equation}
\Pi^{\mathrm{2D}}_{ij}(\mathbf q)
\equiv
\Pi_{iijj}(\mathbf q).
\label{eq:RPA_Pi2D}
\end{equation}
The RPA equation then reduces from an equation in the full four-index space to an ordinary matrix equation in the density indices,
\begin{equation}
W
=
V
\left(
\mathbbm 1+\Pi^{\mathrm{2D}}V
\right)^{-1}.
\label{eq:RPA_density_matrix}
\end{equation}

\subsection{Isotropic interaction as a special case}

We now recover the conventional RPA result for an interaction that is independent of spin, valley, and layer--sublattice labels,
\begin{equation}
\hat V_0
=
\frac{1}{2A}
\sum_{\mathbf q}\sum_{ab}
V_0(\mathbf q)
:\rho_a(\mathbf q)\rho_b(-\mathbf q):.
\end{equation}

Because the bare interaction is proportional to the uniform density channel, the matrix equation collapses to the usual scalar result,
\begin{equation}
\boxed{
V_{\mathrm{RPA}}(q)
=
\frac{V_0(q)}
{1+\Pi_0(q)V_0(q)}
}.
\label{eq:RPA_scalar}
\end{equation}

The scalar polarization function is obtained by contracting the tensor polarization with the uniform density vertices,
\begin{equation}
\Pi_0(q)
=
\sum_{abcd}
\delta_{ab}\delta_{cd}
\Pi_{abcd}(q)
=
\sum_{ac}\Pi_{aacc}(q).
\label{eq:RPA_scalar_contraction}
\end{equation}
For eigenstates that remain diagonal in spin--valley space but may mix layer--sublattice components, this contraction reproduces the polarization obtained directly from the total density operator.

For Landau-level eigenstates,
\begin{equation}
|n\rangle
=
\sum_{s,\alpha}
u_{n,s,\alpha}|s,\alpha\rangle ,
\end{equation}
where \(s\) denotes the internal spin--valley--layer--sublattice state and \(\alpha\) the orbital Landau-level index, the density matrix element may be written
\begin{equation}
\left|
\langle m,X_i|
\rho_{\mathbf q}
|n,X_j\rangle
\right|^2
=
\delta_{X_i,X_j+q_y}
\left|
\sum_{s,a,b}
u^{*}_{m,s,\alpha}
u_{n,s,\alpha}
F_{\alpha \beta}(\mathbf q)
\right|^2 .
\end{equation}
Defining the corresponding band form factor \(F_{nm}(\mathbf q)\), the guiding-center sum gives
\begin{equation}
\Pi_0(\mathbf q)
=
\frac{1}{2\pi \ell_B^2}
\sum_{\substack{n>n_F, m\leq n_F}}
\frac{
|F_{nm}(\mathbf q)|^2
+
|F_{nm}(-\mathbf q)|^2
}
{\epsilon_n-\epsilon_m},
\label{RPA:isotropic}
\end{equation}
where we used$\frac{N_{\mathrm{LL}}}{A}
=
\frac{1}{2\pi l_B^2}$, where $N_{LL}$ is number of electrons within Landau level.

\subsection{Screening of a weak anisotropic perturbation}
\label{sec:RPA_perturbation}

The interaction used in the main calculation consists of an isotropic gate-screened Coulomb interaction supplemented by a weaker symmetry-breaking contribution,
\begin{equation}
\hat V
=
\hat V_0+\Delta\hat V.
\end{equation}
Since the anisotropic part is small compared with the dominant Coulomb interaction, we retain its RPA renormalization only to first order in \(\Delta V\).

Formally,
\begin{equation}
\Delta V_{\mathrm{RPA}}
=
(V_0+\Delta V)
\left[
\mathbbm 1+\Pi(V_0+\Delta V)
\right]^{-1}
-
V_0
\left(
\mathbbm 1+\Pi V_0
\right)^{-1}.
\label{eq:RPA_delta_general}
\end{equation}

Two cases are relevant.

\subsubsection{Density--density perturbation}

First consider
\begin{equation}
\Delta\hat V
=
\frac{1}{2A}
\sum_{\mathbf q}\sum_{ij}
\Delta V_{ij}(\mathbf q)
:\rho_i(\mathbf q)\rho_j(-\mathbf q): .
\end{equation}
Expanding Eq.~\eqref{eq:RPA_delta_general} to first order in \(\Delta V\) gives
\begin{equation}
\boxed{
\Delta V_{\mathrm{RPA}}
=
\left(
\mathbbm 1+V_0\Pi^{\mathrm{2D}}
\right)^{-1}
\Delta V
\left(
\mathbbm 1+\Pi^{\mathrm{2D}}V_0
\right)^{-1}
+O(\Delta V^2)
}.
\label{RPA:rho_rho_form}
\end{equation}
Thus the perturbation is screened independently on its two density vertices. This is the expression used whenever the symmetry-breaking interaction can be represented in density--density form.

\subsubsection{General four-point perturbation}

More generally, the perturbation may have the four-point form
\begin{equation}
\Delta\hat V
=
\frac{1}{2A}
\sum_{\mathbf q}
\sum_{ijmn}
\Delta V_{ijmn}(\mathbf q)
c^\dagger_{i,\mathbf{k}+\mathbf q}
c^\dagger_{j,\mathbf{k}'-\mathbf q}
c_{m,\mathbf{k}'}
c_{n,\mathbf{k}}.
\end{equation}
In this case the full rank-four polarization tensor must be retained. Expanding the general RPA equation to first order in \(\Delta V\) produces the corresponding linear renormalization of the four-point vertex. In the notation used in the main calculation, this may be expressed as
\begin{equation}
V^{\rm RPA}_{djmc}
=
\Delta V_{djmc}
-
\delta_{cd}
\frac{V_0}{1+\Pi_0V_0}
\sum_{in}
\Delta V_{ijmn},
\widetilde{\Pi}_{in}
-
\delta_{jm}
\frac{V_0}{1+\Pi_0V_0}
T_{dc},
\label{RPA:4_point_form}
\end{equation}
with
\begin{equation}
T_{dc}
=
\Delta A_{dc}
-
\frac{\delta_{cd}V_0}
{1+\Pi_0V_0}
\sum_{il}
\Delta A_{il}\widetilde{\Pi}_{il},
\end{equation}
and
\begin{equation}
\Delta A_{il}
=
\sum_{gha}
\Pi_{ghaa}\Delta V_{iaal},
\qquad
\widetilde{\Pi}_{il}
=
\sum_g\Pi_{ggil}.
\end{equation}

This expression is used in Sec.~\ref{sec:realistic_phase_diagram} to obtain the relevant phase diagram.

\subsection{Phenomenological polarization used in the calculation}
\label{sec:RPA_fit}

A direct treatment of all screening Landau levels is numerically impractical in the regime of interest because the relevant single-particle gaps can be small compared with the interaction scale. We therefore use the RPA as an effective description of virtual transitions involving the remote levels.

\begin{figure}
\centering
\includegraphics[width=0.4\linewidth]{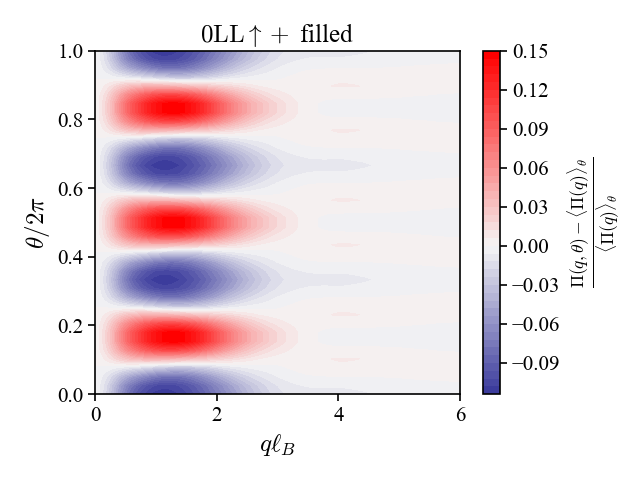}
\includegraphics[width=0.4\linewidth]{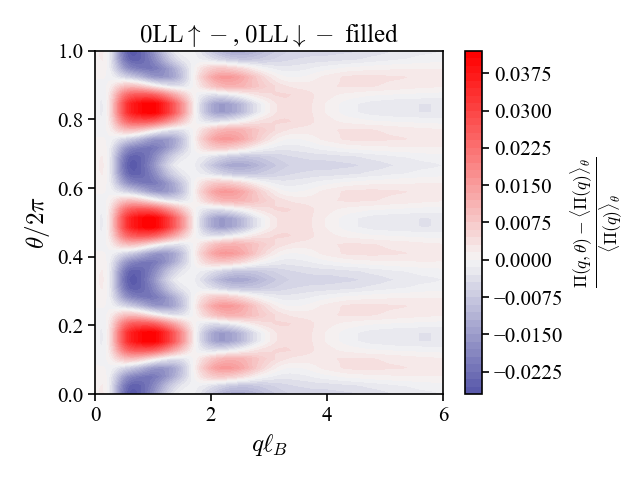}
\caption{Angular dependence of the static RPA polarisation for two representative filling configurations. The colour scale shows the relative deviation of $\Pi_0(q,\theta)$ from its angular average $\langle\Pi_0(q)\rangle_\theta$. The characteristic $C_3$ anisotropy inherited from the lattice is visible in both cases and is strongest at intermediate momenta. Results are shown for $E_C=50~\mathrm{meV}$ and $B=13~\mathrm{T}$.}
\label{fig:RPA_theta}
\end{figure}

Rotational symmetry is not exact in the full microscopic model. Numerically, however, we find the anisotropy of the polarisation to be relatively weak in the parameter regime of interest and therefore approximate
\begin{equation}
    \Pi_0(\mathbf q)\simeq\Pi_0(q).
\end{equation}
To assess this approximation, we parametrise the momentum as
$\mathbf{q}=q(\hat{\mathbf{x}}\cos\theta+\hat{\mathbf{y}}\sin\theta)$,
where $q=|\mathbf{q}|$ and $\theta$ is the momentum-space angle. In
Fig.~\ref{fig:RPA_theta}, we plot the relative deviation of
$\Pi_0(q,\theta)$ from its angular average as a function of $q$ and
$\theta$ for two representative filling configurations. In both cases,
the anisotropy is largest at intermediate momenta, $q\sim\ell_B^{-1}$,
and exhibits the expected $C_3$ symmetry inherited from the lattice.
Its magnitude depends on the filling configuration: for the configuration
shown in the left panel it reaches approximately $10$--$15\%$, whereas
for that shown in the right panel it remains below approximately $4\%$.
Although the larger anisotropy is not negligible, it is comparable to
the uncertainty in determining the overall RPA polarisation, while
retaining the full angular dependence would substantially complicate
the subsequent calculations. We therefore neglect this anisotropy and
use the rotationally averaged polarisation throughout.

The numerically evaluated scalar polarization is well represented by
\begin{equation}
\Pi_0(q)
=
\frac{A_{\Pi}}{E_C\ell^2_B}
\tanh\left(
b\ell^2_Bq^2
\right).
\label{eq:RPA_fit}
\end{equation}
The two coefficients are chosen to reproduce the small-\(q\) and large-\(q\) behaviour of the microscopic polarization. In quantum Hall, the relevant interaction length scale is typically $q\leq \frac{1}{\ell_B}$, regime in which $\Pi_{0}(q)\approx A_{\Pi} b q^2/E_C$, so we expect the behaviour to be controlled by product $A_{\Pi} b$. $A_{\Pi}$ is obtained by fitting the curve in the large-$q$ regime, $q>5\ell_B^{-1}$, using the least-squares method. Once $A_{\Pi}$ has been determined, $b$ is obtained by fitting the small-$q$ region, $q<4\ell_B^{-1}$, using nonlinear least squares.

\begin{figure}
\centering
\includegraphics[width=0.4\linewidth]{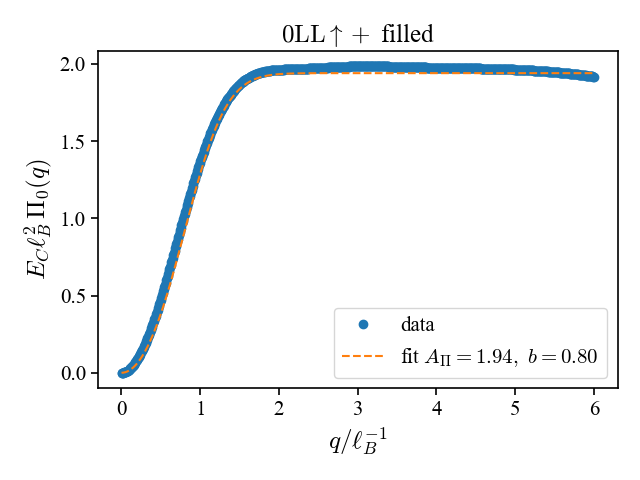}
\includegraphics[width=0.4\linewidth]{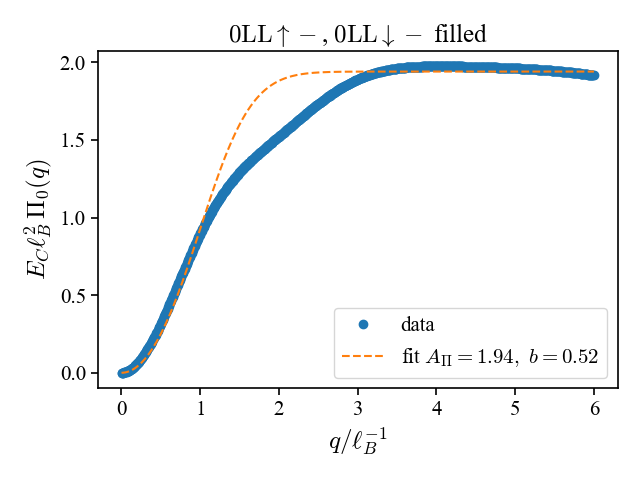}
\caption{Representative fits of the microscopic polarization to Eq.~\eqref{eq:RPA_fit}. Here we used $E_C=50\rm meV, B=13\;\rm T$. $A_{\Pi}$ is obtained by fitting large $q$ behaviour, and $b$ is then obtained by fitting the low $q$ behaviour. }
\label{fig:RPA_fit}
\end{figure}

Table~\ref{tab:RPA} summarizes representative fit parameters near band crossing. The fitted values vary moderately between the different integer-filled configurations. In particular, \(A_{\Pi}\approx 2\) is essentially unchanged in these fits since it is set by remote band behaviour, which should not care about explicit details of dispersion near charge neutrality, while most of the filling dependence is encoded in \(b\). Some of the fits corresponding to band crossings are shown in Fig.~\ref{fig:RPA_fit}.

\begin{table}[h!]
\centering
\begin{tabular}{|c|c|c|}
\hline
\textbf{Bands filled} & $A_{\Pi}$ & $b$ \\
\hline
$0LL\uparrow-$ & 1.94 & 0.65   \\
\hline
 $0LL\uparrow+$ &   1.94 & 0.8\\
\hline
 $0LL\uparrow-,0LL\uparrow+$ &1.94 & 0.66 \\
\hline
 $0LL\uparrow-,0LL\downarrow-$  & 1.94 &0.53\\
\hline
$0LL\uparrow+,0LL\downarrow+$ &  1.94 & 0.8 \\
\hline
\end{tabular}
\caption{
Representative RPA fit parameters near
\(\Delta_1=8\,\mathrm{meV}\)
for trilayer graphene, using the coefficients of
Ref.~\cite{chanda}.
The \(1\mathrm{LL}\uparrow-\) and
\(1\mathrm{LL}\downarrow-\) bands are filled in all cases considered.
The variation of the fits with displacement field is modest over the range relevant to the calculation.
}
\label{tab:RPA}
\end{table}

Across the integer transition at \(\nu=4\), the calculated screening changes by approximately \(16\%\). More generally, the fits obtained from the different nearby integer configurations remain of comparable magnitude. We therefore regard the precise RPA parametrization as a phenomenological uncertainty rather than as a sharply determined microscopic quantity.

The procedure used in the main text is consequently the following: the dominant isotropic Coulomb interaction is screened using Eq.~\eqref{eq:RPA_scalar} with the fitted polarization of Eq.~\eqref{eq:RPA_fit}, while weak symmetry-breaking interactions are renormalized perturbatively using either Eq.~\eqref{RPA:rho_rho_form} or, when necessary, the full four-point expression Eq.~\eqref{RPA:4_point_form}. The polarization tensors used here are obtained from a microscopic calculation using the single-particle dispersion of the tight-binding model described in App.~\ref{app:tight_binding}.

\section{RPA Screening of Short-Range Interactions}
\label{app:RPA_short_range}
As discussed in the main text, RPA includes Hartree-like but not Fock-like diagrams. For short-range interactions, this distinction leads to an ambiguity that we illustrate below and resolve for the interaction considered in this work.

\subsection{Ambiguity of a contact interaction in RPA}

Consider two fermionic species labelled by \(i,j\), interacting through an isotropic, momentum-independent density-density interaction,
\begin{equation}
\hat V
=
\sum_{i,j}\sum_{\mathbf q}
U:\rho_i(\mathbf q)\rho_j(-\mathbf q): .
\end{equation}
In matrix notation, the interaction may be represented as
\begin{equation}
    V^{A}_{ij}=\begin{pmatrix}
        U &U \\ U &U
    \end{pmatrix}
\end{equation}.

For a strictly local interaction, however, the same-species term vanishes identically. Indeed, for a single fermionic flavour,
\begin{equation}
V_0\sum_{\mathbf q}
:\rho_i(\mathbf q)\rho_i(-\mathbf q):
=
V_0\int d^2\mathbf r,
:\rho_i(\mathbf r)\rho_i(\mathbf r):
=0,
\end{equation}
as a consequence of fermionic antisymmetry. The same physical interaction can therefore equivalently be written as
\begin{equation}
    V^{B}_{ij}=\begin{pmatrix}
        0 &U \\ U &0
    \end{pmatrix}
\end{equation}.

In an exact perturbative calculation, in which all Feynman diagrams at a given order are retained, the two representations must give identical results. The distinction becomes important only after restricting the diagrammatic expansion, as is done in RPA.

To see this explicitly, consider a polarization matrix of the form
\begin{equation}
\Pi(\mathbf q)
=
\frac{\Pi_0(q)}{2}\mathbbm 1 .
\end{equation}
Expanding the RPA interaction to second order in \(U\), representation \(A\) gives
\begin{equation}
      V^{\rm{RPA},A}_{ij}=\begin{pmatrix}
        U &U \\ U &U
    \end{pmatrix}-\Pi_{0}(q)U^2 \begin{pmatrix}
        1 &1 \\ 1 &1
    \end{pmatrix}+\dots,
\end{equation}
whereas representation \(B\) gives
\begin{equation}
      V^{\rm RPA,B}_{ij}=\begin{pmatrix}
        0 &U \\ U &0
    \end{pmatrix}-\frac{\Pi_{0}(q)U^2}{2} \begin{pmatrix}
        1 &0 \\ 0 &1
    \end{pmatrix}+\dots
\end{equation}

These two results expose complementary problems with a naive RPA treatment. In representation \(A\), the resummation contains same-species Hartree contributions that should vanish for a contact interaction. In the complete perturbative expansion these terms are cancelled by the corresponding exchange diagrams, but the latter are absent in RPA. The resulting diagonal contributions are therefore spurious. Including $\emph{ all}$ diagrams we obtain:
\begin{equation}
      V^{\rm{RPA},all}_{ij}=\begin{pmatrix}
        0 &U \\ U &0
    \end{pmatrix}-\frac{\Pi_{0}(q)U^2}{2} \begin{pmatrix}
        1 &1 \\ 1 &1
    \end{pmatrix}+\dots
\end{equation}

Representation \(B\) avoids these spurious bare same-species terms. However, once only the Hartree-like RPA series is retained, the summation no longer preserves the manifest \(SU(2)\) symmetry of the original interaction. In other words, although \(A\) and \(B\) are equivalent representations of the same microscopic contact interaction, the restricted RPA approximation is not invariant under the transformation between them.

This issue is specific to the strictly local part of the interaction. If the interaction has a nontrivial momentum dependence, \(U\rightarrow U(\mathbf q)\), then in general
\begin{equation}
\sum_{\mathbf q}
U(\mathbf q)
:\rho_i(\mathbf q)\rho_i(-\mathbf q):
\neq 0,
\end{equation}
and the above operator identity no longer applies. For a long-ranged Coulomb interaction, the direct particle-hole channel is precisely the one naturally captured by the conventional RPA resummation.

\begin{figure}[h]
\centering
\includegraphics[width=0.7\linewidth]{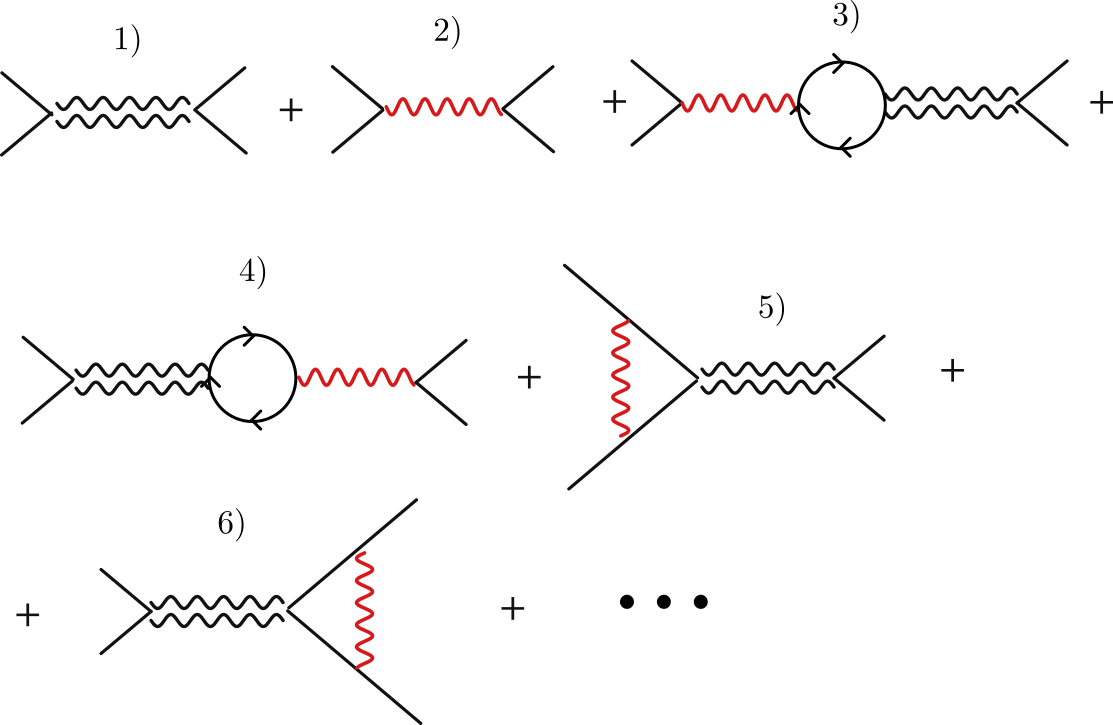}
\caption{Feynman-diagram expansion involving the RPA-screened Coulomb interaction $V_C^{\rm RPA}$ (double wavy lines) and the ultra-short-range interaction $\Delta V_0$ (red wavy lines). Diagrams 1)--4) correspond to the usual RPA interaction expansion, whereas diagrams 5) and 6) represent exchange contributions arising from the ultra-short-range interaction. If the diagonal components of $V_0$ are set to zero, diagrams 5) and 6) vanish. When the diagonal components are retained, however, these exchange diagrams give nonzero contributions that cancel the spurious contribution associated with the diagonal part of $V_0$, thereby recovering the RPA result obtained when the diagonal interaction elements are set to zero.
}
\label{fig:RPA_expansion_first_orderRPA_with_fock}
\end{figure}

\subsection{Perturbative treatment of the short-range interaction}

We seek a consistent prescription for screening the ultra-short-range interaction to first order in $\Delta V$. In particular, the result should preserve SU(2) symmetry whenever the polarization tensor respects this symmetry and should not depend on the same-species component of a momentum-independent interaction $V_0$, which vanishes identically at the operator level. As we show below, to first order in $\Delta V$, both requirements are satisfied by setting the diagonal elements of $\Delta V$ to zero before carrying out the RPA summation.

Accordingly, to first order in $\Delta V$, we adopt the prescription of setting the diagonal elements of the ultra-short-range interaction to zero. We write the interaction as
\begin{equation}
V_{\mathrm{tot}}=V+\Delta V,
\end{equation}
where $V$ is the momentum-dependent Coulomb interaction and $\Delta V$ is the ultra-short-range correction with vanishing diagonal elements. We sum the RPA diagrams to all orders in $V$, while retaining only terms linear in $\Delta V$.

To first order in $\Delta V$, this prescription includes all non-vanishing diagrams involving the ultra-short-range interaction, since the corresponding exchange diagrams vanish identically. The only exchange diagrams omitted at this order are therefore those involving the Coulomb interaction, which are expected to be parametrically smaller than the direct Coulomb contributions. The relevant direct and exchange diagrams are shown explicitly in Fig.~\ref{fig:RPA_expansion_first_orderRPA_with_fock}. Thus, setting the diagonal elements of $\Delta V$ to zero removes the unphysical same-species ambiguity without omitting any non-vanishing contribution involving $\Delta V$ at this order.

For concreteness, consider a two-component system,
\begin{equation}
V_{ij}(\mathbf q)
=
\begin{pmatrix}
V_C(\mathbf q) & V_C(\mathbf q)\\
V_C(\mathbf q) & V_C(\mathbf q)
\end{pmatrix},
\qquad
\Delta V_{ij}
=
\begin{pmatrix}
0 & \Delta V_0\\
\Delta V_0 & 0
\end{pmatrix}.
\end{equation}
For an SU(2)-symmetric polarization function,
\begin{equation}
\Pi(\mathbf q)=\frac{\Pi_0(q)}{2}\mathbbm{1},
\end{equation}
summing the RPA diagrams to all orders in $V_C(\mathbf q)$ and retaining terms only to first order in $V_0$ gives
\begin{equation}
      V^{RPA}_{ij}= \frac{V_C}{1+\Pi_0(\vec q) V_C}\begin{pmatrix}
        1 &1 \\ 1 &1
    \end{pmatrix}+\Delta V_0\begin{pmatrix}
        1 &0 \\ 0 &1
    \end{pmatrix}-\Delta V_0 \frac{V_C \Pi_0(q)(1+V_C \Pi_0(q)/2)}{(1+V_C \Pi_0(q))^2}\begin{pmatrix}
        1 &1 \\ 1 &1
    \end{pmatrix}
\end{equation}
The resulting interaction is isotropic and therefore preserves SU(2) symmetry to this order, explicitly demonstrating that the zero-diagonal prescription is consistent with the symmetry of the underlying problem.

Alternatively, one may retain arbitrary diagonal elements in the bare $V_0$ interaction. In this representation, however, the corresponding Fock diagrams must also be included, and they exactly cancel the unphysical dependence on these diagonal terms to first order in $\Delta V$. Setting the diagonal elements of $\Delta V$ to zero from the outset is therefore an equivalent and simpler prescription at this order, avoiding the need to introduce terms that ultimately cancel.

This prescription is specific to first order in $\Delta V$. Beyond first order, additional diagrams involving the ultra-short-range interaction no longer vanish in general and must be included consistently. We therefore restrict our treatment of $\Delta V$ to first order throughout.

\section{Screening-Length Dependence of the Transition Point}
\label{app:toy_numerics}

As noted in the main text, performing iDMRG calculations with a purely unscreened Coulomb interaction is numerically challenging. We therefore introduce a finite screening length $\zeta$ and consider an interaction of the form
\begin{equation}
    V(r)=\frac{e^2}{4\pi \epsilon_0 \epsilon_{\rm hBN}}\frac{e^{-r^2/(2\zeta^2)}}{r},
\end{equation}
which approaches the unscreened Coulomb interaction in the limit $\zeta\to\infty$. In the DMRG calculations with RPA screening, the screening length $\zeta$ is introduced through
\begin{equation}
V(r)=e^{-r^2/(2\zeta^2)}V_{\mathrm{RPA}}(r),
\qquad
V_{\mathrm{RPA}}(r)
=\mathrm{FT}\left[
\frac{V_C(q)}{1+\Pi_0(q)V_C(q)}
\right]
\end{equation}
where $\mathrm {FT}$ refers to Fourier transform. Thus, the parameter $\zeta$ introduces an additional Gaussian envelope on top of the RPA-screened interaction, controlling its effective spatial range. Since gate distance is larger than screening $\zeta$, we do not include it in the potential.

Ideally, one would perform iDMRG calculations for several values of $\zeta$ and extrapolate the resulting phase boundaries to the limit $\zeta\to\infty$ (as is standard in the literature \cite{Zaletel_2015,papic_dmrg,Zaletel_2013}). Such a scaling analysis is prohibitively expensive given the numerical cost of our iDMRG calculations. Instead, in this appendix we use exact diagonalization (ED) to examine the dependence of the relevant transition points on $\zeta$ and thereby assess their robustness to the finite screening length used in iDMRG.

\begin{figure}[h]
    \centering
\includegraphics[width=0.49\linewidth]{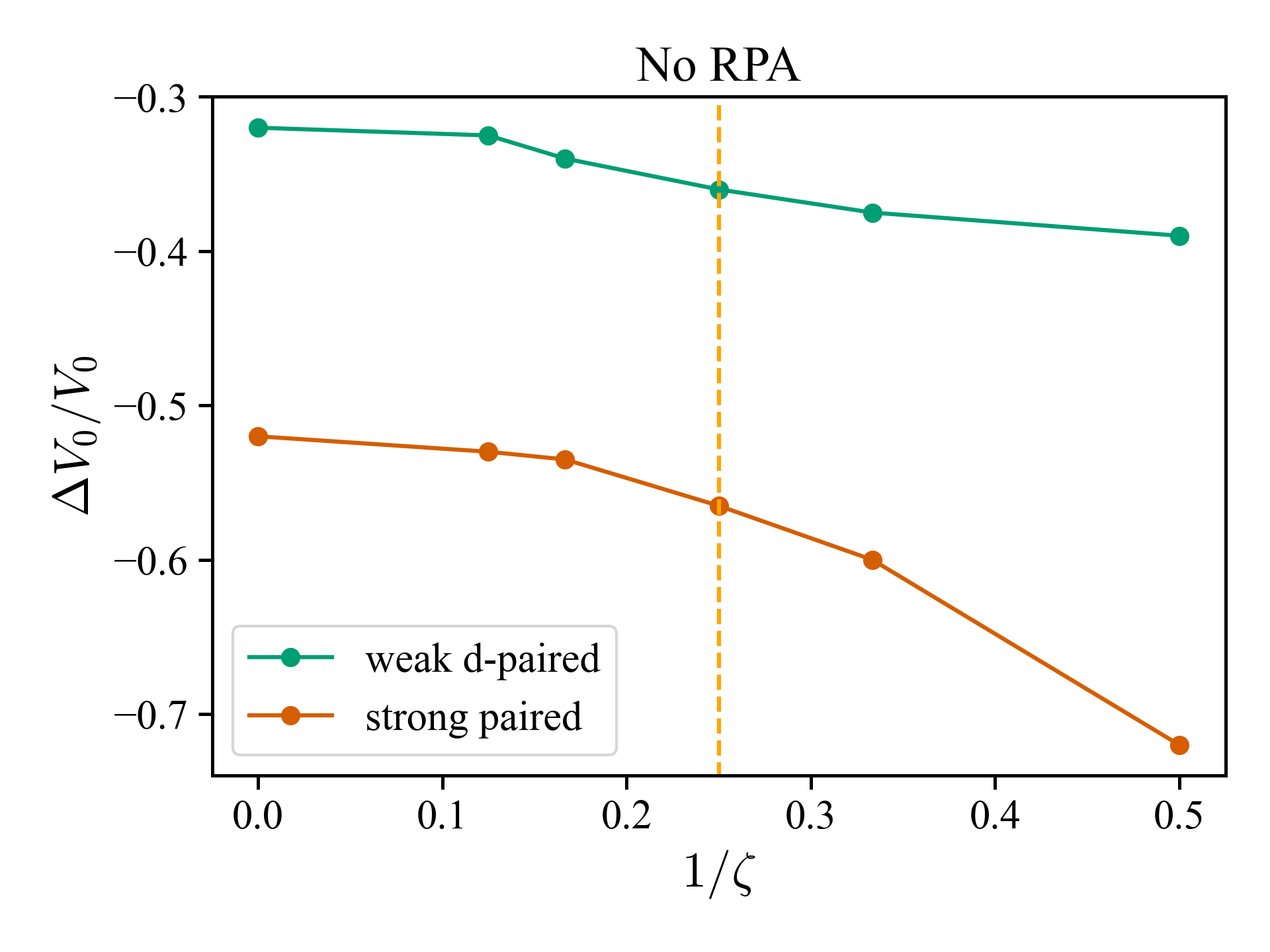}
\includegraphics[width=0.49\linewidth]{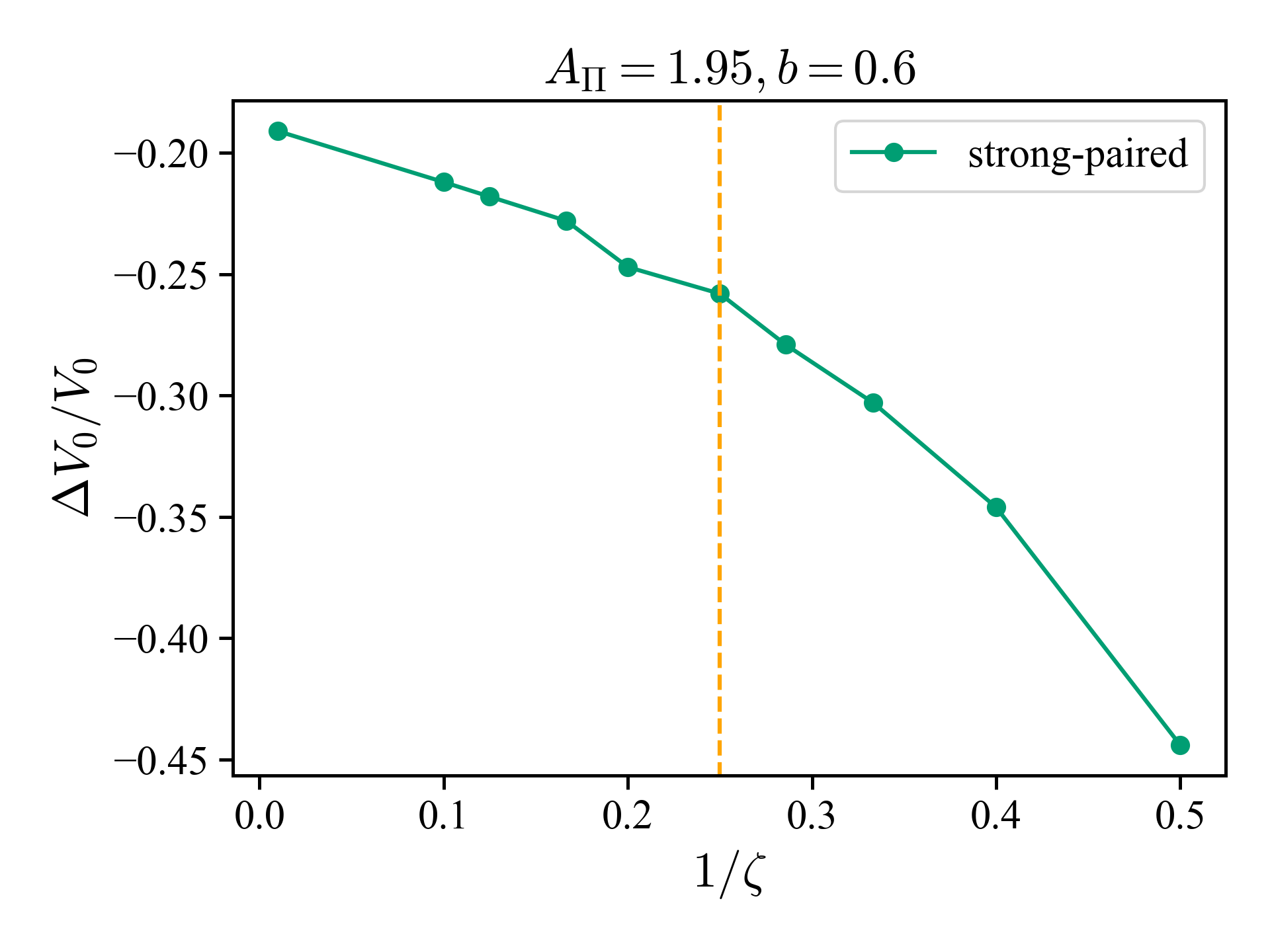}
    \caption{  Transition points $\Delta V_0$ at which the weak- and strong-paired states develop a large overlap ($>0.8$) with the ground state, shown as a function of inverse screening length $1/\zeta$ for $N=10$ electrons without RPA on left, and for $N=8$ electrons with RPA screening on the right. The vertical line indicates $\zeta=4\ell_B$, that is value used in DMRG calculations. }
\label{fig:transition_against_zeta}
\end{figure}

Specifically, for several values of $\zeta$, we calculate the overlaps of the ED ground state with the trial states as a function of $\Delta V_0$, for both the RPA-unscreened and RPA-screened interactions. We identify the transition points from the values of $\Delta V_0$ at which the corresponding trial-state overlap becomes large. The resulting dependence of the transition points on the inverse screening length $1/\zeta$ is shown in Fig.~\ref{fig:transition_against_zeta}.

For all transitions considered, we observe the same systematic dependence on the screening length: decreasing $\zeta$ shifts the transition toward more negative values of $\Delta V_0/V_0$. The magnitude of this shift, however, depends on the transition. For the RPA-unscreened interaction, the weak-to-strong-pairing transition exhibits a substantially stronger dependence on $\zeta$ than the CFL--weak-pairing transition.

We note that the CFL--weak-pairing boundary is intrinsically less sharply defined than weak-strong boundary: the spectral signature of the transition does not coincide precisely with the onset of a large overlap with the weak d-paired trial state. The transition points shown here, which are extracted from the overlap, should therefore be interpreted with this uncertainty in mind.

The same overall trend is observed for the RPA-screened interaction, where the CFL-to-strong-pairing transition shifts toward more negative $\Delta V_0/V_0$ with decreasing $\zeta$. Thus, all three transitions shift in the same direction as the screening length is reduced. This behaviour is expected because shorter-range screening suppresses the higher Haldane pseudopotentials more strongly than $V_0$, requiring a larger reduction of $V_0$ to reach the corresponding transition.

Nevertheless, the difference between the unscreened limit, $\zeta=\infty$, and $\zeta=4$, used in our iDMRG calculations, remains relatively modest. Importantly, extrapolating toward $\zeta=\infty$ suggests that the transitions may occur at smaller $|\Delta V_0|$ than predicted by our finite-$\zeta$ iDMRG calculations.

\section{Additional ED energy spectra data}
\label{app:spectra_data}

As $\Delta V_0$ is decreased further, we also observe a transition from the weak-$d$-paired state to the strong-paired state. Across this transition, approximately between $\Delta V_0/V_0=-0.477$ and $-0.57$, the fifth state in the $K=(0,0)$ sector drops out of the low-energy manifold. This spectral evolution indicates that the transition is second order. Interestingly, a similar trend was observed in Ref.~\cite{kun_yang_weak_to_strong} for pseudopotential interactions. However, the range of zeroth Haldane pseudopotential values over which the fifth state was clearly observable was $\approx 0.5$, substantially larger than in the present case.

\begin{figure}[H]
    \centering
\includegraphics[width=0.65\linewidth]{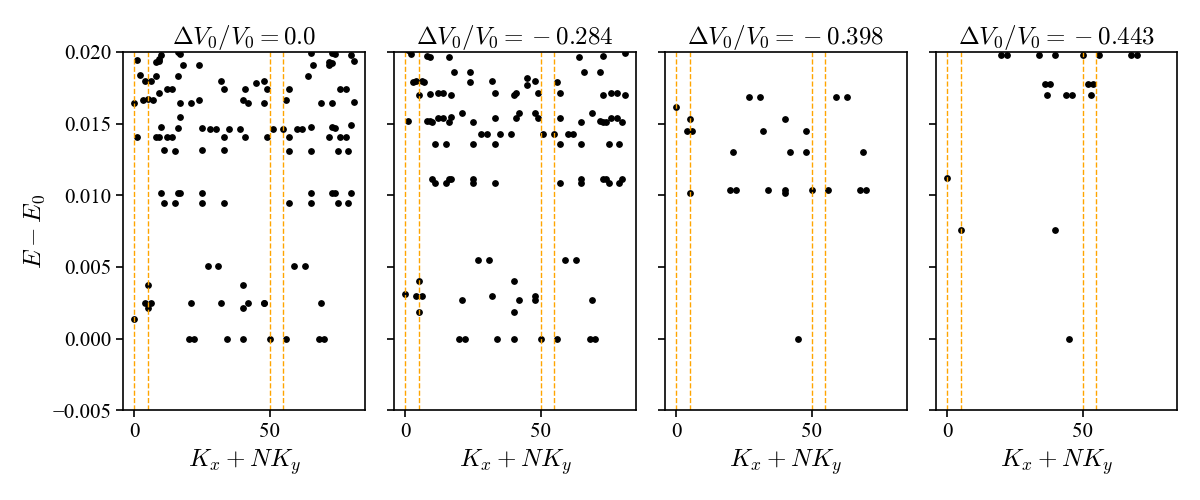}
\includegraphics[width=0.81\linewidth]{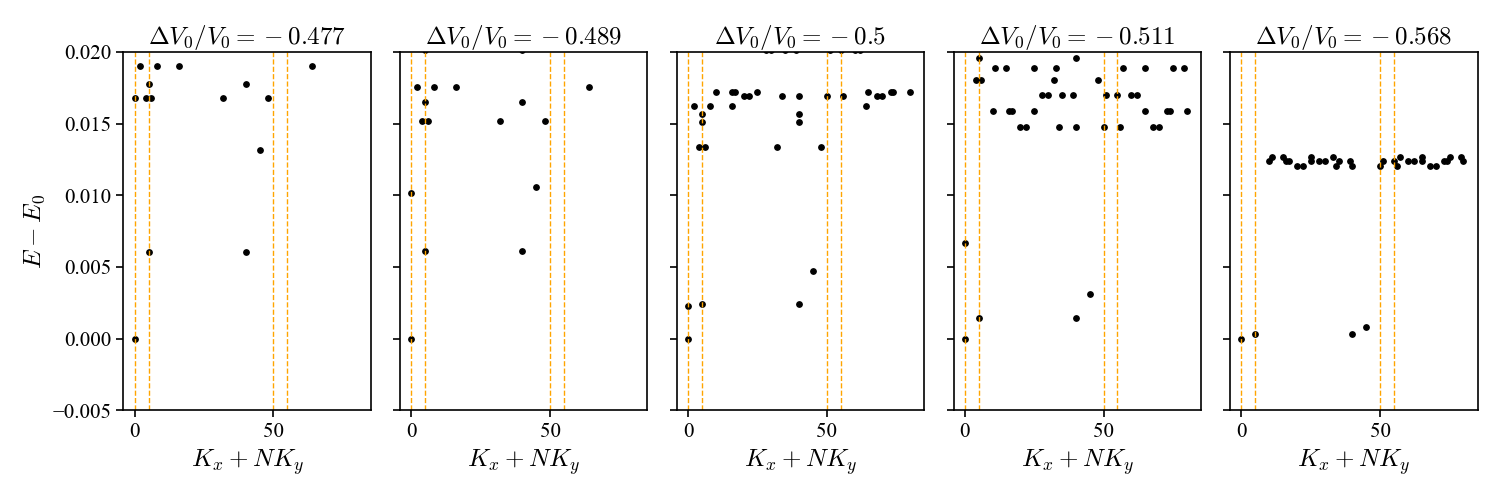}
    
     \caption{Evolution of the low-energy $S_z=0$ spectrum as a function of $\Delta V_0/V_0$ for $N=10$, without RPA screening. The orange dashed lines represent ground state sectors of weak d-paired state and strong paired state. }
    \label{fig:spectra_evolution}
\end{figure}

In this appendix, we present the evolution of the energy spectra as a function of $\Delta V_0/V_0$, shown in Fig.~\ref{fig:spectra_evolution}. At $\Delta V_0/V_0=0$, we observe the degeneracy expected for the CFL~\cite{CFL_N_8}, namely an eightfold ground-state degeneracy. This behavior persists approximately up to $\Delta V_0/V_0 \simeq -0.3$, where the low-energy spectrum starts to mix with states that were previously at higher energies, but no clear topological degeneracy structure is observed. The characteristic degeneracy structure of the weak-$d$-paired state becomes clearly resolved only around $\Delta V_0/V_0 \simeq -0.44$.

\FloatBarrier
\section{Evidence for a Gapless Phase}
\label{app:realistic_numerics}
In this section, we investigate the nature of the ground state in the grey region of the main-text phase diagram, Fig.~\ref{fig:RPA_overlaps}. Using complementary ED and iDMRG calculations, we find evidence that this region is gapless and most likely corresponds to a composite Fermi liquid (CFL) throughout. While ED allows us to probe the broader region of the phase diagram, our DMRG analysis focuses on the representative point $A_{\Pi}=2$, $b=0.6$, and $\Delta V_0=0$.

\begin{figure}
    \centering
\includegraphics[width=0.49\linewidth]{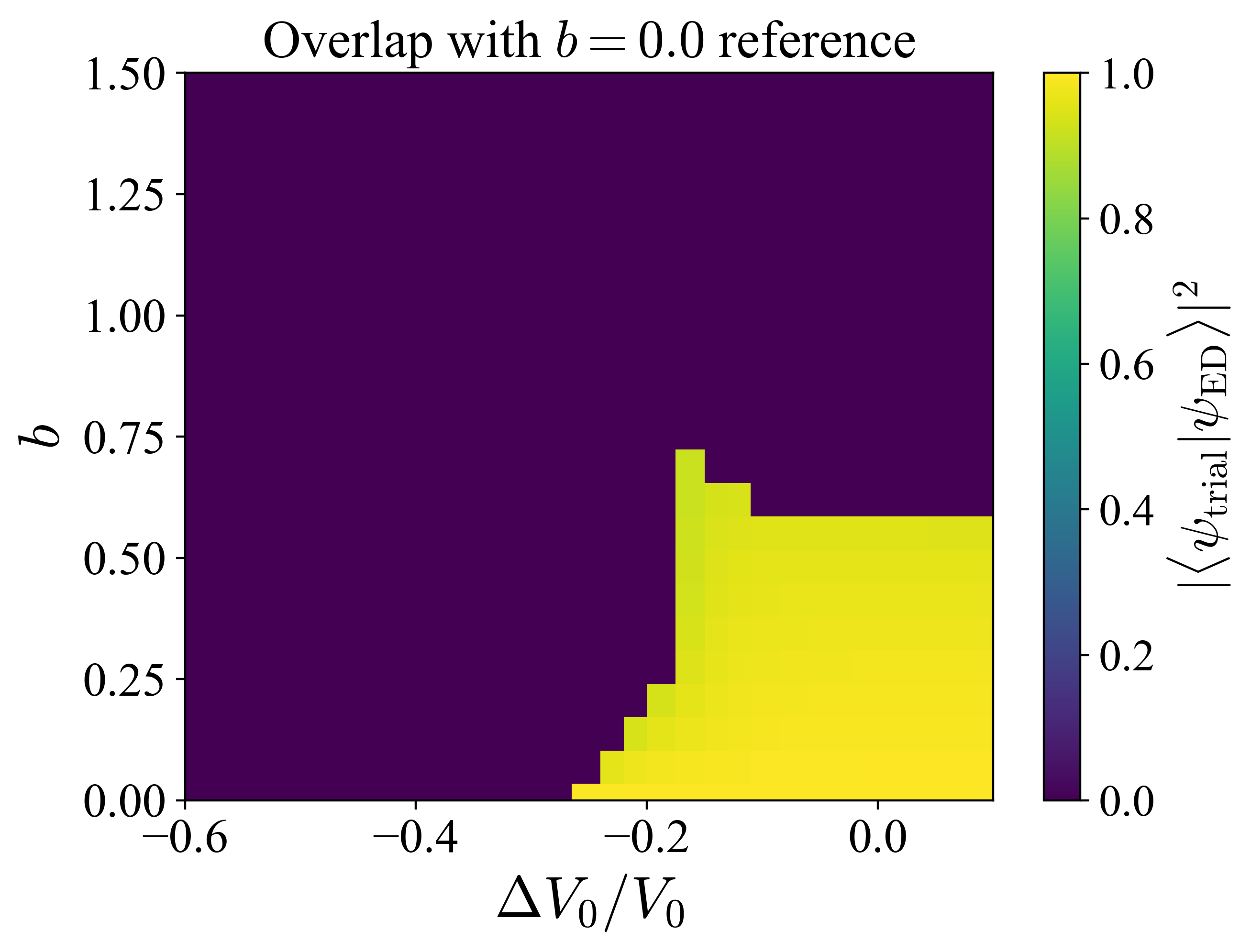}
\includegraphics[width=0.49\linewidth]{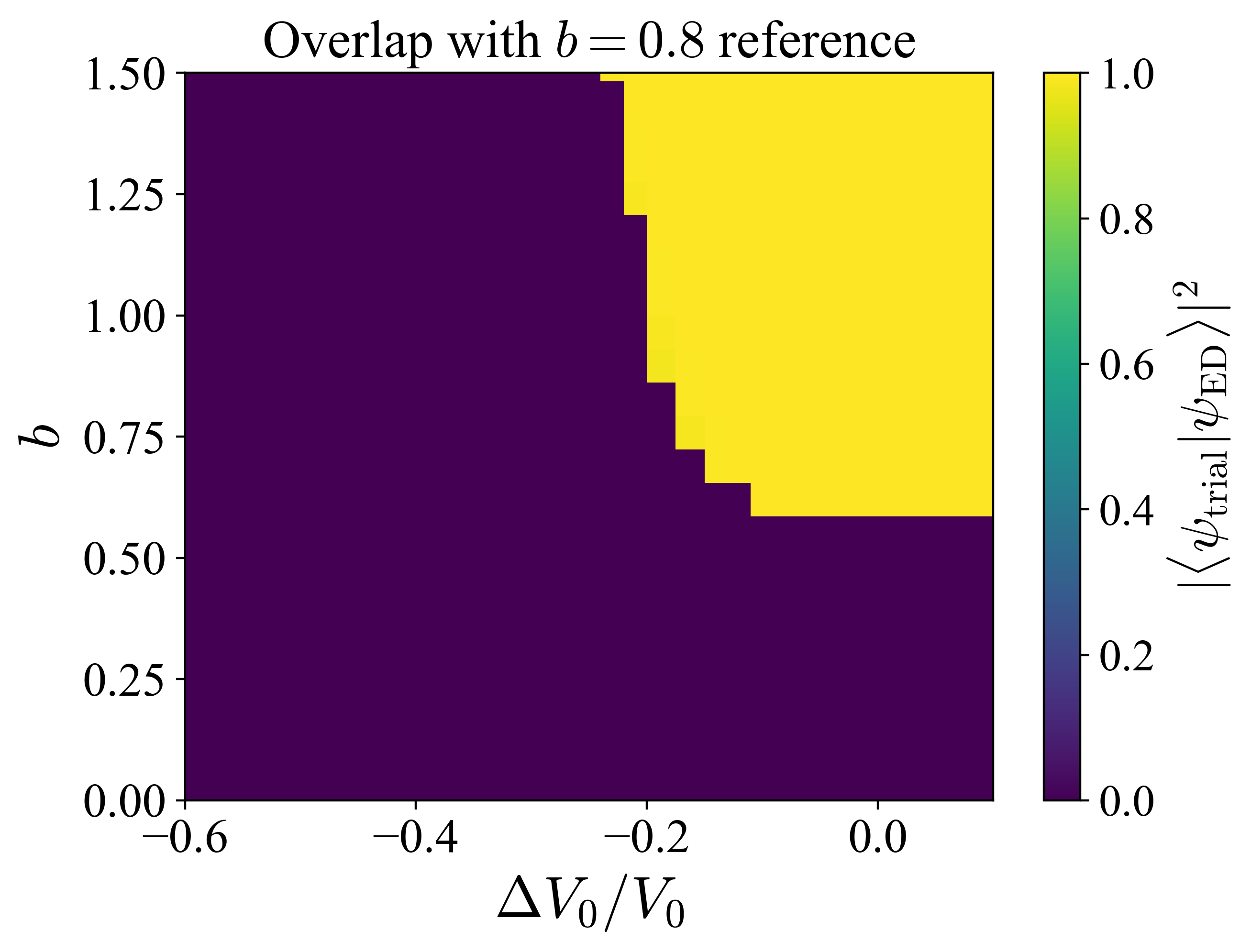}
\caption{Squared overlaps of the ED ground state with reference states in the two gapless regions as a function of $\Delta V_0/V_0$ and the RPA screening parameter $b$, for $A_{\Pi}=2$. \emph{Left:} overlap with the reference CFL state at $b=0$ and $\Delta V_0=0$, corresponding to the unscreened Coulomb interaction. The overlap remains close to unity up to $b\simeq0.6$ and throughout the region extending to the boundary of the weak- or strong-pairing phase. \emph{Right:} overlap with the ground state at $b=0.8$ and $\Delta V_0=0$. The large overlap throughout the region at stronger RPA screening indicates that it is likewise internally adiabatically connected. The abrupt change between the two overlap regions near $b\simeq0.6$ indicates a transition within the gapless part of the phase diagram. The low-energy spectra for both $N=8$ and $N=10$ show no obvious signature of a gapped topological phase on either side of this transition.}
\label{fig:ED_overlap_evidence}
\end{figure}

We begin with the ED results. To test whether the RPA-screened state is continuously connected to the CFL, we calculate its overlap with the ground state at $b=0$ and $\Delta V_0=0$. This point corresponds to the unmodified Coulomb interaction, whose ground state at this filling is well established as a CFL \cite{CFL_N_8,CFL_RezayiRead1994,CFL_Onoda2001}, and therefore provides a natural reference state. The resulting overlap map, analogous to those presented in the main text, is shown in Fig.~\ref{fig:ED_overlap_evidence}. The overlap remains close to unity throughout the region between $\Delta V_0=0$ and the transition into the weak- or strong-pairing phase. On the finite systems accessible to ED, this provides evidence that this region is adiabatically connected to the gapless CFL state, and therefore supports its identification as gapless.

For $b\gtrsim0.6$, the overlap with the reference CFL state abruptly drops to zero. To probe this region, we instead take the ground state at $b=0.8$ and $\Delta V_0=0$ as a reference and calculate the corresponding overlap map. As shown in the right panel of Fig.~\ref{fig:ED_overlap_evidence}, the overlap remains large throughout this region, suggesting that it forms a single phase. We believe that the loss of overlap near $b\simeq0.6$ most likely reflects a transition to a distinct CFL-like state, which remains gapless, as supported by the analysis below. We note that destabilising one CFL ground-state manifold in favour of another can occur relatively easily, since the finite-size spectrum depends sensitively on the shape of the composite-fermion Fermi surface. Consequently, even a small deformation of the Fermi surface can alter the ground-state manifold. In Ref.~\cite{CFL_N_8}, it was shown that introducing a weak mass anisotropy destroys the characteristic ground-state structure of the $N=10$ CFL. The $N=8$ state, on the other hand, may be more robust, since it corresponds to a closed shell of composite fermions.

The low-energy spectrum provides an additional diagnostic of this region. We illustrate its evolution using two representative points, $A_{\Pi}=2$, $\Delta V_0=0$, with $b=0.3$ and $b=0.8$.

At $b=0$, the unscreened case, the ground-state structure is that of the established CFL. For $N=8$ the ground state is non-degenerate, while for $N=10$ we find the characteristic eightfold ground-state manifold followed by a cluster of low-lying excited states, as previously reported \cite{CFL_N_8}. This system-size dependence is naturally understood from the composite-fermion shell-filling structure: the orbital occupations and corresponding Hund-rule filling for the unpolarized CFL at $N=8$ and $N=10$ were worked out in Ref.~\cite{CFL_N_8}, where they were shown to agree with the ground-state multiplicities obtained from ED.

As RPA screening is turned on, this structure begins to deform. For $N=8$, the ground state remains non-degenerate at both $b=0.3$ and $b=0.8$, with a clear gap separating it from the excited spectrum throughout. This is presumably a consequence of the stability of the closed shell of composite fermions at this system size. For $N=10$, the low-lying part of the spectrum already starts to change at $b=0.3$: some of the states that made up the eightfold manifold and the low-lying cluster above it move up in energy while others move down, so that the manifold is no longer exactly degenerate. The spectrum nonetheless remains obviously gapless and retains a structure closely related to the $b=0$ case. 
In particular, if we track a state that forms the sector-$K$ ground state at $b=0$ as $b$ is increased to $0.3$, we typically find that it retains almost $100\%$ overlap with the corresponding $b=0$ state, even though its energy may have changed substantially. This indicates that the low-energy spectrum is being reshuffled, rather than reconstructed from qualitatively different states.
 The same trend continues at $b=0.8$, where the low-lying levels are further redistributed in the same way, but the overall gapless character persists. At this stronger screening, however, the overlap of each sector's ground state with its $b=0$ counterpart is noticeably reduced. The eigenvalues spectra are shown in Fig.~\ref{fig:ED_spectra}.

We additionally find that the finite-size excitation gap is substantially smaller than in the neighbouring paired phases. Although the available system sizes do not allow us to establish the thermodynamic scaling of this gap from ED alone, its small magnitude is consistent with a gapless state. A more detailed characterization would require tracking the evolution of the low-energy spectrum with aspect ratio, which changes the composite-fermion shell-filling structure and provides an additional diagnostic; such an analysis is beyond the scope of the present work.

\begin{figure}
    \centering
    \includegraphics[width=0.49\linewidth]{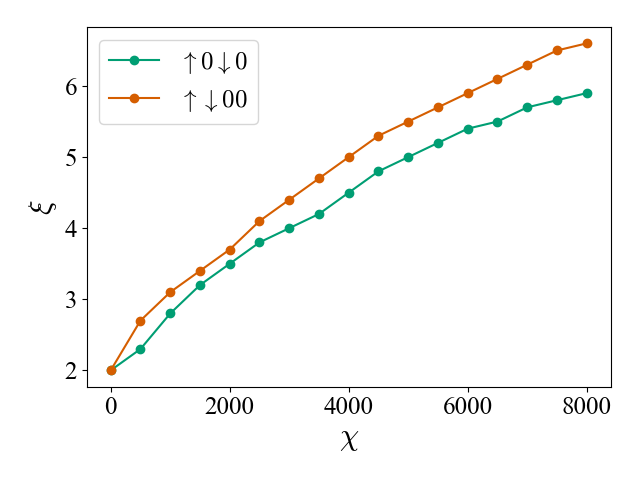}
    \includegraphics[width=0.49\linewidth]{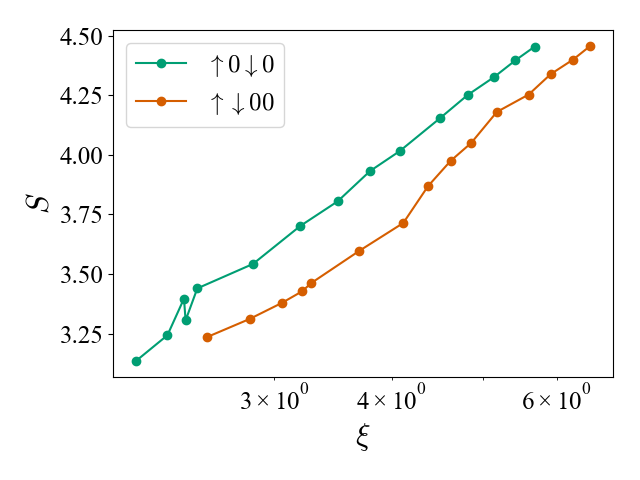}
    \caption{DMRG results at $L_y=15\ell_B$, $A_{\Pi}=2$, $b=0.6$, and $\Delta V_0=0$. \emph{Left:} Correlation length $\xi$ as a function of bond dimension $\chi$ for the two $K$ sectors. The approximately linear growth of $\xi$ with $\chi$, with no indication of saturation over the accessible bond dimensions, is consistent with a gapless state. \emph{Right:} Entanglement entropy $S$ as a function of the correlation length $\xi$. The approximately linear dependence of $S$ on $\log\xi$ is consistent with the finite-entanglement scaling expected for a critical state. We note that the $x$-axis is logarithmic.}
    \label{fig:DMRG_correlation_length}
\end{figure}

\begin{figure}
    \centering
    \includegraphics[width=1\linewidth]{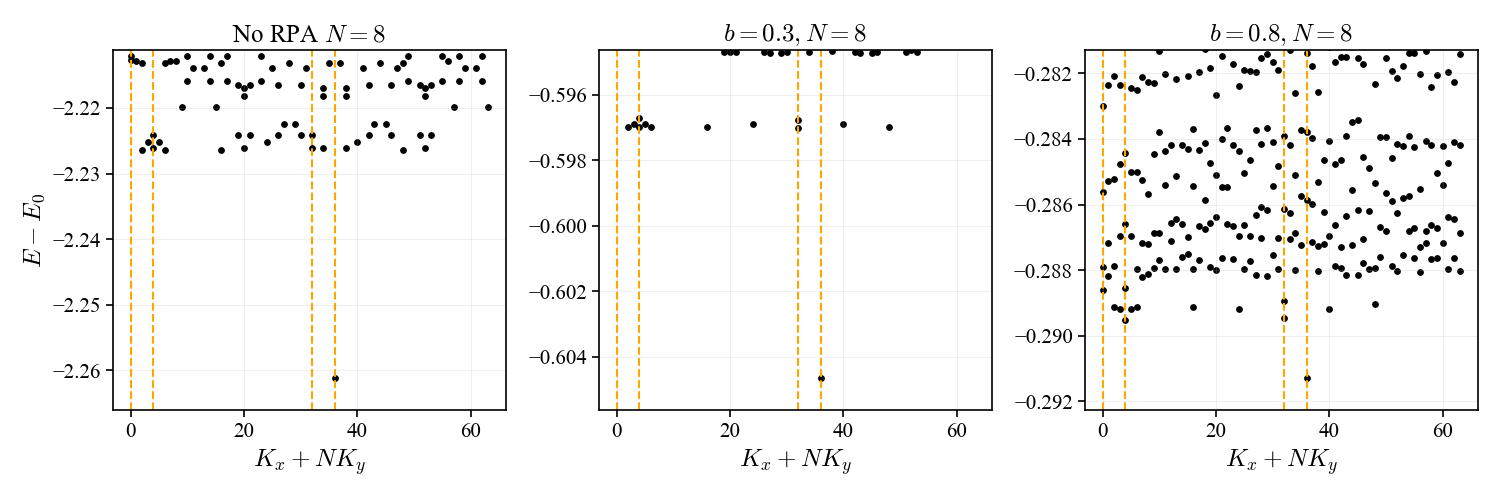}
    \includegraphics[width=1\linewidth]{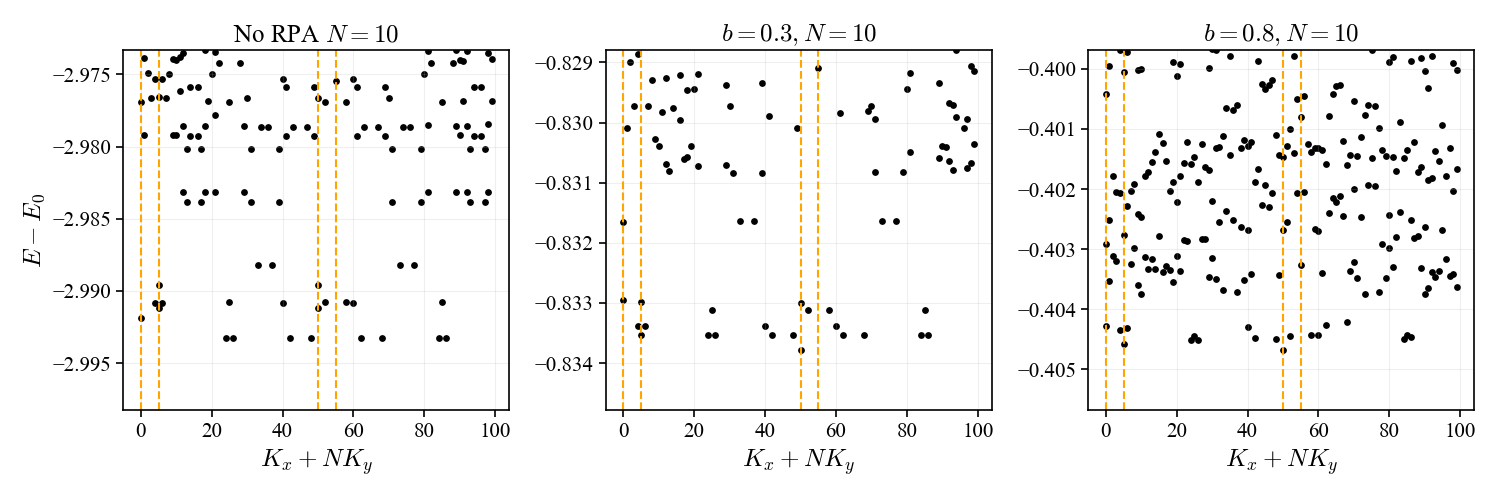}  
   \caption{Low-energy spectra for $N=8$ (top row) and $N=10$ (bottom row) at $A_{\Pi}=2$ and $\Delta V_0=0$, for the unscreened Coulomb interaction (left), $b=0.3$ (middle), and $b=0.8$ (right), plotted against the combined momentum quantum number $K=k_x+Nk_y$. Orange dashed lines mark the momentum sectors corresponding to the topological ground states of the weak $d$-paired and strong-paired phases. For $N=8$, the ground state remains non-degenerate and well separated from the excited spectrum at all values of $b$ shown. For $N=10$, the low-lying energy states change significantly, but the low-lying spectrum shows no apparent topological degeneracy.}
    \label{fig:ED_spectra}
\end{figure}

We next turn to DMRG and focus on the representative point
$A_{\Pi}=2$, $b=0.6$, and $\Delta V_0=0$. We also perform the same analysis at $b=0.8$ and find qualitatively similar behaviour. We first examine the correlation length $\xi$ as a function of the MPS bond dimension $\chi$. Rather than approaching a $\chi$-independent value, $\xi$ increases approximately linearly with $\chi$ over the range of bond dimensions accessible to us, as shown in Fig.~\ref{fig:DMRG_correlation_length}. This absence of saturation is consistent with a critical, rather than a gapped, ground state. Furthermore, the correlation lengths in the two $K$ sectors are distinct, ruling out the Halperin 331 state as well as the weak $d$-paired state and the strong paired state.

Further evidence is provided by the scaling of the entanglement entropy with the correlation length. For a quasi-one-dimensional critical state represented by an infinite MPS, the bipartite entanglement entropy is expected to scale with the finite-entanglement correlation length as
\begin{equation}
    S = \frac{c_{\mathrm{eff}}}{6}\log\xi + s_0,
    \label{eq:finite_entanglement_scaling}
\end{equation}
where $c_{\mathrm{eff}}$ is the central charge of the effective one-dimensional critical theory. Our data exhibit the expected linear dependence of $S$ on $\log\xi$, providing further evidence that the state is gapless. The result is shown in Fig.~\ref{fig:DMRG_correlation_length}.

We emphasize that Eq.~\eqref{eq:finite_entanglement_scaling} by itself establishes criticality but does not uniquely identify the state as a CFL. A more detailed entanglement entropy analysis as a function of cylinder circumference could provide additional information about the underlying gapless modes. For example, Ref.~\cite{Geraedts2016} showed that, for the Dirac CFL on a cylinder, the extracted central charge grows with the number of composite-fermion wires intersecting the Fermi surface, with the emergent gauge field removing one collective mode. Carrying out an analogous mode-counting analysis for the present system would provide a considerably stronger DMRG diagnostic of the CFL, but is beyond the scope of this work.

Finally, we examine the low-lying entanglement spectrum as the system size is varied. In contrast to the neighbouring gapped paired phases, we do not observe a robust low-lying counting structure that remains stable with system size. While the absence of such a pattern is not, by itself, a diagnostic of gaplessness, it is consistent with the critical behaviour inferred from the correlation-length and entanglement-entropy scaling.

The DMRG and ED results provide strong evidence that the grey region in Fig.~\ref{fig:RPA_overlaps} is gapless. We repeat the DMRG analysis at $b=0.3$, $b=0.6$, and $b=0.8$ and find the same gapless behaviour.

\FloatBarrier
\section{Widths of the Incompressible Regions}
\label{app:incompressible_widths}

In this appendix, we estimate the range of band splittings over which the incompressible paired states remain stable. Experimental data for the charge gap as a function of displacement field $D$ at $\nu=7/2$ are available in Ref.~\cite{chanda}. Using the corresponding band dispersion, we convert $D$ first to the interlayer potential $\Delta_1$ and then to the band splitting $\mu$. We find that the incompressible region extends over a total range
\begin{equation}
\Delta\mu \approx 1.25 \,{\rm meV},
\end{equation}
while the full width at half maximum of the charge gap corresponds to
\begin{equation}
\Delta\mu_{\rm FWHM} \approx 0.25 \,{\rm meV}.
\end{equation}
Thus, taking the center of the incompressible region as $\mu=0$, the experimentally observed state persists approximately over $|\mu|<\Delta\mu/2$.

We evaluate this stability range numerically using exact diagonalization (ED). Starting from Eq.~\eqref{eq:hamiltonian_main_ref} we vary the interaction parameter $\Delta V_0$ and the splitting $\mu$ between the two bands. For each fixed value of $\Delta V_0$ within the incompressible phase, we determine the range of $\mu$ over which the unpolarized incompressible state (JSS or strong paired state) remains the ground state.

\begin{figure}[H]
    \centering
    \includegraphics[width=0.49\linewidth]{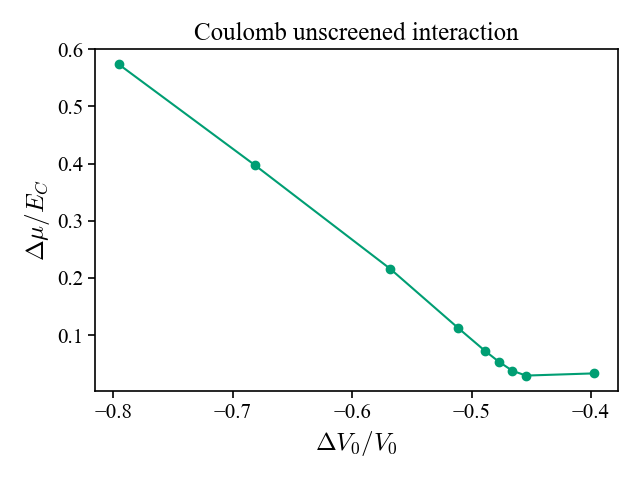}
     \includegraphics[width=0.49\linewidth]{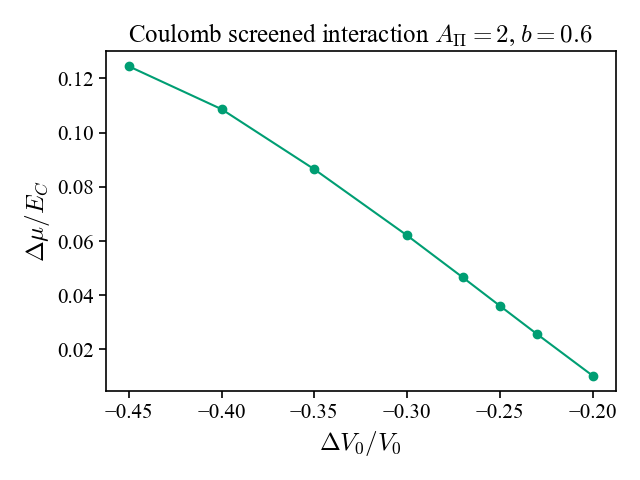}
   \caption{
Stability width $\Delta\mu$ of the unpolarized incompressible paired state as a function of the interaction perturbation $\Delta V_0/V_0$, obtained from exact diagonalization at $N=8$. The width $\Delta\mu$ is determined by the value of the band splitting at which the $S_z=0$ ground state crosses a state in an $S_z\neq0$ sector. 
\emph{Left:} Unscreened Coulomb interaction perturbed by $\Delta V_0$. 
\emph{Right:} RPA-screened Coulomb interaction with $A_{\Pi}=2$ and $b=0.6$, also perturbed by $\Delta V_0$. 
For the unscreened interaction, $\Delta\mu\simeq0.03$--$0.5\,E_C$, while for the RPA-screened interaction the corresponding physical scale is $\Delta\mu\simeq0.5$--$6\,\mathrm{meV}$ for $E_C=50\,\mathrm{meV}$ at $B=12\,\mathrm{T}$. These estimates are of the same order of magnitude as the experimentally inferred stability window.} 
    \label{fig:width_states_ED}
\end{figure}

Since the band splitting is equivalent to adding a term $\mu \hat S_z$ to the Hamiltonian, and $[H,\hat S_z]=0$, the eigenstates within each $S_z$ sector are unchanged as $\mu$ is varied. The only effect of increasing $\mu$ is therefore to shift the energies of different $S_z$ sectors relative to one another, until a state with $S_z\neq0$ crosses the $S_z=0$ ground state. The value of $\mu$ at this crossing determines the half-width $\Delta\mu/2$ of the incompressible region.
We perform this analysis for two interactions: the unscreened Coulomb interaction perturbed by $\Delta V_0$, and the RPA-screened Coulomb interaction with $A_{\Pi}=2$ and $b=0.6$, also perturbed by $\Delta V_0$. The resulting stability width $\Delta\mu$ is plotted as a function of $\Delta V_0/V_0$ in Fig.~\ref{fig:width_states_ED}. 

For the unscreened Coulomb interaction, we find that $\Delta\mu$ varies substantially with $\Delta V_0$, spanning the range $\Delta\mu\approx0.03-0.5\,E_C$. Taking $E_C\approx30\,{\rm meV}$, where we heuristically account for RPA screening by increasing $\epsilon_r$, this corresponds to $\Delta\mu\approx 1-15\,{\rm meV}$. Although this is a broad range, it includes the experimentally estimated value. We note, however, that the upper end of this range corresponds to very large values of $\Delta V_0$ that are unlikely to be physically realised. The physically relevant estimate is therefore expected to lie closer to the lower end of this range.

Similarly, we analyse the RPA-screened interaction. We obtain $\Delta \mu \approx 0.5$--$6\,\mathrm{meV}$, which is of the same order of magnitude as the experimental values. Here, we use the unscreened interaction, $E_C=50\,\mathrm{meV}$ at $B=12\,\mathrm{T}$, since the RPA explicitly accounts for the additional screening.

A more quantitative determination of this stability window would require calculations at larger system sizes, as finite-size effects are expected to be significant for the $N=8$ system considered here. In particular, DMRG calculations as a function of $\mu$ would provide an independent estimate of the extent of the incompressible region and help determine whether a gapped interpolation beyond the states considered here is possible. 

For the Halperin 331 state, a gapped interpolation can be constructed between the unpolarized $S_z=0$ sector and the fully polarized Pfaffian state. We do not know of an analogous interpolation for the singlet paired states considered here. In particular, antisymmetry of the electronic wave function constrains the $d$-wave paired composite-fermion state to be a pseudospin singlet. It nevertheless remains possible that a gapped interpolation to a polarized state exists outside the family of states considered here. If so, the actual range of $\mu$ over which the incompressible phase survives could be larger than the estimate obtained above.

\FloatBarrier
\section{Explicit Landau-Level Mixing}
\label{app:explicit_ll_mixing}

In this appendix, we attempt to account explicitly for Landau-level (LL) mixing to all orders by including four Landau levels in the iDMRG calculation. In particular, we consider a system consisting of the \(\uparrow\) 0LL and \(\downarrow\) 0LL at energy \(0\), together with the \(\uparrow\) 2LL and \(\downarrow\) 2LL at energy \(\mu\), at total filling \(\nu=1/2\). The 2LLs are intended to model the empty bands immediately above the 0LLs in dispersion given by Ref.~\cite{chanda}.

\begin{figure}[h]
    \centering
\includegraphics[width=0.44\linewidth]{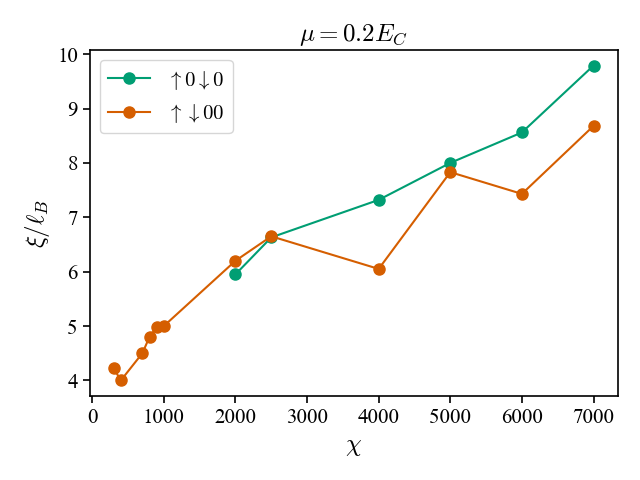}
\includegraphics[width=0.44\linewidth]{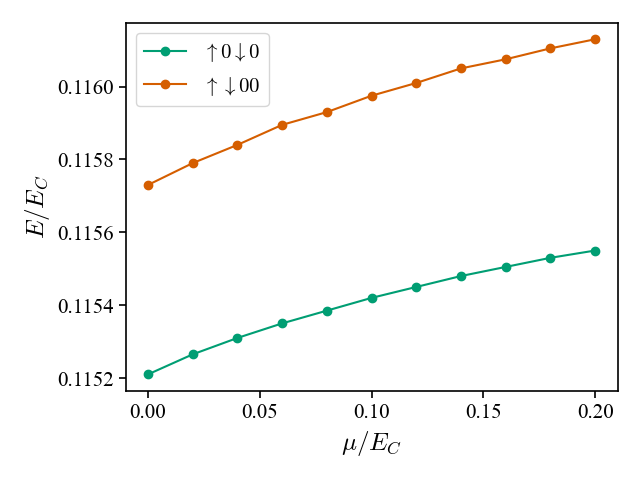}
\includegraphics[width=0.44\linewidth]{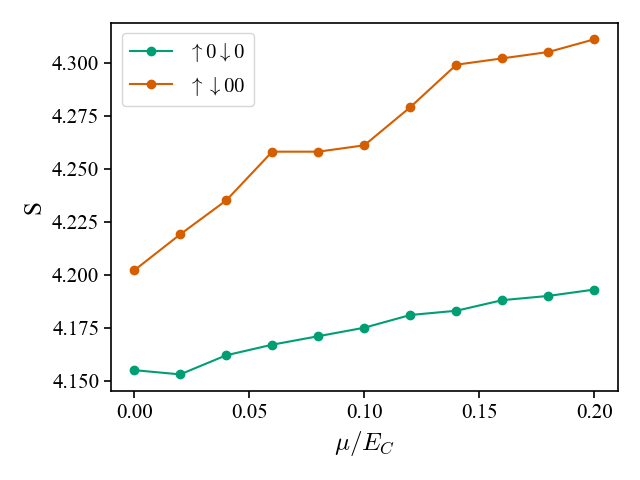}
\includegraphics[width=0.44\linewidth]{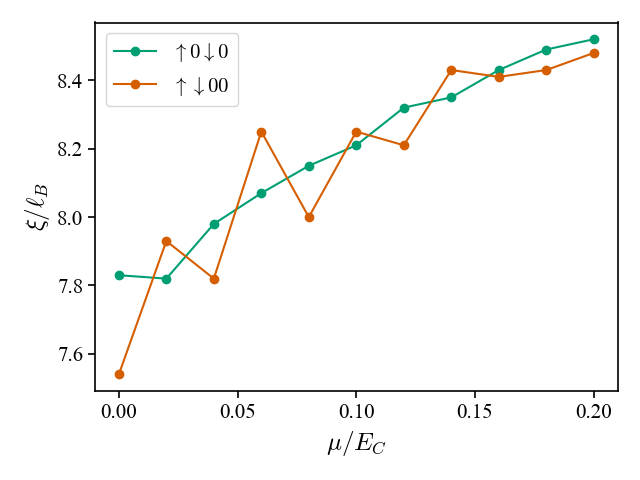}
    \caption{
Diagnostics of Landau-level mixing obtained from iDMRG at $L_y=15\ell_B$ and screening length $\zeta=2\ell_B$. 
Top left: correlation length $\xi$ as a function of bond dimension $\chi$ at fixed $\mu=0.2E_C$. 
Top right, bottom left, and bottom right show, respectively, the ground-state energy, entanglement entropy, and correlation length as functions of $\mu/E_C$ at fixed $\chi=6000$, for the two momentum sectors shown. 
All quantities evolve smoothly with $\mu$, with no clear indication of a phase transition. The correlation length remains large throughout the range studied and increases strongly with $\chi$ at fixed $\mu=0.2E_C$, consistent with gapless behaviour.
}
    \label{fig:LL_mixing_data}
\end{figure}

We fix the cylinder circumference to \(L_y=15\ell_B\) and use a Gaussian Coulomb interaction with \(\zeta=2\ell_B\). Particles are allowed to redistribute between the bands, while the total many-body momentum \(K\) is conserved. We then track several diagnostics as a function of \(\mu\): the entanglement spectrum, correlation length, ground-state energy, and entanglement entropy.

\begin{figure}
    \centering
\includegraphics[width=\linewidth]{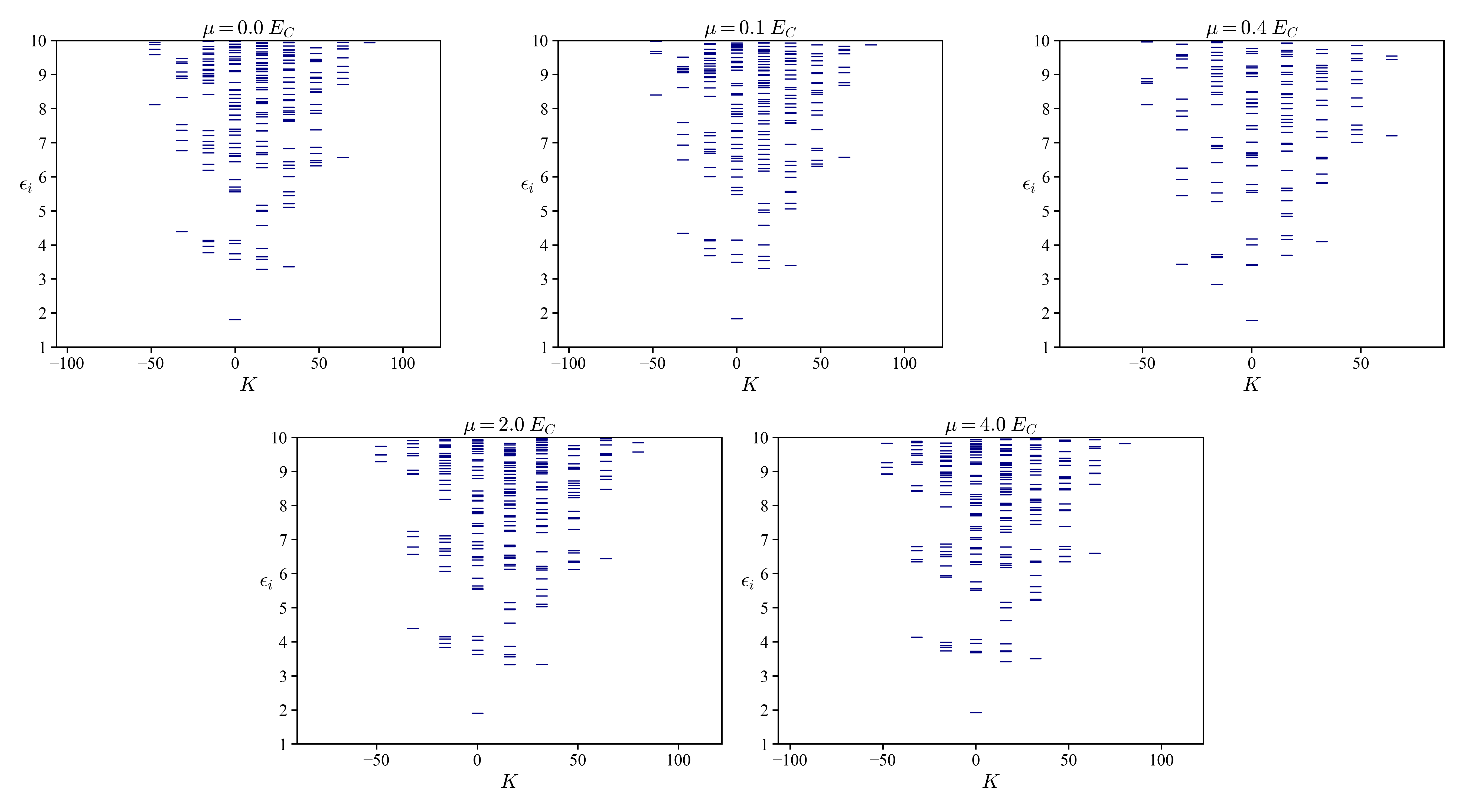}
  \caption{Entanglement spectra at $L_y = 15$ as the chemical potential $\mu$ is varied from $\mu = 0.0\,E_C$ to $\mu = 4.0\,E_C$. The low-lying part of the spectrum shows no discernible change across this range of $\mu$, and the level counting does not resemble any of the gapped trial states.}
    \label{fig:ES_LL_mixing}
\end{figure}

We find no clear evidence of a gapped phase for any values of $\mu$, or any phase transition as $\mu$ is varied. The low-lying structure of the entanglement spectrum changes only weakly (see Fig.~\ref{fig:ES_LL_mixing}), and none of the other diagnostics exhibits discontinuous or diverging behaviour (see Fig.~\ref{fig:LL_mixing_data}). Furthermore, the correlation length remains large, \(\xi \gtrsim 6\ell_B\), throughout the parameter range studied. At fixed \(\mu=0.2 E_C\) (i.e. extremely large Landau level mixing), \(\xi\) also appears to increase rapidly with bond dimension $\chi$, suggesting that the state is gapless.

These calculations are numerically demanding, with individual runs requiring approximately two weeks, and we reach bond dimensions of up to \(\chi=6000\). Consequently, we cannot rule out a phase transition that becomes apparent only at larger bond dimension or for larger screening length \(\zeta\). 

\FloatBarrier

\section{Monolayer Graphene}
\label{app:monolayer}
In this appendix, we summarise the microscopic and experimental estimates of the anisotropy parameters entering our two-band description. 

First, we provide more detailed estimates of the microscopic short-range corrections derived in Appendix~\ref{app:short_range} and projected onto the two-band crossing; the results are shown in Table~\ref{tab:MLG_first_principles}. We use $B=30\,{\rm T}$ and the bare Coulomb interaction. The lattice-scale correction is dominant, while the electron--phonon contribution is comparatively small. As noted in the main text, these microscopic estimates are considerably smaller than the correction required to reach the strong-pairing phase.

\begin{table}[t]
\centering
\begin{tabular}{lcc}
\hline
Perturbation
& $\Delta V_0/V_0$ (unscreened)
& $\Delta V_0/V_0$ (RPA-screened)
\\ \hline
Lattice-scale
& $-0.034$
& $-0.017$
\\
Electron--phonon coupling
& $0.003$
& $0.004$
\\ \hline
\end{tabular}
\caption{First-principles estimates of the correction to the zeroth
pseudopotential in monolayer graphene, before and after RPA screening.
The RPA screening is evaluated using the bare dispersion.}
\label{tab:MLG_first_principles}
\end{table}

We therefore turn to the values of $g_z$ and $g_\perp$ inferred from experiment. These values can be extracted by tracking the positions of the different phase transitions in the $\nu=0$ phase diagram; the relevant candidate states are summarised in Table~\ref{tab:MLG_phases}, originally calculated in Ref.~\cite{kharitonov}. The relevant phases are the ferromagnetic (F), charge-density-wave (CDW), canted antiferromagnetic (CAF), and partially spin-polarised (PSP) states. In addition, at $\nu=1/2$, $g_z$ can be extracted from the data by identifying the magnetic field at which the incompressible state appears. The extracted values are summarised in Table~\ref{tab:MLG_experimental}. We use the notation $u_{\perp,z}:=g_{\perp,z}(a_0/\ell_B)E_C$, while $\epsilon_Z$ denotes the Zeeman energy. Here onwards, we use $\epsilon_{\rm hBN}=4.3$ as value for hBN, and assume it ranges $4-4.6$, when calculating range of $g_z,g_{\perp}$.

\begin{table}[h]
\centering
\renewcommand{\arraystretch}{1.6}
\setlength{\tabcolsep}{6pt}
\begin{tabular}{|c|c|c|}
\hline
phase & energy & condition \\
\hline
F & $-2\epsilon_Z - 2u_\perp - u_z$ & \\
\hline
CDW & $u_z - 2\Delta_{AB}$ & \\
\hline
CAF & $-u_z - \dfrac{\epsilon_Z^2}{2|u_\perp|}$ & $0 < -\dfrac{\epsilon_Z}{2u_\perp} < 1$ \\
\hline
PSP & $u_\perp - \dfrac{\Delta_{AB}^2}{u_z + |u_\perp|}$ & $\dfrac{\Delta_{AB}}{u_z + |u_\perp|} < 1$ \\
\hline
\end{tabular}
\caption{Energies of the relevant phases of monolayer graphene at $\nu=0$, adapted from Ref.~\cite{Zibrov2018}.}
\label{tab:MLG_phases}
\end{table}

First, Ref.~\cite{Young2014} presents the F--CAF transition at $\nu=0$, which gives us a direct estimate of $g_\perp$. Equating the corresponding energies in Table~\ref{tab:MLG_phases} gives
\begin{equation}
\epsilon_Z=-2u_\perp=-2g_\perp\frac{a_0}{\ell_B}E_C.
\end{equation}
The transition is extracted from Fig.~1 of Ref.~\cite{Young2014}, which shows the longitudinal conductivity as a function of the in-plane magnetic field $B_{\parallel}$ for several fixed values of the perpendicular field $B_{\perp}$. For each $B_{\perp}$, the transition is identified from the saturation of the conductivity as $B_{\parallel}$ is increased. We extract the corresponding transition fields from the different curves and summarise the resulting values in Table~\ref{tab:MLG_g_perp}.

\begin{table}
\centering
\begin{tabular}{|c|c|c|}
\hline
$B_{\perp}$ (T) & $B_T$ (T) & $g_\perp$ \\
\hline
0.75 & 24.0 & -9.12 \\
1.00 & 25.0 & -7.13 \\
1.40 & 29.5 & -6.01 \\
1.60 & 30.5 & -5.43 \\
2.00 & 31.0 & -4.42 \\
\hline
\end{tabular}
\caption{Values of $g_\perp$ extracted from the F--CAF transition in Ref.~\cite{Young2014}, using $\epsilon_{\rm hBN}=4.3$. The transition field $B_T$ is determined from the saturation of the conductivity as a function of $B_{\parallel}$ for fixed $B_{\perp}$.}
\label{tab:MLG_g_perp}
\end{table}

At $\nu=1/2$, the crossing condition given in the main text,
\begin{equation}
\Delta_{AB}
=
g_z\frac{a_0}{\ell_B}E_C,
\label{eq:gz_extraction}
\end{equation}
allows $g_z$ to be extracted from the magnetic field at which the incompressible state appears. For Ref.~\cite{Zibrov2018}, using $\Delta_{AB}=35\,{\rm meV}$ and $B=30\,{\rm T}$ gives $g_z\approx9.3$. Similarly, Ref.~\cite{magnon} gives $\Delta_{AB}=3.8\,{\rm meV}$ and $B=8.4\,{\rm T}$, yielding $g_z\approx4.2$. Varying $\epsilon_{\rm hBN}$ over the range used in experiments, we estimate $g_z \approx 3.9$--$4.5$.

Finally, Ref.~\cite{magnon} also probes the $\nu=0$ state by increasing the magnetic field, observing a transition from CDW through PSP to CAF. Since neither the CDW nor PSP phase supports magnon transport, the onset of magnon transport signals the PSP--CAF transition. Equating the corresponding energies in Table~\ref{tab:MLG_phases} gives
\begin{equation}
u_\perp-\frac{\Delta_{AB}^2}{u_z+|u_\perp|}
=
-u_z-\frac{\epsilon_Z^2}{2|u_\perp|}.
\end{equation}
The transition is observed in the range $B=5$--$7\,{\rm T}$, giving $g_\perp\in[-2.15,-0.4
]$ for $g_\perp<0$. Here, we cut the solution off at the point where condition $0<-\frac{\epsilon_Z}{2u_{\perp}}<1$ breaks down.

It is furthermore possible that the PSP phase is completely skipped, such that the system undergoes a direct CDW--CAF transition. Using the CDW--CAF boundary gives a fairly similar value of $g_\perp$. Importantly, we note that a lot of the pairs of values directly obtained from experiments in Ref.~\cite{magnon} are not consistent with PSP-CAF or CDW-CAF transition at all. We discard such numbers.

Finally, we discuss the RPA parameters used for the phase diagram in Fig.~\ref{fig:phase_diagram_monolayer}. We use $A_\Pi=0.5$ and $b=0.4$, obtained from the fit to the bare dispersion. The corresponding zeroth pseudopotential of the screened Coulomb interaction is $V_0\approx0.5$. We furthermore use
\begin{equation}
\frac{\Delta V_0}{V_0}=-0.27,
\end{equation}
as a conservative estimate for the onset of incompressible behaviour. Stronger screening would be more favourable, as it reduces both the Coulomb pseudopotential and the magnitude of the short-range correction required to reach the strong-pairing transition.

\begin{table}[t]
\centering
\begin{tabular}{|l|c|c|}
\hline
Experiment
& Observable
& Extracted parameter
\\ \hline
Young et al.~\cite{Young2014}
& F--CAF transition at $\nu=0$
& $g_\perp=[-9.8,-4.1]$
\\
Zibrov et al.~\cite{Zibrov2018}
& $\nu=1/2$ incompressibility
& $g_z=[8.7-10]$
\\
Magnon experiment~\cite{magnon}
& $\nu=1/2$ incompressibility
& $g_z=[3.9-4.5]$
\\
Magnon experiment~\cite{magnon}
& PSP--CAF transition at $\nu=0$
& $g_\perp=[-2.15,-0.4]$
\\ \hline
\end{tabular}
\caption{Values of the anisotropy parameters $g_z,g_{\perp}$ extracted from experiments.}
\label{tab:MLG_experimental}
\end{table}

\end{document}